\documentclass[letterpaper,11pt]{article}
\usepackage{jhepmod} 

\def\la{\langle}
\def\ra{\rangle}

\def\be{\begin{equation}}
\def\ee{\end{equation}}
\def\ben{\begin{eqnarray}}
\def\een{\end{eqnarray}}

\def\oh{\bf \hat\Omega}

\def\bk{{\bf k}}

\def\rD{{\rm D}}

\def\bk{{\bf k}}

\def\bl{{\bf l}}
\def\bx{{\bf x}}
\def\2p{{(2\pi)^2}}

\def\bl{{\bf l}}
\def\be{\begin{equation}}
\def\ee{\end{equation}}
\def\beq{\begin{equation}}
\def\eeq{\end{equation}}
\def\ben{\begin{eqnarray}}
\def\een{\end{eqnarray}}
\def\bes{\begin{subequations}}
\def\ees{\end{subequations}}

\def\oh{{\hat\Omega}}

\newcommand{\beqa}{\begin{eqnarray}}
\newcommand{\eeqa}{\end{eqnarray}}
\newcommand{\rP}{{P}}

\def\ikap0{{\cal J}_{\theta_0}(r)}

\def\one1{\langle \kappa_{(i)}\kappa_{(j)} \rangle}
\def\one{{[\bar \xi^{(ij)}]}}

\def\ba{\begin{eqnarray}}
\def\ea{\end{eqnarray}}

\def\bk{{\bf k}}
\def\bq{{\bf q}}
\def\rL{{\rm L}}
\def\rD{{\rm D}}

\def\bk{{\bf k}}

\def\bl{{\bf l}}

\def\bx{{\bf x}}

\def\2p{{(2\pi)^2}}

\def\bl{{\bf l}}

\def\be{\begin{equation}}
\def\ee{\end{equation}}

\def\beq{\begin{equation}}
\def\eeq{\end{equation}}

\def\ben{\begin{eqnarray}}
\def\een{\end{eqnarray}}

\def\oh{{\hat\Omega}}

\def\bk{{\bf k}}

\def\bl{{\mathbf l}}

\def\bx{{\bf x}}

\def\2p{{(2\pi)^2}}

\def\bl{{\bf l}}
\def\neweq{\stackrel{\text{sq}}{\approx}}

\def\kp{{k_{\perp}}}
\def\qp{{q_{3\perp}}}

\def\bl{{\bf{l}}}

\def\xd{{\rm d}}

\def\xxi{{\rm i}}

\def\bkl{{\bf l}}

\def\bql{{\bf l}}

\def\muq{{\,\mu_{13}\,}}
\def\musq{{\,\mu^2_{13}\,}}

\usepackage{booktabs}
\usepackage[english]{babel}
\usepackage{amsmath,amssymb,amsbsy,amstext, amsthm, simplewick, amsfonts}
\usepackage{hyperref}
\usepackage{graphicx}
\usepackage{subcaption}

\usepackage{siunitx}
\usepackage{upgreek}
\usepackage{framed}
\usepackage{wrapfig}
\usepackage{multirow}
\usepackage{bbm}
\usepackage[svgnames,dvipsnames,x11names]{xcolor}
\usepackage[utf8x]{inputenc}
\usepackage{selinput}
\usepackage{bm}
\usepackage{float}
\usepackage{yfonts}
\usepackage{mathtools}
\usepackage{enumitem}   
\usepackage{longtable}
\usepackage{tabularx}

\usepackage{xcolor}
\usepackage{soul}

\usepackage{xcolor}
\def\cb{\textcolor{black}}

\def\ctom{\textcolor{black}}
\def\dtom{\textcolor{black}}
\def\dtom{\textcolor{black}}

\def\xref{\textcolor{black}}

\def\ssb{\textcolor{black}}

\usepackage{tikz}
\usetikzlibrary{arrows,chains,matrix,positioning,scopes}
\usepackage{tikz-cd}
\usetikzlibrary{matrix, calc, arrows}

\def\sraise{\;\raise1.0pt\hbox{$'$}\hskip-6pt\partial}
\def\slower{\;\overline{\raise1.0pt\hbox{$'$}\hskip-6pt\partial}}

\usepackage{colortbl}

\makeatletter
\newlength{\apb@width}
\newcommand{\autoparbox}[2][c]{\settowidth{\apb@width}{#2}\parbox[#1]{\apb@width}{#2}}

\makeatother

\def\0{{\boldsymbol{0}}}

\def\beq{\begin{equation}}
\def\eeq{\end{equation}}
\def\be{\begin{equation}}
\def\ee{\end{equation}}

\usepackage[framemethod=default]{mdframed}
\newmdenv[skipabove=7pt,
skipbelow=7pt,
rightline=false,
leftline=false,
topline=false,
bottomline=false,
backgroundcolor=gray!10,
linecolor=gray,
innerleftmargin=5pt,
innerrightmargin=5pt,
innertopmargin=5pt,
innerbottommargin=5pt,
leftmargin=0cm,
rightmargin=0cm,
linewidth=4pt]{eBox}

\title{Estimating the Integrated Bispectrum from \\ Weak Lensing Maps}

\author{D. Munshi$^a$, J. D. McEwen$^b$, T. Kitching$^c$
  P. Fosalba$^{d,e}$, R. Teyssier$^{f}$, J. Stadel$^{g}$}
\affiliation{$^{a,b,c}$Mullard Space Science Laboratory, University College London, Holmbury St Mary, Dorking, Surrey RH5 6NT, UK 
  \quad $^{d}$Institute of Space Sciences (ICE, CSIC), Campus UAB, 
  Carrer de Can Magrans, s/n, 08193 Barcelona, Spain, \quad
  $^{e}$Institut d'Estudis Espacials de Catalunya (IEEC), 
  E-08034 Barcelona, Spain
  \quad $^{f,g}$Institute for Computational Science, University of Zurich,
  Winterthurerstrasse 190, 8057 Zurich, Switzerland}
  \emailAdd{D.Munshi@ucl.ac.uk, $^b$Jason.McEwen@ucl.ac.uk, $^c$t.kitching@ucl.ac.uk, $^d$fosalba@ieec.uab.es,
  $^{f}$romain.teyssier@uzh.ch, $^{g}$stadel@physik.uzh.ch}

\abstract{We use a recently introduced statistic called {\em Integrated Bispectrum} (IB) to probe the gravity-induced
  non-Gaussianity at the level of the bispectrum from weak lensing convergence or $\kappa$ maps.
  We generalize the concept of the IB to spherical coordinates,
  {This result is next connected to the response function approach.} 
  Finally, we use the Euclid Flagship simulations to compute the IB as a function of redshift and wave number.
  We also outline how the IB can be computed
  using a variety of analytical approaches including the ones based on {\em Effective Field Theory} (EFT), 
  {\em Halo models} and models based on the {\em Separate Universe approach} \xref{in projection or two-dimension (2D)}.
  \ssb{Comparing these results against simulations we find that the existing theoretical models tend to
    over-predict the numerical value of the IB}.
  We emphasize the role of the finite volume effect in numerical estimation of the IB.
  \xref{We introduced the concept of {\em squeezed and collapsed tripsectrum} for 2D $\kappa$ maps.
  We derive the IB for many paramterized theories of {\em modified gravity}
  including the Horndeskii and beyond-Horndeskii theories specifically for the non-degenerate scenarios that
  are also known as the \ssb{Gleyzes}-Langlois-Piazza-Venizzi or GPLV theories. In addition the cosmological models
  with clustering quintessence and models
  involving massive neutrinos are also derived.}
}

\begin{document}

\maketitle
\flushbottom

\section{Introduction}
\label{sec:intro}
%
%
Thanks to recently completed Cosmic Microwave Background (CMB) experiments, 
such as the \textit{Planck} Surveyor\footnote{{\href{http://http://sci.esa.int/planck/}{\tt Planck}}}\citep{Planck1}, we now 
have a standard model of cosmology. 
However there are many outstanding questions that remain unclear including, but not limited to, the nature of dark matter (DM) and dark energy (DE); or possible modifications of General Relativity (GR) on cosmological scales \citep{MG1,MG2}. In addition the sum of the neutrino \cb{masses} \citep{nu} remains unknown. 
Fortunately it is expected that the operational \xref{large scale structure (LSS)} surveys including
CFHTLS\footnote{\href{http://www.cfht.hawai.edu/Sciences/CFHLS/}{\tt http://www.cfht.hawai.edu/Sciences/CFHLS}},
Dark Energy Survey\footnote{\href{https://www.darkenergysurvey.org/}{\tt https://www.darkenergysurvey.org/}}(DES)\citep{DES}, Dark Energy Spectroscopic Instrument (DESI)\footnote{\href{http://desi.lbl.gov}{\tt http://desi.lbl.gov}},
Prime Focus Spectrograph\footnote{\href{http://pfs.ipmu.jp}{\tt http://pfs.ipmu.jp}},
KiDS\citep{KIDS} and  near-future Stage-IV LSS
surveys such as \textit{Euclid}\footnote{\href{http://sci.esa.int/euclid/}{\tt http://sci.esa.int/euclid/}}\citep{Euclid},
LSST\footnote{\href{http://www.lsst.org/llst home.shtml}{\tt {http://www.lsst.org/llst home.shtml}}}\citep{LSST_Tyson} and
WFIRST\citep{WFIRST}  will provide answers to many of the questions that cosmology is facing.

Weak lensing, or the minute distortions in the images of the distant galaxies by the intervening large-scale structure
allow us to extract information about clustering of the intervening  mass distribution in the Universe \citep{review}.
The weak lensing surveys are complementary to spectroscopic galaxy redshift surveys such as
BOSS\footnote{\href{http://www.sdss3.org/surveys/boss.php}{\tt http://www.sdss3.org/surveys/boss.php}}\citep{SDSSIII}
or WiggleZ\footnote{\href{http://wigglez.swin.edu.au/}{\tt http://wigglez.swin.edu.au/}}\citep{WiggleZ} as they
provide an unbiased picture of the underlying dark matter distribution, whereas the galaxies and other
tracers provide a biased picture \citep{bias_review}. 

One challenge for weak lensing is that 
observations are sensitive to smaller scales where clustering is nonlinear and non-Gaussian \citep{bernardeau_review}, and therefore 
difficult to model and measure. A second challenge is that the statistical estimates of cosmological parameters based on power spectrum 
analyses are typically degenerate in particular
cosmological parameter combinations e.g.\ $\sigma_8$ and $\Omega_{\rm M}$; to overcome these degeneracies external data sets (e.g.\ CMB), and the addition of 
tomographic or 3D \citep{3D} information is typically used. However to address both of these challenges an alternative procedure
is to use higher-order statistics that probe the (quasi)nonlinear regime \citep{higher1,higher2,higher3,MunshiBarber1,MunshiBarber2,MunshiBarber3}.

Previously it has been noted that even in the absence of any primordial non-Gaussianity \citep{Inflation},
the gravitational clustering induces mode coupling that results in a secondary non-Gaussianity which
is more pronounced on smaller scales. Thus a
considerable amount of effort has been invested in understanding the gravity induced secondary non-Gaussianity
from weak lensing surveys. These statistics include the lower-order cumulants \citep{MunshiJain1} and their correlators \citep{MunshiBias};
the multispectra including the skew-spectrum \citep{AlanBi} and kurtosis spectra \citep{AlanTri} 
as well as the entire PDF \citep{MunshiJain2} and
the statistics of hot and cold spots. Future surveys such as
\textit{Euclid} will be particularly interesting in this regard.
With its large fraction of sky-coverage it will be able to detect the gravity induced non-Gaussianity with a very
high signal-to-noise (S/N). It is also worth mentioning here that, in addition to breaking the degeneracy
in cosmological parameters, higher-order statistics are also
important for a better understanding of the covariance of lower-order estimators \citep{MunshiBarber4}.

In this paper we will focus on a set of estimators that are particularly sensitive to the
squeezed state of the bispectra also known as the {\em Integrated Bispectra} (IB). These estimators
are particularly interesting because of their simplicity, and ease of implementation.
While in previous work such estimators have been used in 3D for quantifying galaxy clustering \citep{SDSSII}
as well as in 1D to probe Lyman-$\alpha$ absorption features \citep{Lyman_bispec,chiang2}, our main 
aim here is to develop these estimators for probing future 2D projected surveys with
a focus on weak lensing surveys, and in particular \citep{Euclid}. 

This paper is organized as follows. In \textsection{\ref{sec:sq}} we introduce the position-dependent power spectrum
for flat-sky as well as well as in the all-sky limit. We also present the correspondence between these two cases. The analytical models are discussed in \textsection{\ref{sec:model}}.
The simulations are presented in \textsection{\ref{sec:sims}}.
An approximate error-analysis is presented in \textsection\ref{sec:error}.
The discussions are presented respectively in \textsection\ref{sec:disc} and \textsection\ref{sec:conclu}.

In Appendix \textsection\ref{sec:beyond} we derive perturbative expressions of squeezed
bispectrum for many extensions of standard $\Lambda$CDM model.
A very brief summary of Eulerian Perturbation Theory is given in Appendix-\textsection\ref{sec:brief}.
This will help our discussion next to go beyond bispectrum to trispectrum.
In Appendix-\textsection\ref{sec:Tri} we derive the squeezed and collapsed limits of weak lensing
trispectrum. Finally, in Appendix-\textsection{\ref{sec:los}} we introduce an approximation that can simplify
the evaluation of IB and other projected statistics.

%

%
\section{\bf Position-Dependent Power Spectrum from Convergence Maps}
\label{sec:sq}
%
A relatively new observable has been developed in recent years (see e.g.\ \cite{Chiang_original, PhD_chiang}) that relies on the fact that the power spectrum measured from
a survey sub-volume correlates with the mean of the same observable in the same sub-volume. This correlation
gives a direct estimate of the bispectrum in the {\em squeezed limit}.
Extensions of these results were presented for the divergence of velocity,
and in the context of 2D or projected surveys, as well as at higher-order at the level of the squeezed trispectrum \citep{Integrated}.
The IB was also derived in redshift-space for a class of modified gravity theories in \citep{MG_IB}.
In this section we will present the analytical results relevant to the position-dependent power spectrum
from 2D surveys with an emphasis particularly on 2D weak lensing surveys. 
%
\subsection{Flat-Sky Treatment}
\label{subsec:flat_sky}
%
Measurement of the bispectrum from weak lensing surveys is difficult due to non-ideal sky-coverage caused by masking of
regions close to bright objects as well as irregular survey boundaries. The IB
proposed here tries to bypass these complexities by concentrating on the squeezed limit
of the bispectrum, which can be estimated by techniques developed for estimation of the power spectrum.
We will concentrate on the projected survey but generalization to tomographic bins is straight-forward.

In this section we develop the recently introduced statistics of position-dependent power spectrum to
the case of weak lensing \citep{Integrated}.
We consider 2D weak lensing surveys but extension to 2D projected galaxy surveys can be done
in a straight-forward manner. We will use the small angle approximation before generalising
to the all-sky case in \textsection\ref{sec:all_sky}.

Let us consider a weak lensing convergence map $\kappa(\bm\theta)$ at a angular
position of the sky $\bm\theta$ defined over a patch of the sky $\bm\theta=({\vartheta},{\varphi})$,
where $\vartheta$ and $\varphi$ are
R.A. and declination respectively. 
We will divide the entire patch into equal area sub-patches. We will focus on
one such {\em sub-patch} centered around the angular position $\bm\theta_0=(\vartheta_0,\varphi_0)$.
The {\em local} average of $\kappa$ on a sub-patch of the sky can differ from its global value of
zero, where for a sub-patch the local average is:
\bes\ben
&& \bar\kappa({{\bm\theta}_0}) := {1 \over {\alpha}}\int \xd^2{\bm\theta}\,
  \kappa_{}({\bm\theta})\,W({{\bm\theta}-\bm{\theta}_0});
\;\; {\alpha} := \int  W({{\bm\theta}-\bm{\theta}_0}) \xd^2{\bm\theta}; \\
&&  W({\bm\theta}) := \Theta_{}(\vartheta-\vartheta_0)\Theta_{}(\varphi-\varphi_0).
\een\ees
Here, $W$ describes the sky-patch and $\Theta_{}$ represents the \xref{one-dimensional top-hat function i.e.}
$\Theta_{}(\vartheta-\vartheta_0)=1$ if  $|\vartheta-\vartheta_0|<\vartheta_{\rm S}$ and zero otherwise
and similarly $\Theta_{}(\varphi-\varphi_0)=1$ if  $|\varphi-\varphi_0|<\varphi_{\rm S}$ and zero otherwise.
In our notation, $\vartheta_{\rm S}$ and $\varphi_{\rm S}$ represent half-width of a sub-patch along
the $\vartheta$ and $\varphi$ directions while \ctom{$\alpha$ is the effective area of a
  sub-patch. We will assume that all sub-patches are of the same size and $\alpha$ is
independent of ${\bm\theta_0}$.}
In 2D, we will denote the Fourier wave-number as $\mathbf l$ and use the following convention for Fourier Transform:
\ben
&& \kappa({\mathbf l}) := 
\int \xd^2{\bm\theta} \exp(-\xxi\, {\bf l} \cdot \bm\theta) \kappa(\bm\theta); \;\;
\kappa(\bm\theta) :=  \int {\xd^2\mathbf l \over \xref{(2\pi)^2}  }\exp(\xxi\,{\bf l}\cdot {\bm\theta}) \kappa({\bf l}).
\label{eq:transform_convention}
\een
The power spectrum $P^{\kappa}$ and bispectrum $B^{\kappa}$ in 2D are defined using the following conventions:
\ben
&& \la\kappa({\bl}_{1})\kappa({\bl}_{2})\ra :=  
(2\pi)^{2}\delta_{\rm 2D}({\bl}_{1}+{\bl}_{2})P^{\kappa}(l_{1}); \quad l=|{\bl}|;
\label{eq:flat_sky_power}\\
&& \la\kappa({\bl}_{1})\kappa({\bl}_{2})\kappa({\bl}_{3})\ra 
:= (2\pi)^{2}\delta_{\rm 2D}({\bl}_{1}+{\bl}_{2}+{\bl}_{3})B^{\kappa}({\bl}_{1},{\bl}_{2},{\bl}_{3}).
\label{eq:flat_sky_bispec}
\een
The angular brackets represent the ensemble average. Here, $\delta_{\rm 2D}$ is the Dirac delta function in 2D.
The window $W$ describing a patch can be used to extend the limits of angular integration.
Thus, the flat-sky (local) Fourier transform takes the following form:
\bes
\ben
\kappa({\bl}_{}; {\bm\theta}_0) :=
\int \xd^2\bm\theta \;\kappa(\bm\theta) W(\bm\theta-\bm\theta_0)\exp(-\xxi\bl\cdot\bm\theta) \\
= \int {\xd^2{\bl}_1\over (2\pi)^2 }\, \kappa({\bl-\bl_1})\,
W({\bl_1})\, \exp(-\xxi{\bl_1}\cdot{\bm\theta}_0).
\een
\ees
\ctom{We use $W({\bl})$ to denote the Fourier transform of $W(\bm\theta)$.}
Notice that the Fourier coefficient $\kappa({\bl}_{}; {\bm\theta}_0)$ for $\bl=0$ (monopole) is identical to $\bar\kappa(\bm\theta_0)$ defined above.
The {\em local} convergence power spectrum  $P^{\kappa}({\bl}_{}; {\bm\theta}_0)$ in this
fraction of sky is given by \ctom{(we will denote the global power spectrum
as $P^{\kappa}({\bl}_{})$)}:
\ben
&& P^{\kappa}({\bl}_{}; {\bm\theta}_0) = {1 \over {\alpha}_{}} \int {\xd^2 {\bl}_{1} \over (2\pi)^2} 
\int {\xd^2 {\bl}_{2} \over (2\pi)^2}
\kappa({\bl}_{}-{\bl}_{1}) \kappa(-{\bl}_{}-{\bl}_{2}) \nonumber \\
&&\quad\quad\quad\quad \times \exp [-\xxi\, ({\bl}_{1} +
  {\bl}_{2})\cdot {\bm\theta}_0] W({\bl}_{1}) W({\bl}_{2}).
\label{eq:definePS}
\een
The resulting IB is defined by cross-correlating the local estimate
of the power spectrum and the local average of the projected field:
\ben
&& {\cal B}^{\kappa}({\bl}_{}) := \la P^{\kappa}({\bl}_{};{\bm\theta}_0)\bar\kappa({\bm\theta}_0) \ra \nonumber
\label{eq:IB_def1}; \\
&& {\cal B}^{\kappa}({\bl}_{}) = {1 \over {\alpha}^2_{}}  
\int {\xd^2{\bm\theta}_0 \over 2\pi}  \int {\xd^2 {\bl}_{1} \over (2\pi)^2} \int {\xd^2 {\bl}_{2} \over (2\pi)^2}\int {\xd^2 {\bl}_{3} \over (2\pi)^2}
\la \kappa({\bl}_{}-{\bl}_{1}) \kappa (-{\bl}_{}-{\bl}_{2})\kappa(-{\bl}_{3}) \ra\nonumber \\
&& \hskip1cm \times \;W({\bl}_{1}) W({\bl}_{1}+{\bl}_{3}) W({\bl}_{3})
\exp [ -\xxi\, ({\bl}_{1}+{\bl}_{2}+{\bl}_{3})\cdot{\bm\theta}_0 ].
\label{eq:IB_def2}
\een
The power spectrum $P^{\kappa}$ and the average $\bar\kappa$ used in Eq.(\ref{eq:IB_def1}) are estimated from the same patch of the sky.
Many such patches are created by dividing the entire survey area. The computation of the average
and the power-spectrum from a patch is far simpler than estimating the bispectrum directly.
However, it is worth mentioning that such a simplification comes at a price, as the IB can only extract information about 
the squeezed limit of the bispectrum if we focus on wave numbers ${l}$
much larger than the wave numbers that correspond to the fundamental mode of the patch.

Next, we will specialize our discussion for weak lensing surveys. The weak lensing convergence $\kappa$ is a line-of-sight projection of the 3D density contrast $\delta({\bf r})$:
\ben
&& \kappa({\bm\theta}) := \int_0^{r_s} \xd r\, w(r) \delta(r,{\bm\theta}); \quad
w(r) := {3 \Omega_{\rm M} \over 2} {H_0^2 \over c^2} a^{-1} {d_A(r) d_{A}({r_s-r}) \over d_A(r_s)}.
\label{eq:def_kappa}
\een
\ctom{Here, $r$ is the comoving radial distance, $a$ represents the scale factor,
$H_0$ the Hubble parameter, $\Omega_{\rm M}$ the cosmological matter density parameter and $d_A(r)$ is the {comoving} angular diameter distance.}
We will ignore the source distribution and assume them to be localized on a single source plane
\ctom{at a distance $r=r_s$}, we will also ignore photometric redshift errors.
\ctom{However, such complications are essential to link predictions to observational data and will
be included in our analysis in a future publication. }
Fourier decomposing $\delta$ along and perpendicular to the line-of-sight direction we obtain: 
\ben
\kappa(\bm\theta) = \int_0^{r_s} \xd\, r\, \omega(r) \int {\xd k_{\parallel} \over 2\pi} \int {\xd^2{\bf k}_{\perp} \over (2\pi)^2} \exp [\xxi (r\,k_{\parallel} + d_A(r)\;{\bm\theta}\cdot{\bf k}_{\perp})]\delta({\bf k}; r).
\label{eq:los}
\een
\ctom{In our notation,} $k_{\parallel}$ and ${\bf k}_{\perp}$ are the components of
the wave vector ${\bf k}$ along and perpendicular to
the line-of-sight direction and $\delta({\bf k})$ is the Fourier transform of the 3D density contrast $\delta$.

We have used the following convention for the 3D FT and its inverse:
\ben
&& \delta({\bf k}) := 
\int \xd^3{\bf x} \exp(-\xxi {\bf k} \cdot{\bf x}) \delta({\bf x}); \quad
\delta({\bf x}) :=  \int {\xd^3{\bf k}\over \xref{(2\pi)}^3 }\exp(\xxi\,{\bf x}\cdot {\bf k}) \delta({\bf k}).
\een
\ctom{We have introduced ${\mathbf x} = (r,{\bm\theta})$ as the comoving distance
and ${\mathbf k}$ as the corresponding wave number. We will use $\delta_{\rm 3D}$ to
denote the 3D Dirac delta function.}
The 3D power spectrum and bispectrum for the density contrast $\delta$ are:
\ben
&& \la\delta({\bk}_{1})\delta({\bk}_{2})\ra :=
(2\pi)^{3}\delta_{\rm 3D}({\bk}_{1}+{\bk}_{2})P_{\rm 3D}(k_{1}); \quad k=|{\bf k}|;\\
&& \la\delta({\bk}_{1})\delta({\bk}_{2})\delta({\bk}_{3})\ra 
:= (2\pi)^{3}\delta_{\rm 3D}({\bk}_{1}+{\bk}_{2}+{\bk}_{3})B_{\rm 3D}({\bk}_{1},{\bk}_{2},{\bk}_{3}).
\een

Using the Limber approximation
the convergence power spectrum $P^{\kappa}(k)$ and bispectrum $B^{\kappa}({\bf k}_1,{\bf k}_2,{\bf k}_3)$ can be expressed
respectively in terms of the 3D power spectrum $P_{\rm 3D}(k)$ and bispectrum  $B_{\rm 3D}({\bf k}_1,{\bf k}_2,{\bf k}_3)$:
\bes
\ben
&& P^{\kappa}({l}) = \int_0^{r_s} \xd r\, {\omega^2(r) \over d_A^2(r)}
P_{\rm 3D}\left ({{l} \over d_A(r)}; r \right ); \label{eq:defPower}\\
&& B^{\kappa}({\bl}_{1},{\bl}_{2},{\bl}_{3}) = \int_0^{r_s} \xd r\, {\omega^3(r) \over d_A^4(r)}
B_{\rm 3D}\left ({{\bl}_{1} \over d_A(r)},{{\bl}_{2} \over d_A(r),},
{{\bl}_{3} \over d_A(r)}; r \right).
\een
\ees
The expression for $B_{\rm 3D}$ is deferred until Eq.(\ref{eq:PT_matter_bispec}). 
\ctom{In 2D the angular average of the IB (denoted as $\bar{\cal B}^{\kappa}(l)$) can be defined as follows:}
\ben
\bar{\cal B}^{\kappa}(l) := \int {\xd\,\varphi_l\over 2\pi}\;
    {\cal B}^{\kappa}({\bl}); \quad l=|{\bl}|.
    \een
\ctom{Here $\varphi_l$ is the angle between the vector $\bl$ and the $\vartheta$ (R.A.) direction.   
  Next, carrying out the ${\bm\theta_0}$ integral in Eq.(\ref{eq:IB_def2}) and using the resulting
  2D delta function to perform the $\bl_2$ integral leaves us with the following expression:}
\ben
&&{\bar{\cal{B}}}^{\kappa}({\bf l}) = \int {\xd \varphi_l \over 2\pi}\int {\xd^2{\bf l}_{1} \over (2\pi)^2 }
\int {\xd^2{\bl}_{3} \over (2\pi)^2 }
B^{\kappa}({\bf l}_{}-{\bf l}_{1}, -{\bf l}_{}+{\bf l}_{1}+{\bf l}_{3}, -{\bf l}_{3}) \nonumber \\
&& \hspace{2cm} \times W(\bl_{1})W(\bl_{1}+\bl_{3}) W(\bl_{3}).
\label{eq:def2D_esimator}
\een
\ctom{Notice we have also used the definition of the convergence bispectrum in Eq.(\ref{eq:flat_sky_bispec}).}
To simplify this further, we will use the following property of the window function $W$:
\ben
W^2(\bm\theta)=W(\bm\theta); \quad W({\bl}_{3}) = \int {\xd^2 {\bl}_{1} \over (2\pi)^2}
W(-{\bl}_{1}-{\bl}_{3})W({\bl}_{1}), 
\een
such that in the {\em squeezed limit} the tree-level perturbative matter bispectrum in the small angle approximation
takes \cb{the} following form \citep{Chiang_original}:
\ben
&& B_{\rm 3D}\left ({\bl/d_A(r)}_{}-{{\bl}_{1}/d_A(r)}, -
    {\bl/d_A(r)}_{}+{\bl}_{1}/d_A(r)+{\bl}_{3}/d_A(r), -{\bl}_{3}/d_A(r) \right )\nonumber \\
&& = \left [ {13 \over 7} + {8 \over 7}\left ( {{\bl}_{}\cdot {\bl}_{3} \over l_{} l_{3}} \right )^2 -
  \left ( {{\bl}_{}\cdot {\bl}_{3} \over l l_3} \right )^2 {\xd\ln P_{\rm 3D}(l_{}) \over \xd\ln l_{}}  \right ] P_{\rm 3D}\left ({l_{}\over d_A(r)} \right)P_{\rm 3D}\left ({l_{3}\over d_A(r)} \right)+ \cdots
\label{eq:PT_matter_bispec}
\een
The terms which are higher-order in terms of $(l_{1}/l_{})$ or $(l_{3}/l_{})$
(where, $l_{i}=|{\bl_{i}}|$) are ignored as we take the
limiting case that $l_{}\gg l_{i}$.
Using the fact that the circular average of {$\langle \left ( {\bl} \cdot {{\bl}}_{i}/ l l_i \right )^2\rangle$ is $[{1/2}]$
(which is in contrast to the derivation of results applicable in 3D where 
spherical average of {$\langle \left ( {\bl} \cdot {{\bl}}_{i}/ l l_i \right )^2\rangle$ is $[{1/3}]$), we arrive at the final expression:
\bes
\ben
&& \bar{\cal B}^{\kappa}(l) = R_2 \left [ {24 \over 7} - {1 \over 2}
  {\xd\ln l^2 P_{\rm 3D}(l) \over \xd \ln l_{}} \right ] P^{\kappa}(l)\sigma_L^2; \quad \sigma_L^2=\la\bar\kappa^2\ra; \label{eq:define_B} \\
&& R_2 = \int_0^{r_s} \xd\,r {w^3(r)\over d_A^{4+2n}(r)} \bigg / \left ( \int_0^{r_s} \xd\,r{ w^2(r) \over d_A^{2+n}(r)} \right )^2, 
\label{eq:ib_def}
\een
\ees
where we have approximated the power spectrum with a power law $P_{\rm 3D}(k)\propto k^n$. The normalized IB,
\ctom{denoted as ${\cal B}^{\prime}(l_{})$}, is defined as:
\ben
   {\cal B}^{\prime}(l_{}) =  {1 \over P^{\kappa}(l_{})\sigma_L^2}{{\cal\bar B}^{\kappa}(l_{})}
     =  R_2 \left [ {24 \over 7} - {1 \over 2} {d\ln l_{}^2 P_{\rm 3D}(l_{}) \over d\ln l_{}} \right ].
    \label{eq:define_normIB}
    \een
The above expression was derived using tree-level perturbation theory.\footnote{\ctom{This result is similar to the result obtained in
    Ref.\citep{Bernardeau95} using a vertex generating function approach.
    If we identify the second-order vertex $\nu_2=12/7$ computed in Ref.\citep{Bernardeau95} using a 2D spherical dynamics
    to $Q_2$, we get ${\cal B}^{\prime}(l_{})= 24/7\, R_2$.
    Which of course can also be obtained in our approach in the no-smoothing limit, from Eq.(\ref{eq:define_normIB}),
    by substituting $n=-2$. However, it is important to realize that despite the formal mathematical similarity the
    statistics introduced in  Ref.\citep{Bernardeau95} or its two-point generalisations 
    and the IB computed here are not the same and their physical
    interpretation is completely different. Indeed, this formal similarity of mathematical expressions too are
    only valid at the level of second-order.}}. 
Using very similar arguments we can show that if we assume a hierarchical {\em ansatz} for the
underlying 3D bispectrum \citep{bernardeau_review}:
\ben
{B}_{\rm 3D}({\bf k}_1,{\bf k}_2,{\bf k}_3) = Q_3[P_{\rm 3D}({\bf k}_1)P_{\rm 3D}({\bf k}_2)+ \xref{P_{\rm 3D}({\bf k}_2)P_{\rm 3D}({\bf k}_3)}
  + P_{\rm 3D}({\bf k}_3)P_{\rm 3D}({\bf k}_1)].
\een
Here, $Q_3$ is the hierarchal amplitude
of three point correlation function which can be computed using \dtom{Hyperextend Perturbation Theory(HEPT)}\citep{Hyper}.
The corresponding IB is given by\citep{Integrated}:
\ben
\bar{\cal B}^{\kappa}(l) = 2R_2 \,Q_3\, P^{\kappa}(l)\sigma_L^2;\quad\quad Q_3 = {4-2^n\over 1+2^{n+1}}
\een
\dtom{Here $n$ denotes the spectral index for the linear 3D power spectrum assumed locally to be
power law $P_{\rm lin}(k)\propto k^n$}. Similar results in the highly nonlinear regime can also be derived using the halo model, loop level corrections,
effective field theory \citep{MunshiRegan}, phenomenological fitting functions \citep{Gil-Marin} or
separate Universe models \citep{chiang2,chiang4,chiang5}. We will not include the primordial non-Gaussianity
(of local type \citep{Bartolo}) in our analysis but this can also be included as a straight-forward expansion \citep{chiang3}. \dtom{Most of these theories including the fitting functions can be seen as
(physically motivated) interpolation of the
results obtained by PT and HEPT which are the two limiting cases.}
%
\subsection{Zeldovich Approximation}
The \ctom{lowest-order in Lagrangian perturbation theory commonly known as the} Zeldovich approximation (ZA) can be used to simplify many aspects of gravitational clustering. Using the same line of analysis as above
the IB can be calculated for the ZA. {\cb The squeezed 
  bispectrum in this approximation takes the following form}:
\bes
\ben
&& {B}_{\rm ZA} \neweq \left [ 1+ 2 \left ( {\bl_{}\cdot \bl_{3} \over l_{} l_{3}} \right )^2  -
\left ( {\bl_{}\cdot \bl_{3} \over l_{} l_{3}} \right )^2 {\xd\ln P_{\rm 3D}(l_{}) \over \xd\ln l_{}} \right ] P_{\rm 3D}\left ({l\over d_A(r)}\right )P_{\rm 3D}\left ({l_{3}\over d_A(r)}\right).
\label{eq:z2}
\een
\ees
\cb{Next, going through
  the expressions outlined in Eq.(\ref{eq:define_B})-Eq.(\ref{eq:define_normIB}) we arrive at the
  corresponding result for the normalized IB using ZA:}
\ben
    {\cal B}^\prime_{\rm ZA}(l) = R_2 \left [ 3 - {1 \over 2} {d\ln l_{}^2 P_{\rm 3D}(l_{}) \over d\ln l_{}} \right ] =
    R_2\left [ 3 - {1\over2} (n+2) \right ].
    \label{eq:ib_za_approx}
    \een
\ctom{The last equality holds only for a power law $P_{\rm 3D}(k)$.
  Comparing with Eq.(\ref{eq:define_normIB}), we can see that the ZA under predicts the normalized IB.
  This is consistent with the well-known fact that the normalised skewness parameter $S_3$ for the ZA is
  lower than its SPT value \citep{MSS}}.
\subsection{Response Function Approach}
The (linear) response function approach \citep{Sherwin,YinLi,Baldauf,Fabian,Bareira_response1,PhD_chiang} to IB takes advantage of the fact that the
the bispectrum in its squeezed configuration
can be interpreted as a response of the small scale power spectrum to a long wavelength modulation of $\bar\kappa$.
In this scenario an over(under)dense region is treated as a separate Universe, and the over(under)density is absorbed in a redefinition of the
background cosmology.
To connect with the response function approach and the modelling used in separate Universe approaches, we expand the convergence power spectrum as follows.
We expand the 2D convergence power spectrum estimated from a patch as a function of the average $\kappa$,
which is a result of the long wavelength fluctuations of the $\kappa$ field: 
\ben
P^\kappa(l;\bm\theta_0) = P^\kappa(l;\bm\theta_0)|_{\bar\kappa(\bm\theta_0)=0} +
{\xd P^\kappa(l;\bm\theta_0) \over \xd {\bar\kappa}(\bm\theta_0)} \Big{|}_{\bm\bar\kappa(\bm\theta_0)=0}{\bar\kappa(\bm\theta_0)} + \dots.
\label{eq:def_response_function}
\een
By correlating $\bar\kappa(\bm\theta_0)$ with $P^{\kappa}(l;\bm\theta_0)$ and ignoring terms of ${\cal O}(\bar\kappa^3)$ we arrive at the following expression:
\bes\ben
&& \bar{\cal B}^{\kappa}(l)\equiv \la\bar\kappa(\bm\theta_0) P^{\kappa}(l;\bm\theta_0)\ra =
    {\sigma_L^2}{\xd\ln P^{\kappa}(l;\bm\theta_0) \over \xd\bar\kappa(\bm\theta_0)} P^{\kappa}(l;\bm\theta_0); \quad
    \sigma_L^2 := \la\bar\kappa^2\ra. \\
&& {\cal B}^{\prime}(l)= {\xd \ln P^\kappa(l;\bm\theta_0) \over \xd {\bar\kappa}(\bm\theta_0)} \Big{|}_{\bm\bar\kappa(\bm\theta_0)=0}
\een\ees

The gravity induced bispectrum at tree-level, that we have used above in our derivation given in Eq.(\ref{eq:PT_matter_bispec}), 
can be replaced by the primordial bispectrum of inflationary origin (local type \citep{Bartolo}) to estimate the corresponding IB. Indeed, the
primordial non-Gaussianity is severely constrained by recent data from \textit{Planck} \citep{Planck_ng,Planck_mg}, although
it is expected that future Stage -IV surveys such as \textit{Euclid} will further tighten such constraints \citep{DETF}.
Similarly results can also be obtained for non-Gaussianity induced by cosmic strings \citep{strings}.
The generalization of the above result derived for 2D or projected surveys to 3D can be accomplished by cross-correlating
$\bar\kappa$ from one redshift bin to the power spectrum estimated from another and vice versa.

Indeed generalization of IB to include external data sets can also be incorporated in a straight-forward manner 
by cross-correlating the convergence power spectrum in Eq.(\ref{eq:definePS}) against any other projected field
e.g.\ the tSZ $y$-parameter maps, can provide an estimator for the Integrated mixed bispectrum involving
$\kappa$ and $y$ \citep{Integrated}. 
%
%
\subsection{All-Sky Formulation}
\label{sec:all_sky}
%
The discussion in the previous section was based on a flat-sky treatment, and indeed many of the
recent surveys are small enough so that a flat-sky treatment should be adequate.
However, the next generation surveys will cover a considerable fraction of the sky
making an all-sky treatment necessary. For a review of all-sky formulations of
non-Gaussianity see \citep{Bartolo}. Certain aspects of consistency relations
for the galaxy bispectrum in the all-sky limit are discussed in  \citep{all_sky_sq}, 
and the all-sky versus flat-sky correspondence is analyzed in \citep{Hu}. 
We start by defining the spherical harmonic decomposition of the convergence map $\kappa(\bm\theta)$
with and without a mask:
\ben
\kappa_{\ell m} := \int\,{\xd^2\bm\theta}\,\kappa(\bm\theta)\,Y_{\ell m}(\bm\theta); \quad\quad
\tilde\kappa_{\ell m} := \int\,{\xd^2\bm\theta}\,W({\bm\theta})\,\kappa(\bm\theta)\,Y_{\ell m}(\bm\theta).
\een
We use $Y_{\ell m}$ to denote the spherical harmonics of degree $\ell$ and order $m$.
The all-sky power spectrum is defined as ${\cal C}^{\kappa}_{\ell}
= {1\over(2\ell+1)}\sum_{m} \kappa_{\ell m}\kappa^*_{\ell m}$
and for high-$\ell$ identical to its flat-sky counterparts defined in Eq.(\ref{eq:flat_sky_power})
i.e. $P^{\kappa}(l) = {\cal C}^{\kappa}_{\ell}$. \ctom{In an analogus manner, we introduce the
pseudo power spectrum $\tilde{\cal C}^{\kappa}_{\ell}$s constructed from the $\tilde\kappa_{\ell m}$, i.e.
${\cal\tilde C}^{\kappa}_{\ell} = {1\over(2\ell+1)}\sum_{m} \tilde\kappa_{\ell m}\tilde\kappa^*_{\ell m}$.}
using angular braces e.g. $\langle {\cal{C}}_{\ell}\rangle$}. 
The angle-averaged bispectrum $B^{\kappa}_{\ell_1\ell_2\ell_3}$ in spherical harmonic space is:
\bes
\ben
&& \langle \kappa_{\ell_1 m_1} \kappa_{\ell_2 m_2} \kappa_{\ell_3 m_3}\rangle :=
\la B^{\kappa}_{\ell_1\ell_2\ell_3} \ra\;
\left ( \begin{array}{ c c c }
     \ell_1 & \ell_2 & \ell_3 \\
     m_1 & m_2 & m_3
\end{array} \right); \\
&& B^{\kappa}_{\ell_1\ell_2\ell_3} = \sum_{m_i}\left ( \begin{array}{ c c c }
     \ell_1 & \ell_2 & \ell_3 \\
     m_1 & m_2 & m_3
\end{array} \right) \kappa_{\ell_1 m_1} \kappa_{\ell_2 m_2} \kappa_{\ell_3 m_3}, 
\een
\ees
where the matrix denotes the Wigner $3j$ symbol.
Since $\ell_1$, $\ell_2$, and $\ell_3$ form a triangle,
$B^{\kappa}_{\ell_1\ell_2\ell_3}$ satisfies the triangle condition, $|\ell_i-\ell_j | \le \ell_k \le \ell_i +\ell_j$
for all permutations of indices; and parity invariance of the angular correlation function
demands $\ell_1 + \ell_2 + \ell_3$ = even. 
For a spherical sky we introduce the reduced bispectrum $b^\kappa_{\ell_1\ell_2\ell_3}$ by the following expression:  
\bes
\ben
&& \langle \kappa_{\ell_1 m_1}\kappa_{\ell_2 m_2}\kappa_{\ell_3m_3}\rangle =
b^{\kappa}_{\ell_1\ell_2\ell_3} {\cal G}_{\ell_1\ell_2\ell_3}^{m_1m_2m_3}; \label{eq:def_reduced} \\
&& {\cal G}_{\ell_1\ell_2\ell_3}^{m_1m_2m_3} := \int\, d^2{\bm\theta}\, Y_{\ell_1m_1}(\bm\theta)
Y_{\ell_2m_2}(\bm\theta)\, Y_{\ell_2 m_3}(\bm\theta).
\label{eq:gaunt_integral}
\een
\ees 
this reduced bispectrum $b^{\kappa}_{\ell_1\ell_2\ell_3}$ is important in deriving the flat-sky limit.
\cb{The symbol ${\cal G}$ introduced in Eq.(\ref{eq:gaunt_integral})
is also known as {\em Gaunt} integral which represents the
coupling of three spherical harmonics.}
To relate Eq.(\ref{eq:def_reduced}) and Eq(\ref{eq:flat_sky_bispec}) we note that 
the Gaunt integral introduced in Eq.(\ref{eq:gaunt_integral}) becomes a Dirac delta function
in the flat-sky limit 
${\cal G}^{m_1m_2m_3}_{\ell_1\ell_2\ell_3} \approx (2\pi)^2 \delta_{\rm 2D}({\bf l}_1+{\bf l}_2+{\bf l}_3)$, which leads 
us to identify the reduced bispectrum as the flat-sky bispectrum \citep{Hu}:
$b^{\kappa}_{\ell_1\ell_2\ell_3} \approx B^{\kappa}({\bf l}_1, {\bf l}_2, {\bf l}_3)$; and similarly
Eq.(\ref{eq:flat_sky_coupling}) and Eq.(\ref{eq:all_sky_coupling}).

Next, to define the IB we will introduce a mask or a window $W(\bm\theta,\bm\theta_0)$
whose harmonics are defined by
$W_{\ell m}(\bm\theta_0) = \int\, d^2\bm\theta\, W(\bm\theta,\bm\theta_0) Y_{\ell m}(\bm\theta)$.
\ctom{Individual patches are identified with specific value of $\bm\theta_0$. However,
to simplify notation, we will suppress the $\bm\theta_0$ dependence in $W_{\ell m}$ and the
resulting coupling matrix $\rm M$ defined below.}
The all-sky harmonics
and their partial sky counterparts are related through a coupling matrix $\rm M$ \citep{Master}:
\ben
&& \tilde \kappa_{\ell_1 m_1} 
= \sum_{\ell_2 m_2} \, {\rm M}_{\ell_1m_1,\ell_2m_2} \kappa_{\ell_2m_2}; \quad
{\rm M}_{\ell_1m_1,\ell_2m_2} 
= \sum_{\ell m} W_{\ell m} {\cal G}^{m_1m_2m}_{\ell_1\ell_2\ell}.
\label{eq:all_sky_coupling}
\een
Using the flat-sky convention introduced in Eq.(\ref{eq:transform_convention}) 
the flat-sky equivalent of Eq.(\ref{eq:all_sky_coupling}) takes the following form:
\ben
{\tilde \kappa} (\bl_1) = \int \, d^2\bl_2\,  M_{\bl_1\bl_2}\, \kappa(\bl_2);  \quad\quad
M_{\bl_1\bl_2} = \int d^2{\bf l} \; W({\bf l})\, \delta_{\rm 2D}({\bf l}+\bl_1-\bl_2 ).
\label{eq:flat_sky_coupling}
\een
Here, $M_{\bl_1\bl_2}$ is the flat-sky counterpart of  ${\rm M}_{\ell_1m_1,\ell_2m_2}$.

To construct an all-sky estimator we use the fact that the average of $\kappa$ over a region $\bar\kappa$ is given by the monopole
$\kappa_{00}$: $\bar\kappa = \tilde\kappa_{00} = \sum_{\ell_1m_1} M_{00,\ell_1 m_1} \kappa_{\ell_1 m_1}$.
The estimator for the IB is formed by cross-correlating
the pseudo power spectrum, denoted as $\tilde {\cal C}^{\kappa}_{\ell}$, with $\bar\kappa$:
\ben
&& \bar{\cal B}^{\kappa}_{\ell} := \langle{\bar\kappa}\tilde{\cal C}_{\ell} \rangle =
{1 \over 2\ell+1}\sum_m\langle \tilde \kappa_{00}\tilde \kappa_{\ell m}\tilde \kappa^*_{\ell m} \rangle \nonumber \\
&& = {1 \over 2\ell+1} \sum_{m}\sum_{\ell_i m_i}
\left ( \begin{array}{ c c c }
     \ell_1 & \ell_2 & \ell_3 \\
     m_1  & m_2  & m_3
\end{array} \right) 
      {\rm M}_{\ell m, \ell_1 m_1}{\rm M}_{\ell m, \ell_2 m_2}{\rm M}_{00, \ell_3 m_3}  B^{\kappa}_{\ell_1\ell_2\ell_3}
\label{eq:defined_est}
\een
Next, using the correspondence discussed between flat-sky and all-sky before, it is straight-forward to show that in the limit of high-$\ell$
this reduces to the 2D estimator in Eq.(\ref{eq:def2D_esimator}). 

We have focused on the convergence $\kappa$ which is a spin-0 or a scalar field.
Similar results can be obtained by using an E/B decomposition of shear maps \citep{hikage, Marika} and corresponding squeezed $\rm{EEE}$ and
$\rm{BBB}$ bispectra; these results will be presented elsewhere. Finally, many Bayesian estimators have 
been recently developed for power-spectrum analysis e.g. \citep{HamiltonSampling,Wandelt}; 
using such an estimator for ${\cal C}^{\kappa}_\ell$ in Eq.(\ref{eq:defined_est}) would pave the way for
the development of the first Bayesian estimation of the bispectrum.

%



\section{Analytical Modelling of IB}
\label{sec:model}

In this section we will introduce different approaches that we will use for the theoretical
prediction of the IB. Indeed, a detailed understanding of gravitational
clustering in the nonlinear regime is lacking, though is of paramount importance for cosmology, nevertheless these approximations 
should be considered only as an illustrative proxy for the true system exhibiting much richer dynamics. 
\subsection{Halo Model}
\label{subsec:halo}
The `halo model' \citep{halo} is arguably the most popular model for predicting the clustering properties of dark matter.
It is a phenomenological model that is based on the assumption that all matter is contained within
spherical halos of properties that are completely determined by their mass distribution, radial profile and clustering
properties. Variants of the halo model, used to predict dark matter clustering, differ in fine detail with respect to these ingredients.
See \citep{Chiang_original} for the derivation in the case of 3D density field.
Using the small-angle approximation the halo model can be encapsulated by: 
\ben
&& P_{\rm 3D}(k_{\perp},a) = P^{\rm 1h}(k_{\perp},a) + P^{\rm 2h}(k_{\perp},a); \label{eq:defHM_Pk} \\
&& P^{\rm 2h}(k_{\perp},a) = \left [ I_1^1(k_{\perp}) \right ]^2 P_{\rm lin}(k_{\perp},a); \quad
P^{1h}(k_{\perp},a) = I_2^0(k_{\perp},k_{\perp}); \label{eq:2h}\\
&& {\xd P_{\rm 3D}(k_{\perp},a) \over \xd \bar\delta}\Big{|}_{\bm\bar\delta=0} =
\Big [ \left ({24 \over 7} -{1\over 2}
  {\xd\ln k_{\perp}^2P(k_{\perp},a) \over \xd\ln k_{\perp}}\right ) P^{\rm 2h}(k_{\perp},a) \nonumber \\
&&   \hspace{2cm}   + 2I^2_1(k_{\perp})I_1^1(k_{\perp})P_{\rm lin}(k_{\perp},a) + I^{1}_{2}(k_{\perp},k_{\perp})\Big ]. \label{eq:response}
\een
As before we use the small angle approximation i.e. $k_{\perp}>> |k_{\parallel}|$ where
$k_{\parallel}$ and $\bk_{\perp}$ are components parallel and perpendicular to the line-of-sight.
\xref{ The angular averaging for the coefficient of the two-halo term is done in 2D.}
Where, $P_{}^{\rm 1h}$ and $P^{\rm 2h}$ are the one- and two-halo contributions to the total
power spectrum $P_{\rm 3D}(k,a)$
at a redshift $z$, $a=1/(1+z)$ and  wavenumber $k$. The nonlinear power spectrum $P_{\rm 3D}(k,a)$
depends on the linear power spectrum $P_{\rm lin}(k,a)$ through the two halo contribution.
We have also used the following notation in Eq.(\ref{eq:2h})-Eq.(\ref{eq:response}):
\ben    
I_m^n(k_{1\perp},\cdots,k_{m\perp}):= \int \xd\ln M n(\ln M) \left( {M\over \rho} \right )^m b_n(M)
\Pi^m_{i=1}u(M|k_{i\perp}),
\label{eq:kernels}
\een
where $u(M|k)$ is the Fourier transform of the halo radial profile. 
The (higher-order) bias functions $b_N$ for halos of mass $M$ that appear in Eq.(\ref{eq:kernels})
are defined as a response of the halo number density $n(\ln M)$ to change in $\bar\delta$
\cb{(average density contrast in a finite patch)}:
\ben
b_N(M) := {1 \over n(\ln M)} {\partial^N n(\ln M)\over \partial^N \bar\delta }.
\een
We can use Eq.(\ref{eq:response}) in Eq.(\ref{eq:def_response_function}) to compute the IB. Using these ingredients we can finally write the normalised IB in
the halo model as:
\bes\ben
  && {\cal B}^{\prime}({l}) \approx {1 \over P^{\kappa}(l)}{dP^{\kappa} \over d\bar\kappa} = {1 \over |\kappa_{\rm min}| P^{\kappa}(l)}
  \int {w^2(r) \over d^2_A(r)} {\xd \over \xd\bar\delta}
  P_{\rm3D} \left ({l \over  d_A(r)}; r \right ) dr; \label{eq:how2computeIB1}\\
  && \kappa_{\rm min} = -\int_0^{r_s} dr \, w(r).
  \label{eq:how2computeIB}
  \een\ees
  {In our above evaluation, we have used the mapping
    $\delta \rightarrow \kappa/|\kappa_{\rm min}|$ to arrive
    at $\bar\delta \rightarrow \bar\kappa/|\kappa_{\rm min}|$
    (see Appendix-\textsection\ref{sec:los} for
    more detailed discussion) which gives
    ${\xd/\xd\bar\kappa} = {1/|\kappa_{\rm min}|}{\xd/\xd\bar\delta}$.
    \xref{This is a very simple approximation but has been
    used with a remarkable success in the past to compute not just the normalised
    high-order statistics i.e. cumulant and cumulant correlators  as well as the entire one- and two-point PDFs.
    Mathematically it amounts to replacing the integrands in $\kappa_{\rm min}$, $P^{\kappa}$
    and $dP^{\kappa}/ d\kappa$ with their values at a median redshift. More rigorous evaluation is
    carried out in \citep{Takada1,Takada2,Takada3}.}
  We have also implcitly assumed that $\bar\delta$ only changes the power spectrum not the geometry e.g. $d_A(r)$
  and the lensing kernel $w(r)$.} Evaluation of the power spectrum in
  the halo model is done
  using Eq.(\ref{eq:defHM_Pk})
  in association with Eq.(\ref{eq:defPower}).
  The derivative of the power spectrum is given in Eq.(\ref{eq:response}). If we ignore
  the last two terms in Eq.(\ref{eq:response}) we will recover the perturbative
  results in Eq.(\ref{eq:define_B}) and Eq.(\ref{eq:ib_def}).
\citep{Takada1,Takada2,Takada3}
  %
\subsection{Effective Field Theory (EFT)}
%
The EFT is based on incorporating small-scale effects
where standard perturbations theory (SPT) fails \citep{Baumann,Senatore}.
This is done through exploitation
of symmetries inherent in the system by modifying the ideal fluid description to
a non-ideal fluid through inclusion of counter-terms that are related to non-zero pressure,
viscosity and thermal conductivity terms. This also alleviates the UV divergent integrals
that appear at loop level in SPT which often dominate their high-$k$ behavior.
This leads to inclusion of additional parameters
that are calibrated using numerical simulations.
The bispectrum in the EFT was recently discussed in \citep{EFTBi}(
see \citep{MunshiRegan} for more detailed discussion and
complete list of references). 

\ctom{For defining the squeezed limit of the 3D bispectrum we will follow
the following notation:}
\ben
B_{\rm 3D}(\bk_1,\bk_2,\bk_3) \neweq \lim_{\bq_1,\bq_3\rightarrow 0} B_{\rm 3D}(\bk-\bq_1, -\bk+\bq_1+\bq_3, -\bq_3). 
\een
\noindent
\dtom{We have used the parametrization
  $\bk_1= \bk-\bq_1$, $\bk_2=-\bk+\bq_1+\bq_3$ and $\bk_3=-\bq_3$ and taken the
  limit $\bq_1,\bq_3\rightarrow 0$ to recover the squeezed limit \citep{MunshiRegan}.}
\ctom{The superscript $\rm sq$ above represents the squeezed limit.
  Various contributions to the  EFT bispectrum are lsited below:}
\ben
{B^{\text{EFT}}_{\text{2D}} = B^{\text{SPT}}_{\text{2D}} + B_{\delta_c^{(1)}} + B_{\delta_c^{(2)},\delta} + B_{\delta_c^{(2)},e}  + B_{\delta_c^{(2)},\alpha\beta}},
\een
where the arguments to various $B$ terms have been suppressed for clarity. The various terms that contribute to bispectrum at the squeezed limit are then as follows \citep{MunshiRegan}:
\ben
\label{eq:proj1}
&& B^{\text{SPT}}_{\text{2D}}\neweq\left [ {17 \over 7}-{n\over 2} \right ] P_{\rm 3D}(\kp)P_{\rm 3D}(\qp)\,,\\
&&B_{\delta_c^{(1)}}\neweq -\xi \left [\frac{20-7 n}{14}\right ]k_{\perp}^2 P_{\rm 3D}(\kp)P_{\rm 3D}(\qp)\,,\\
&&B_{\delta_c^{(2)},\delta}\neweq -\xi M_d \left [\frac{48-7 n}{14}\right ] k_{\perp}^2 P_{\rm 3D}(\kp)P_{\rm 3D}(\qp)\,,\\
&& B_{\delta_c^{(2)},e} \neweq -\xi M_d \left [ 4{\epsilon_1\over \xi}+{2\over 3}{\epsilon_2\over \xi}+ {1\over 3}{\epsilon_3\over \xi}\right ] k_{\perp}^2 P_{\rm 3D}(\kp)P_{\rm 3D}(\qp)\,,\\
&& B_{\delta_c^{(2)},\alpha\beta} \neweq  -\xi\,
      {2\left [ (56+44 m_d+8 m_d^2) - (11 + 4 m_d) n\right ]\over (2+m_d)(9 + 2 m_d)} k_{\perp}^2
      P_{\rm 3D}(\kp)P_{\rm 3D}(\qp)\,.
\label{eq:proj4}
\een
\ctom{We have used the small-angle approximation in our derivation.
  As before, we have assumed a power-law power spectrum $P_{\rm 3D}(k)\propto k^n$. 
To connect with results derived using SPT in \textsection{\ref{subsec:flat_sky}}
  we use $k_\perp = l/d_A(r)$
  and ${\qp} = l_3/d_A(r)$ and perform the line-of-sight integration similar to Eq.(\ref{eq:def2D_esimator}).}
We will take $\xi = [1.5\pm 0.03]h^{-2}{\rm Mpc}^2$  \citep{EFTBi} \dtom{which is the scale
at which EFT contributions start to dominate and $m_d$ is related to the temporal evolution
of EFT contribution assumed to be $\propto D_+^{m_d}$
($D_+$ being the linear growth factor)} and $M_d$ is defined as:
\ben
M_d := \left [ (m_d+1)(2m_d+7) \over (m_d+2)(2m_d+9) \right ], 
\een
\ctom{where}
\ben
&& { \epsilon_1 \over \xi}= {3466 \over 14091}\,, \quad { \epsilon_2\over \xi} = {7285 \over 32879}\,,
\quad {\epsilon_3 \over \xi}={41982 \over 52879}\,.
\label{eq:params}
\een
Here the counter-terms added to the {\rm SPT} results are denoted by the superscript $\rm SPT$.
The final expression for the normalised IB takes the following form:
\bes\ben
&&    {\cal B}^{\prime}(l) = {\cal B}^{\prime}_{\rm SPT}(l) + R_2^{\rm EFT}\;  l^2 \xi B^{\prime}_{\rm EFT}(l)\\
&&    R_2^{\rm EFT} :=  \int_0^{r_s} \xd\,r {w^3(r)\over d_A^{6+2n}(r)} \bigg / \left ( \int_0^{r_s} \xd\,r{ w^2(r) \over d_A^{2+n}(r)} \right )^2, \\
&& B^{\prime}_{\rm EFT}(l) := -\left [\frac{20-7 n}{14}\right ] - M_d \left [\frac{48-7 n}{14}\right ] - M_d \left [ 4{\epsilon_1\over \xi}+{2\over 3}{\epsilon_2\over \xi}- {1\over 3}{\epsilon_3\over \xi}\right ] \nonumber \\
&& \hspace{2cm} - {2\left [ (56+44 m_d+8 m_d^2) - (11 + 4 m_d) n\right ]\over (2+m_d)(9 + 2 m_d)}. \label{eq:correction}
\een\ees
The SPT value for ${\cal B}^{\prime}$ computed in Eq.(\ref{eq:define_B}) is being denoted here as
${\cal B}^{\prime}_{\rm SPT}$. The prefactor $R_2$ is defined in Eq.(\ref{eq:ib_def}) and its numerical values are tabulated
in Table-\ref{table:model_prediction}.
As explained before, this result
depends crucially on the assumption of locally power-law power spectrum with index $n$.
\ctom{Using a dimensional analysis of Eq.(\ref{eq:correction}) can show
  that the corrections from EFT to SPT scales typically as -$\xi l^2/[d^2_A|\kappa_{\rm min}|]$.
The origin of $\kappa_{\rm min}$ in the correction term stems from using the approximation discussed in
Appendix-\textsection\ref{sec:los}.}


Notice that the parameters defined above can not be computed within the EFT framework.
They are estimated using numerical simulations. Currently these parameters are
known only for $z=0$ and are expected to be valid up to $k\approx 0.22h \rm Mpc^{-1}$ at z=0 \xref{\citep{EFT_bispec}.
  However, this result was derived at the one-loop level and expected to improve when two or more
  loops are included, e.g., the power spectrum one-loop EFT gives sub-percent accuracy up to $k\approx 0.24 h \rm Mpc^{-1}$
  and inclusion of two-loops extends the validity to $k\approx 0.6 h \rm  Mpc^{-1}$ \citep{EFT_power}}.
Derivation of the above expressions also require
the assumptions of self-similarity and the EdS (Einstein-de Sitter) parameter $m_d$ is fixed at $m_d=5/3$.
A different value of $m_d=1$ is also considered.
It is generally assumed that these numerical values will continue to hold for other cosmologies.
\xref{We will perform a detailed comparision of the EFT results and results based on a new fitting function \citep{BiHalo} for
the bispectrum in a separate publication.}
i.e. for values of $M_d$ that we will use in our calculation, $m_d=1$ and $m_d = (1-n)/(3+n)$ where $n$ is the slope of linear power spectrum.
%
\ctom{Here $B^{sq}_{2D}$ is the 2D bispectrum in the {\em squeezed} limit. To separate the projection effect from the dynamical effect we also define a 2D IB and its normalised
counterpart as below:}
\bes\ben
&& \bar{\cal B}_{\rm 2D}(k_{\perp}) = {B^{sq}_{\rm 2D}(k_{\perp}) P_{\rm 3D}(k_{\perp}) \sigma^2_{\rm 2D}};
\quad \sigma^2_{\rm 2D} := {1 \over 2\pi}\int q_{3\perp} \xd q_{3\perp} P_{\rm 3D}(q_{3\perp}); \\
&& {B}^{sq}_{\rm 2D}(k_{\perp}) := {1 \over P_{\rm 3D}(k_{\perp}) \sigma^2_{\rm 2D}} \bar{\cal{B}}^{}_{\rm 2D}(k_{\perp}).
\label{eq:sq_EFT}
\een\ees
\ctom{Comparing this result with Eq.(\ref{eq:define_normIB}) we can check that
  except the factor of $R_2$, 
  ${B}^{sq}_{\rm 2D}$ is identical to ${\cal B}^{\prime}$.
  The geometric effect is encoded in $R_2$ and results from line-of-sight projection.} 

\subsection{Linear Response Function and Separate Universe Approach}
\begin{figure}
  \begin{center}
    \includegraphics{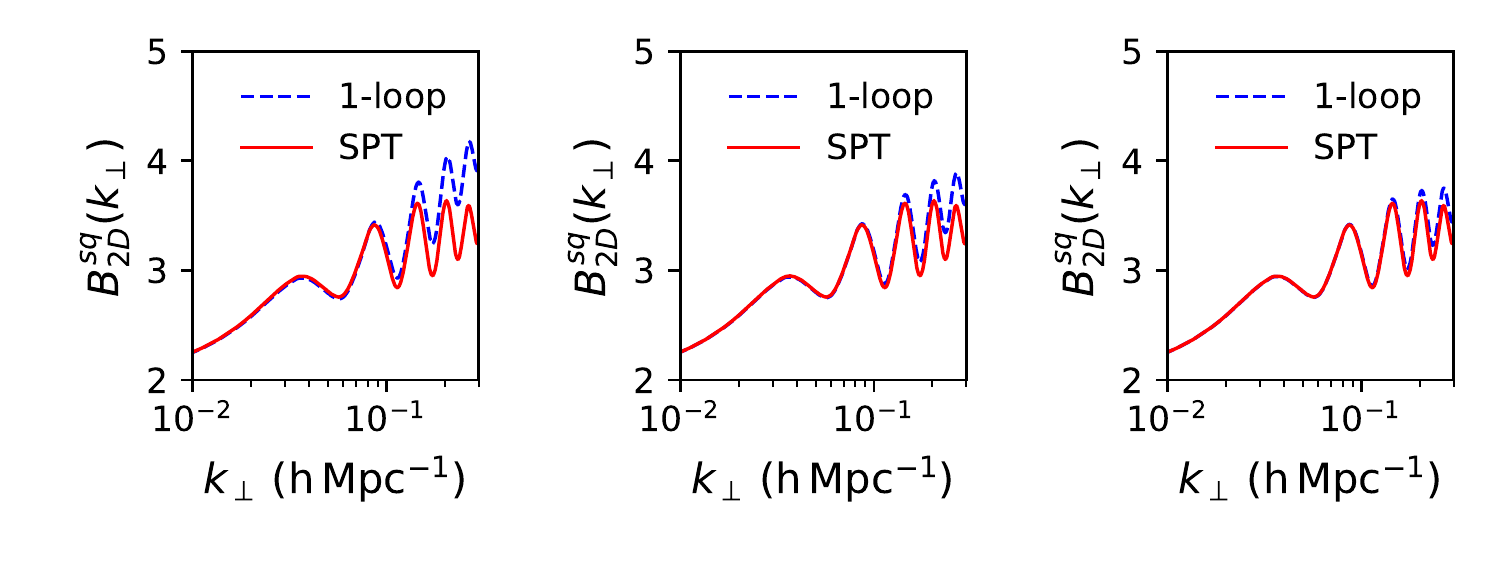}
    \caption{ 
      The normalised 2D IB ${B}^{sq}_{2D}(k_{\perp})$
      defined in eq.(\ref{eq:sq_EFT})
      is plotted as a function of
      the wave-number $k_{\perp}$ for SPT. From left to \cb{right}
      we show redshifts $z=0,1$ and $2$. 
      The expression
      for one loop correction is give in Eq.(\ref{eq:loops})}.
    \label{fig:loops}
  \end{center}
\end{figure}
%
The response function is typically used in conjunction with a separate Universe approach \citep{Sherwin,YinLi,Baldauf,Fabian}.
In the separate Universe approach, the coupling of small-wavelength and long wave-length modes are
treated by assuming each sub volume (sub-patches) actually evolves as a separate Universe.
This approach has been used in various contexts \citep{chiang2,chiang4, chiang5}.
The modulation in background density due to long wavelength perturbation is absorbed in the
redefinition of the background density. This formalism when combined with the perturbative approach results
in the following expressions \citep{bernardeau_review}:
\ben
&& P_{\rm 3D}(k,a) = P_{\rm lin}(k,a) + 2P_{13}(k,a) + P_{22}(k,a) + \cdots; \\
&& P_{22}(k,a) = 2 \int {\xd^3 {\bf q} \over (2\pi)^3} P_{\rm lin}(q,a)P_{\rm lin}(|\bk-\bq|,a) [ F_2(\bq, \bk-\bq) ]^2; \label {eq:loops}\\
&& P_{31}(k,a) = 6 \int {\xd^3 {\bf q} \over (2\pi)^3} P_{\rm lin}(k,a)P_{\rm lin}(q,a) [ F_3(\bq, \bk, -\bq) ]^2.
\een
The kernels $F_2$ and $F_3$ introduced above encapsulates mode coupling at second and third order in perturbation
theory \dtom{(for more discussion about the kernels $F_2, F_3$ as well
as their higher order counterparts and related recursion relation see ref.\citep{bernardeau_review})}.
The response function can be expressed using the  linear power spectrum $P_{\rm lin}(k,a)$ and the loop level corrections
$P_{13}(k,a)$ and $P_{22}(k,a)$ as follows:
\ben
&&{ {\xd\ln P_{\rm 3D}(k_{\perp},a) \over \xd\bar\delta} = {24\over 7} -{1\over 2}{\xd\ln k_{\perp}^2 P_{\rm lin}(k_{\perp},a) \over \xd\ln k_{\perp}} + {10 \over 7}{ 2P_{13}(k_{\perp},a) + P_{22}(k_{\perp},a)
      \over P_{\rm 3D}(k_{\perp},a)}}.
\label{eq:loops1}
\een
\xref{The derivation is very similar to the derivation given in \citep{PhD_chiang}}  for a 3D density field.}. 
The above expression includes loop-level correction at one loop but can be improved by including higher-order loop terms.
However, the entire perturbative series fails to converge at smaller scales and lower redshifts where the
variance is comparable to unity. In this case a nonlinear fitting function is typically used which
reproduces the numerical simulations (e.g. \citep{Salman}). Indeed, such fitting functions, in general, do not provide
an insight to the response to changes in cosmological parameters.
The response function in terms of nonlinear
power spectrum has the following form:
\ben
{{\xd\ln P_{\rm 3D}(k_{\perp},a) \over \xd \bar\delta} = 2+ {10\over 7}{\xd\ln P_{\rm 3D}(k_{\perp},a) \over \xd\ln \sigma_8} -
  {1\over 2} {\xd\ln k_{\perp}^2 P_{\rm 3D}(k_{\perp},a) \over \xd\ln k_{\perp}}}.
\label{eq:separate}
\een
\xref{The derivation follows the same arguments given in \citep{PhD_chiang}. with the angular averages and the resulting numerical
coefficients replaced by their 2D values.}
\ctom{Substituting  ${\xd P_{\rm 3D}/ \xd\bar\delta}$ from
Eq.(\ref{eq:separate}) or Eq.(\ref{eq:loops1}) in Eq.(\ref{eq:how2computeIB1}-\ref{eq:how2computeIB})
we can compute the IB in these approaches as discussed in \textsection\ref{subsec:halo} in the context of halo model.}
Indeed, it is possible to replace $F_2$ in Eq.(\ref{eq:loops}) with an effective  $F_2^{\rm eff}$ that is effectively a fitting function
to numerical simulation \citep{roman,Gil-Marin}. 
\begin{table*}
  \caption{The normalized IB defined in Eq.(\ref{eq:define_normIB})
    as a function of spectral index $n$ and source redshift $z_s$.
    We have used the SPT prescription valid at low values of $\ell$
    presented in Eq.(\ref{eq:define_normIB}) to compute the IB.}
\begin{center}
\begin{tabular}{|c|c |c|c|c|c }
  \hline
  & $n=-1.0$ & $n=-1.5$ & $n=-2.0$ \\
 \hline
 \rowcolor[gray]{0.8} $z_s=0.5$ & 406 & 407 & 442 \\
\hline
 $z_s=1.0$ & 112 & 114 & 124 \\
\hline
\rowcolor[gray]{0.8} $z_s=1.5$  & 56 & 58 & 64 \\
\hline
$z_s=2.0$ & 36 & 37 & 41\\
\hline
\end{tabular}
\end{center}
\label{table:model_prediction}
\end{table*}
 
\begin{figure}[t!]
\begin{center}
\includegraphics[width=50mm]{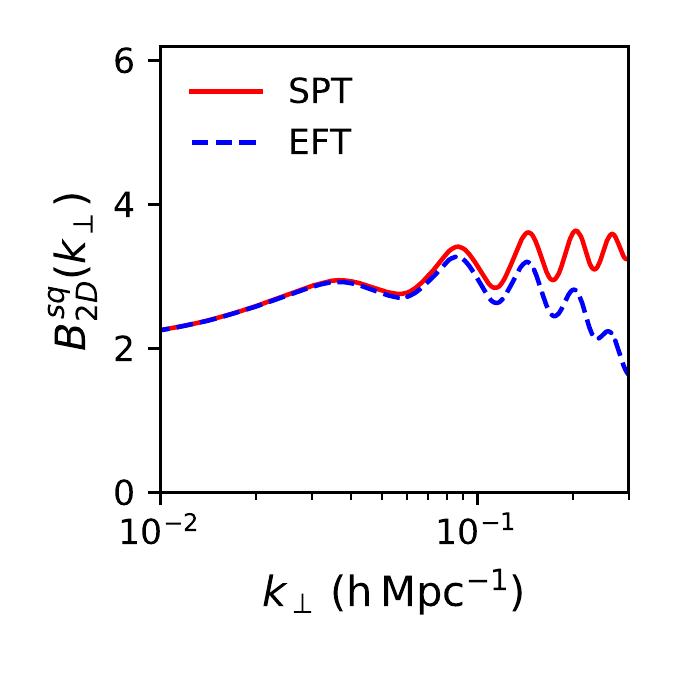}
    \caption{The 2D squeezed bispectrum defined in eq.(\ref{eq:sq_EFT}) is plotted
      as a function of projected wave number (2D) $k_{\perp}$. }
\label{fig:eft}
\end{center}
\end{figure}

The results of comparison of SPT predictions and 1-loop corrections
are presented in Figure-\ref{fig:loops}. Similar results for SPT and EFT
are shown in Figure-\ref{fig:eft}.
%


%
%
%

\section{\textit{Euclid} Flagship Simulations}
\label{sec:sims}
\xref{The Euclid Flagship Simulation~\citep{Potter:17} features a simulation box of 3780 $h^{-1}$Mpc on a side with $12,600^3$ particles, leading to a mass resolution of $m_p = 2.4 \times 10^9$ $h^{-1}$M$_{\odot}$. A softening length of $6 \rm h^{-1}kpc$ was used.
  This 2 trillion particle simulation is largest N-body simulation performed to date and represents a significant computational and man--power investment to the Euclid mission. It was performed using PKDGRAV3~\citep{Potter:16} on the Piz Daint supercomputer at the Swiss National Supercomputer Center (CSCS) in 2016.
The simulations were started at a redshift of $z_{\rm Start} = 49$ and a 2nd-order Lagrangian Perturbation Theory
was used for initial particle displacement. The
CAMB\footnote{\href{https://camb.info}{\tt CAMB}}
transfer function was set at $z=0$. It took runtime of $80$ hours on $4000$ Nodes ($8$ core + K20X GPU).}

\xref{An agreed upon reference cosmology, close to Plank 2015 values, was used with the following parameters: $\Omega_m = 0.319, \Omega_b = 0.049, \Omega_{\rm CDM} = 0.270, \Omega_{\Lambda} = 0.681, w = -1.0, h = 0.67, \sigma_8 = 0.83, n_s = 0.96$. A contribution to the energy density from relativistic species in the background was ignored ($\Omega_{{\rm RAD},\nu} = 0$). Using this Euclid Reference Cosmology allows comparison to many other smaller simulations from N-body codes as well from approximate techniques that also use these reference values within the collaboration. The main data product was produced on-the-fly during the simulation and is a continuous full-sky particle light cone (to $z=2.3$), where each particle was output exactly when the shrinking light surface sweeps by it. This resulting ball of particles contains 10 trillion particle positions and peculiar velocities (240 TB), and it was used to compute the halos and lensing maps (HealPix) for the Flagship Galaxy Catalogue.}

\xref{Following the approach presented in~\cite{Fosalba:08} and~\cite{Fosalba:15b}, we construct a lightcone simulation by replicating the simulation box (and translating it) around the observer.  Given the large box-size used for the Flagship simulation, $L_{box}$=3780 $h^{-1}$Mpc,  this approach allows us to build all-sky lensing outputs without repetition up to $z_{max}=2.3$.}

\xref{Then we decompose the dark-matter lightcone into a set of all-sky concentric spherical shells, of given width $\Delta_r$, around the observer, what we call the ``onion universe''. Each dark-matter ``onion shell'' is then projected onto a 2D pixelized map using the Healpix tessellation.  By combining the dark-matter ``onion shells''  that make up the lightcone, we can easily derive lensing observables, as explained in \cite{Fosalba:08} and \cite{Fosalba:15a, Fosalba:15b}.  This approach, based on approximating the observables by a discrete sum of  2D dark-matter density maps multiplied by the appropriate lensing weights, agrees with the much more complex and CPU time consuming ray-tracing technique within the Born approximation, i.e., in the limit where lensing deflections are calculated using unperturbed light paths, see \cite{Hilbert:20}.}
\xref{The maps we used were generated from fully continious light-cones produced
during the simulations.}

\begin{figure}[htb]
  \centering
  \includegraphics[width=75mm]{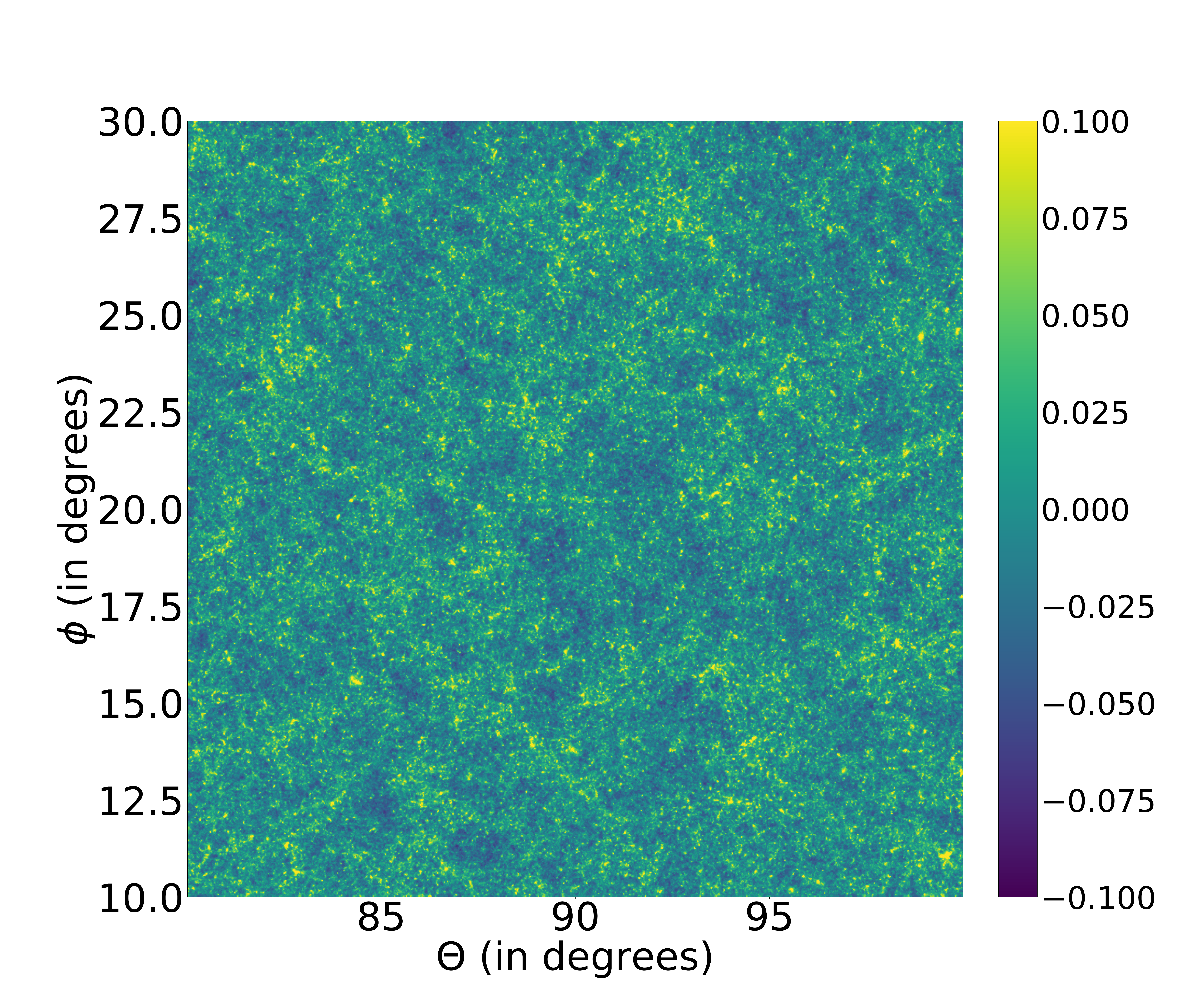}
  \includegraphics[width=75mm]{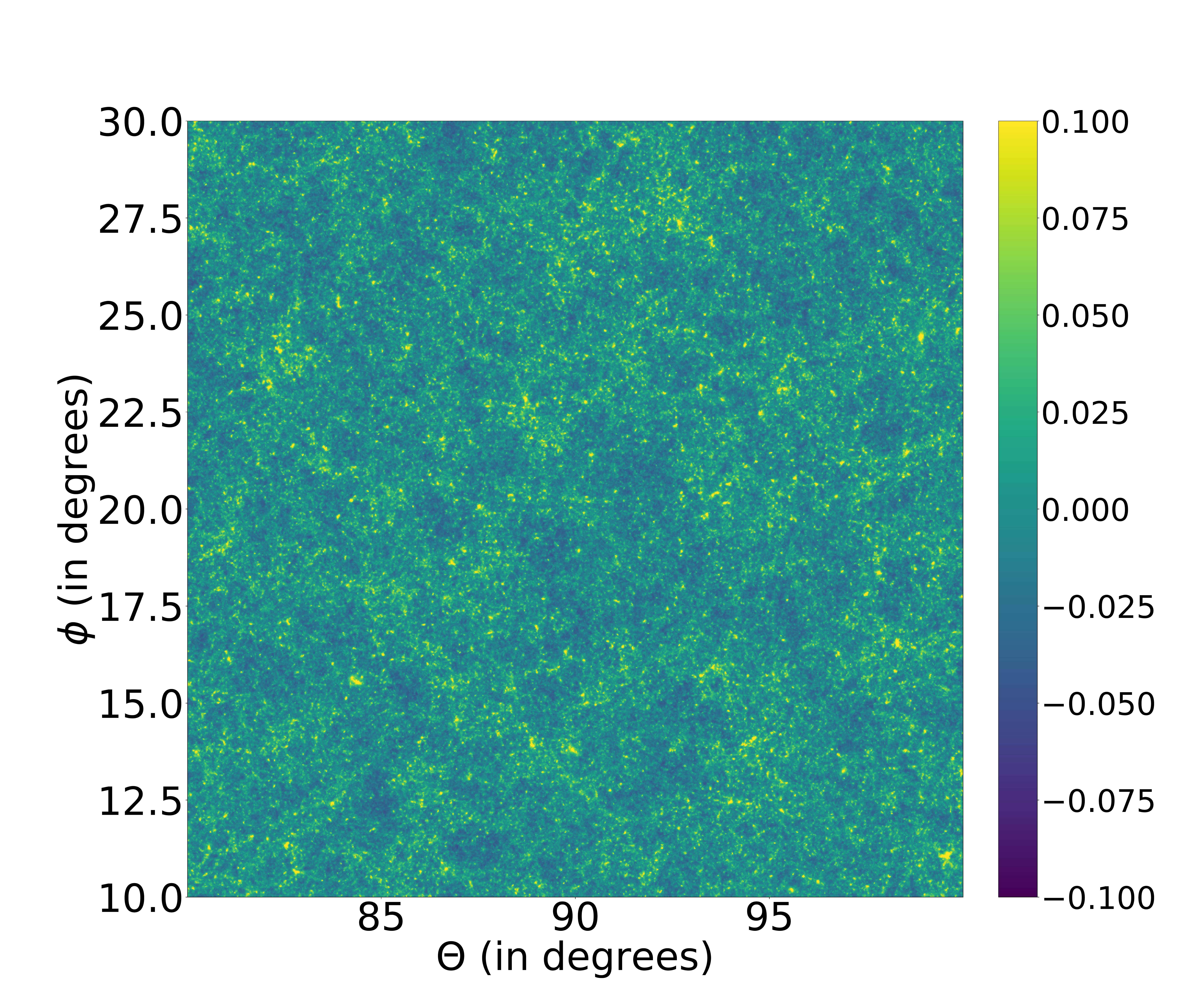}
  \hspace{0cm}
  \caption{The convergence $\kappa$ maps on 2D patches of the sky generated using the all-sky Flagship
    Euclid simulations are presented.
    The patches are $20^{\circ}\times 20^{\circ}$ in size and are created on a grid $1024\times 1024$ grid.
    The original all-sky maps were generated using $\ell_{max}=2N_{side}$ with $N_{\rm side}=2048$. Two different
    panels correspond to different
    source redshifts. The left panel correspond to $z=0.5$ and the right panel correspond to $z=1.0$
    (see text for more details).}
  \label{fig:figs1}
\end{figure}

\begin{figure}[htb]
  \centering
   \includegraphics[width=75mm]{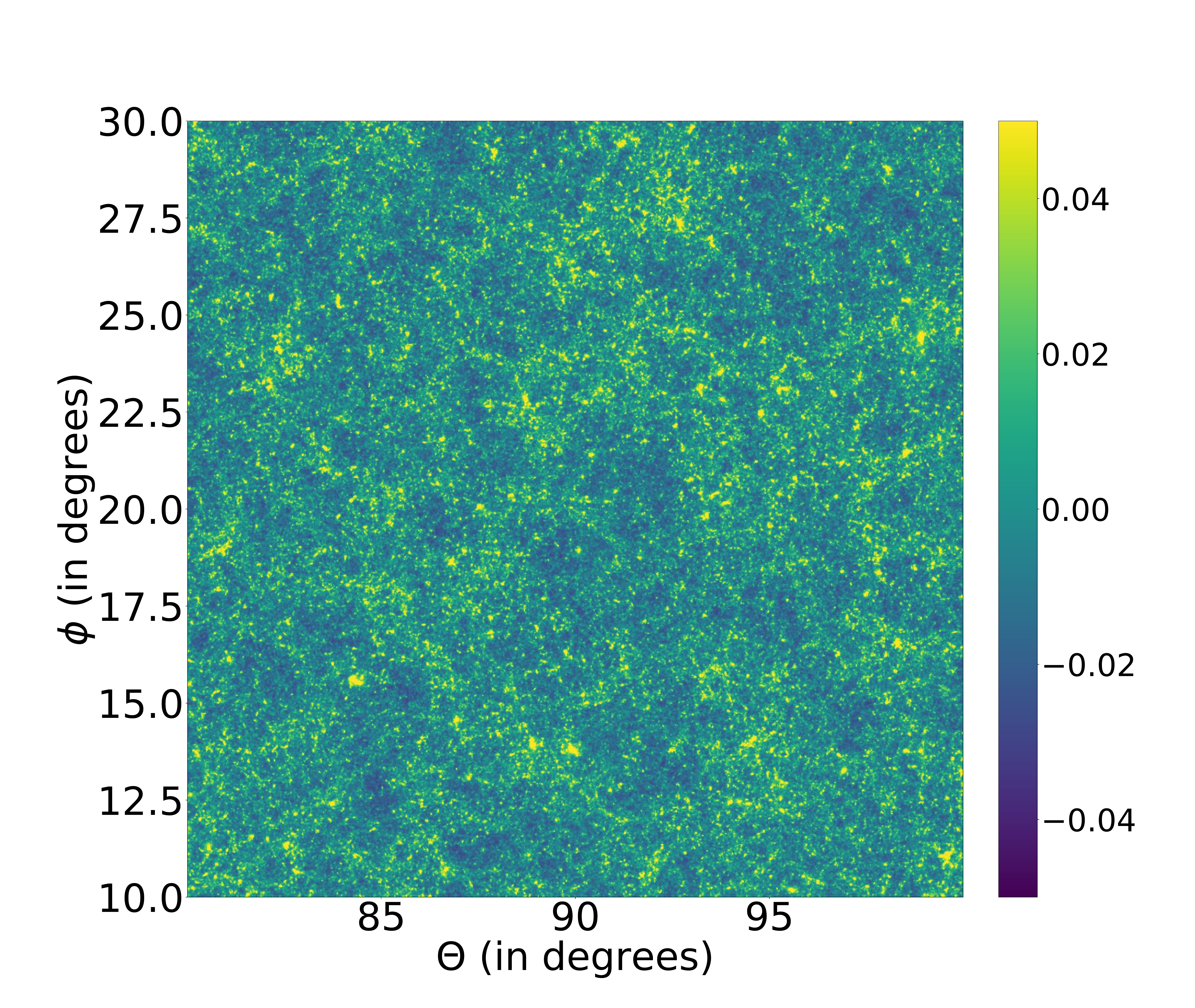}
  \includegraphics[width=75mm]{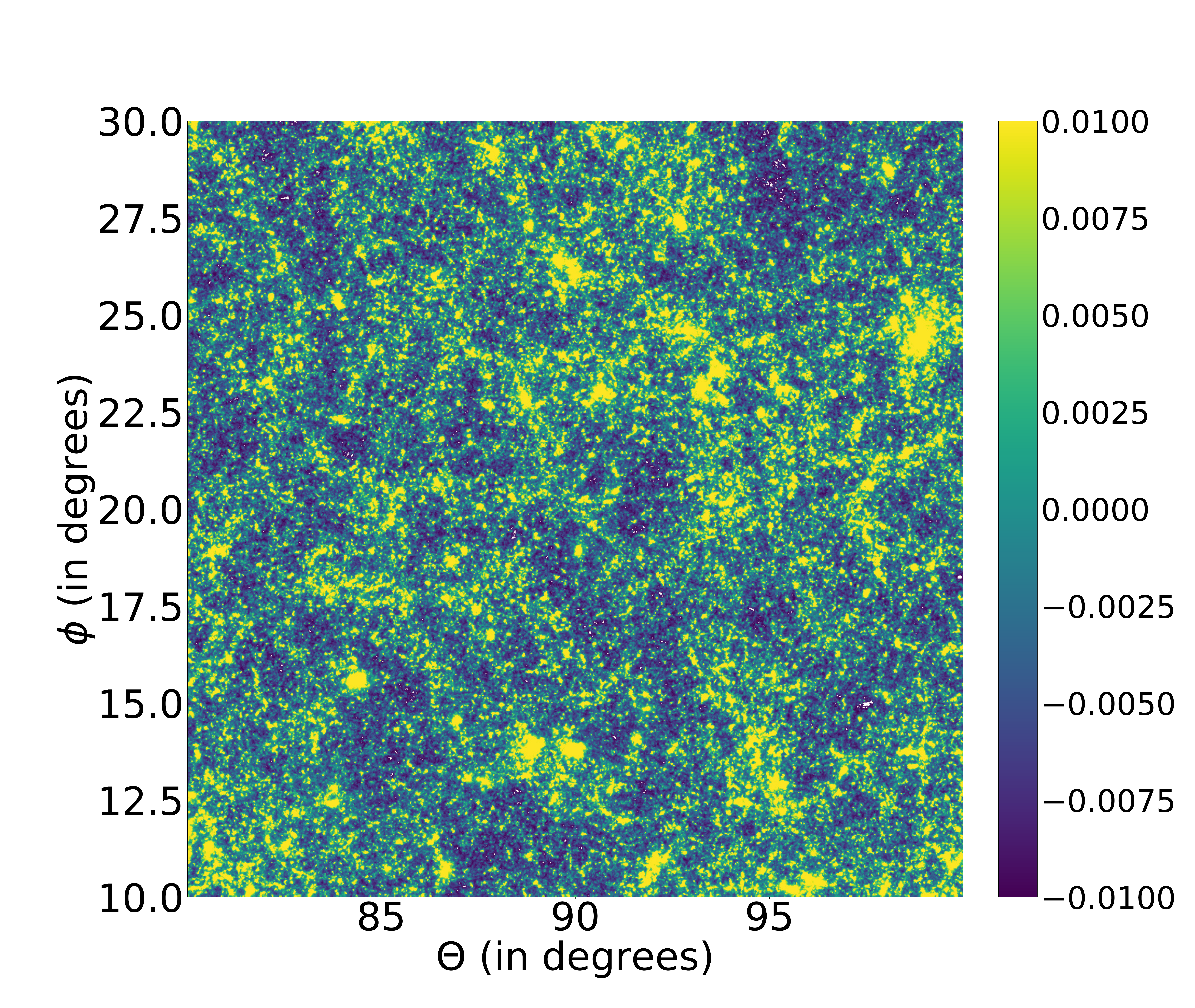}
  \caption{Same as Figure-\ref{fig:figs1}. The left (right) panel corresponds to $z=1.5(2.0)$.}
  \label{fig:figs2}
\end{figure}

\cb{We note that for the analysis presented here,
we have degraded the original {\tt HEALPix} maps to a lower resolution,
  $N_{\rm side}=2048$ which corresponds to a pixel scale of $1.7$ arcmin.
The square patches generated for our study were created using
a recatngular grid and interpolating the value of the convergence map
from the nearest {\tt HEALPix} pixel.}
Few examples of convergence maps
generated for our study are presented in Figure -\ref{fig:figs1}
and Figure -\ref{fig:figs2}.

\xref{These simulations were also used to produce mock galaxy catalogs.
The galaxy catalogs were validated extensively against many
cosmological observables, including redshift space clustering
multipoles, number counts of H-alpha emission line galaxies,
and distributions as a function of redshift to name a few.}

%

\section{Error Budget for IB}
\label{sec:error}
%
%


\subsection{Bias}
The finite volume of a survey makes the estimators of the higher-order correlation functions
or the associated spectra biased. In case of two-point correlation function
(or equivalently the Power Spectrum), this is also known as the \xref{{\em integral constraint} (see \citep{LamHui} and references therein)}.
In the weakly non-linear regime this is directly proportional to average of the two-point
correlation function $\bar\xi_2(L)$. At smaller scales, correction term also depends
on the squeezed limit of the very bispectrum (convolved with survey window)
which we are trying to estimate (see Eq.(413) in \citep{bernardeau_review}).
Such corrections however play an important role in estimation of the
IB at low redshift e.g. $z=0.5$.
\xref{In \citep{LamHui} a systematic prescription was developed to compute the bias in
of (volume-averaged) three-point correlation functions and other high-order statistics, which can be specialised for the case
of squeezed three-point correlation function and its Fourier transform the IB.
A detailed analysis will be presented elsewhere.}

\subsection{Scatter}
Estimations of bispectra and its covariance are rather difficult to compute as they are functions of three different
wave vectors that define a specific triangular configuration. The estimation of bispectra
for all possible triangular configuration can be expensive. The IB
on the other hand involves only wave vectors as it focuses only on the squeezed limit of the
bispectrum. However, quantifying bispectrum and its covariance in the squeezed limit
probes coupling of short and long wavelength modes. 
Direct evaluations of bispectra in the
squeezed limit require simulations with high dynamical range, which can resolve both
high and low wave numbers reasonably well. In general separate universe models are
implemented where the long wave fluctuations are absorbed in the background evolution with a modified
cosmology which allows smaller boxes to be used for resolving the small wavelength modes independently.

The covariance calculation we employ here is a simple order of magnitude calculation
based on the counting of modes based on the survey size, and the smaller patches created from it.
\xref{We follow the derivation presented in \citep{PhD_chiang} for the 3D case.}
For the purposes of
covariance estimatation it is assumed that the underlying distribution is Gaussian.
\bes
If we take a 2D survey (simulation) with an area denoted as $A_{s}$ and focus on patches of area $A_{p}$.
the number of patches is $N_{p} = {A_{s}/A_{p}}$. The number of Fourier modes in the interval
$({\bl}_{}- {\Delta {\bl}_{}/2})$ and $({\bl}_{} +{\Delta {\bl}_{}/2})$ is denoted as $N_{L}$.
The variance of the convergence bispectrum defined in Eq.(\ref{eq:define_B}) is defined as:  
\ben
&& \sigma^2[B^{\kappa}({\bl}_{})] =
\langle [B^{\kappa}({\bl}_{})- \la B^{\kappa}({\bl}_{})\ra]^2 \rangle =
\langle B^{\kappa}({\bl}_{})^2\rangle - \langle B^{\kappa}({\bl}_{})\rangle^2\\
&& \sigma^2[B^{\kappa}({\bl}_{})]= {A_{p} \over A_{s} N_{kL}}\sigma_L^2 [P_L^{\kappa}({\bl}_{})]^2
\een
The variance of the normalized bispectrum defined in Eq.(\ref{eq:define_normIB}) on the other hand is given by:
\ben
\sigma^2[{\cal B}^{\kappa}({\bl}_{})] = {A_{p} \over A_{s} N_{L}}{1\over \sigma_L^2}, 
\label{eq:norm}
\een
where we have introduced the following quantities:
\ben
&& \sigma^2_L := {1 \over V^2} \int {\xd^2\bl_{}\over (2\pi)^2}
P^{\kappa}_{lin}({\bl}) |W({\bl})|^2 ; \\
&& P_L^{\kappa}({\bl}_{}) := {1 \over V^2} \int {\xd^2\bl^\prime\over (2\pi)^2}
P^{\kappa}_{lin}({\bl}_{})W(|{\bl}_{}-{\bl}^\prime|).
\een
\ees

$P^{\kappa}_L({\bl})$ is the power-spectrum convolved with window for the sub-patches.
The above derivations ignore all higher-order correlations.
The expression in Eq.(\ref{eq:norm}) assumes that the variance is dominated by the
Gaussian contribution thus including only the disconnected terms in the calculation.
Higher-order contributions beyond bispectrum contributes to the variance
and can be equally dominant in the low source redshift where the underlying
distribution is highly nonlinear. However following \citep{PhD_chiang}, these contributions are not included.

\xref{A more general derivation will be presented elsewhere which will
  include higher-order contribution to the covariance \citep{szapudi1,szapudi2, szapudi3, Error_Munshi_Coles}}

\section{Results and Discussion} 
\label{sec:disc}
%
\begin{figure}
  \begin{center}
    \includegraphics[width=100mm]{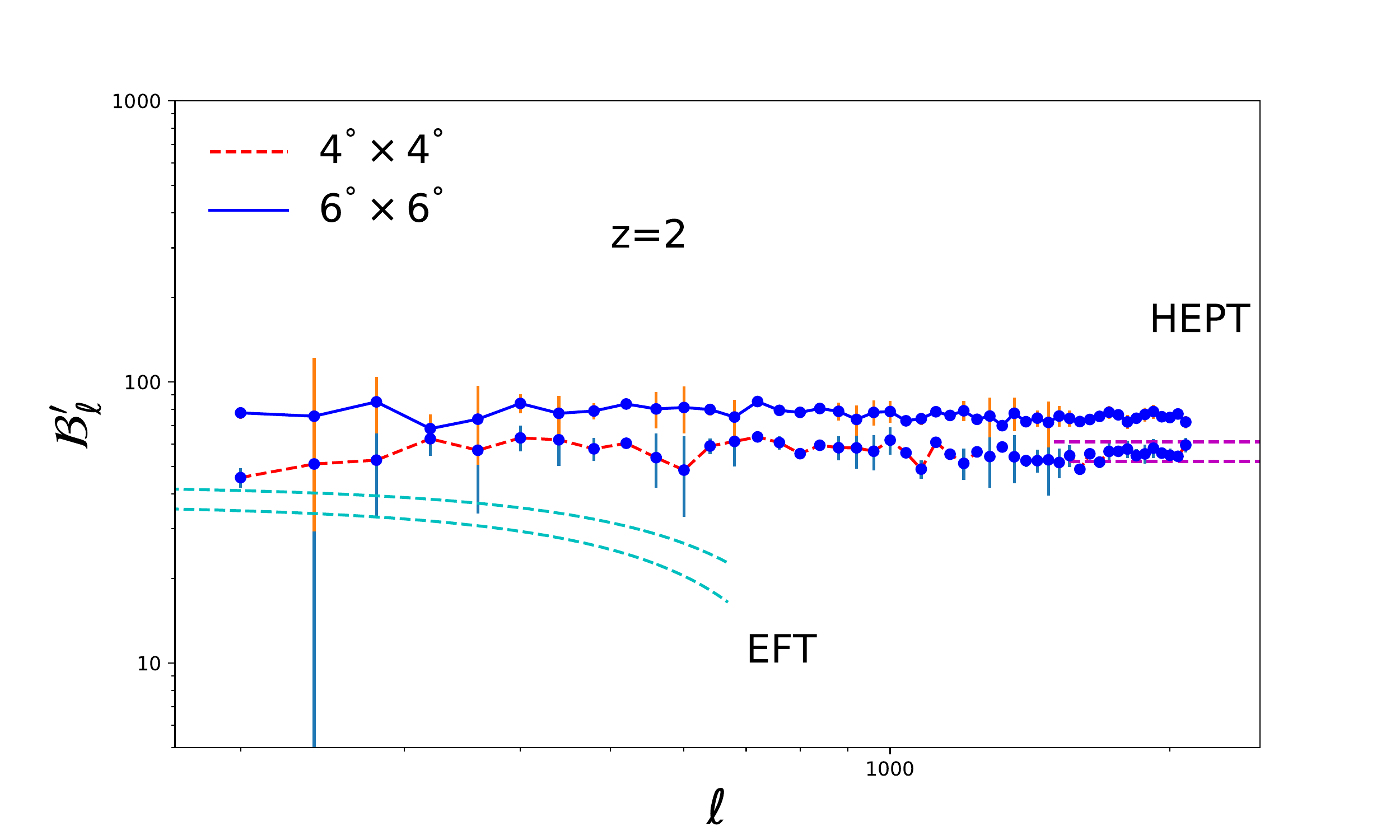}
    \vspace{0.cm}
    \caption{We show the estimated IB or ${\cal B}^\prime_\ell$ defined
      in Eq.(\ref{eq:define_normIB}) as a function of the multipoles $\ell$. A total of 30
      logarithmic bins were
      used to estimated the bispectrum. The red (blue) dots represent a box-size
      of $4^{\circ}$($6^{\circ}$). A total of more than 350 patches were used.
      The original all-sky map was degraded from $N_{\rm side}=2048$ to $N_{\rm side}=1024$.
      The 2D patch-sky maps were constructed on a $1024\times 1024$ grid. We do not
      include noise. The convergence map considered for this plot correspond to
      a redshift of $z=2.0$. The error-bars were constructed from approximately
      300 patches created for our study.
      \dtom{The dashed-lines shown at low $\ell$ are
        the predictions from EFT for spectral slope of $n=-1.5$ (bottom curve)
        and $-2.0$ (top curve) 
      as tabulated in Table-{\ref{table:model_prediction}} 
      The dashed lines at high-$\ell$ are predictions from HEPT with exactly same spectral slope.
      We have truncated the EFT predictions when they become $50\%$ of the SPT predictions
      as they are developed at 1-loop and not likely to be valid beyond this point.
      The HEPT is not expected to valid at low $\ell$ which probes mainly quasi-linear regime
    (see text for more details).}} 
    \label{fig:IB_z2d0}
  \end{center}
\end{figure}
%
\begin{figure}
  \begin{center}
    \includegraphics[width=100mm]{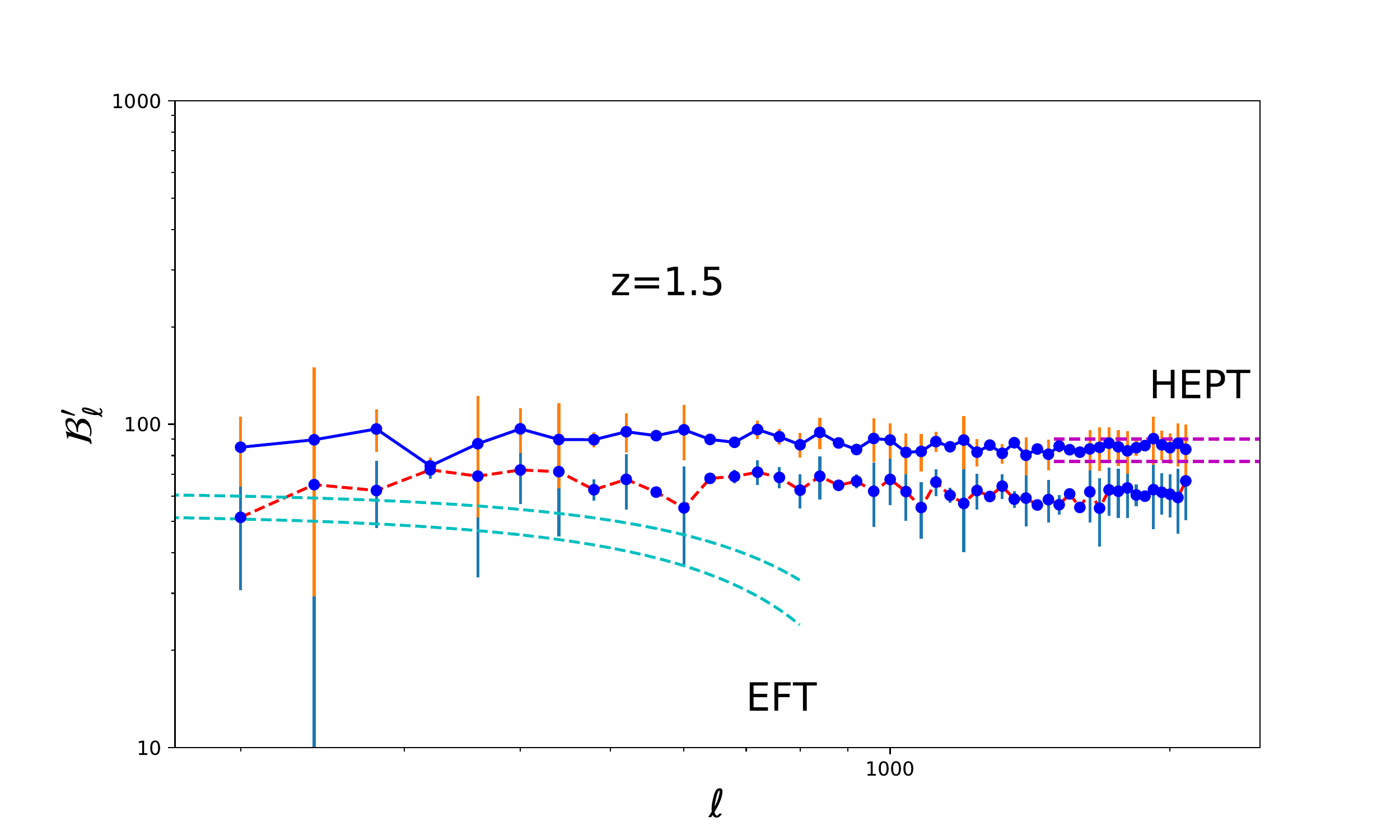}
    \caption{Same as Figure-\ref{fig:IB_z2d0} but for $z_s=1.5$. \dtom{Notice that the value of the normalized IB
        increases with decrease in source redshift $z_s$. This is due to increase in the value of $R_2$ defined in
        Eq.(\ref{eq:ib_def})}. We show results for $\ell \ge 200$ as for smaller values
      of $\ell$, corrections due to the departure from Limber approximation, used in our derivation,
      can no longer be ignored.}
    \label{fig:IB_z1d5}
  \end{center}
\end{figure}
%
\begin{figure}
  \begin{center}
    \includegraphics[width=100mm]{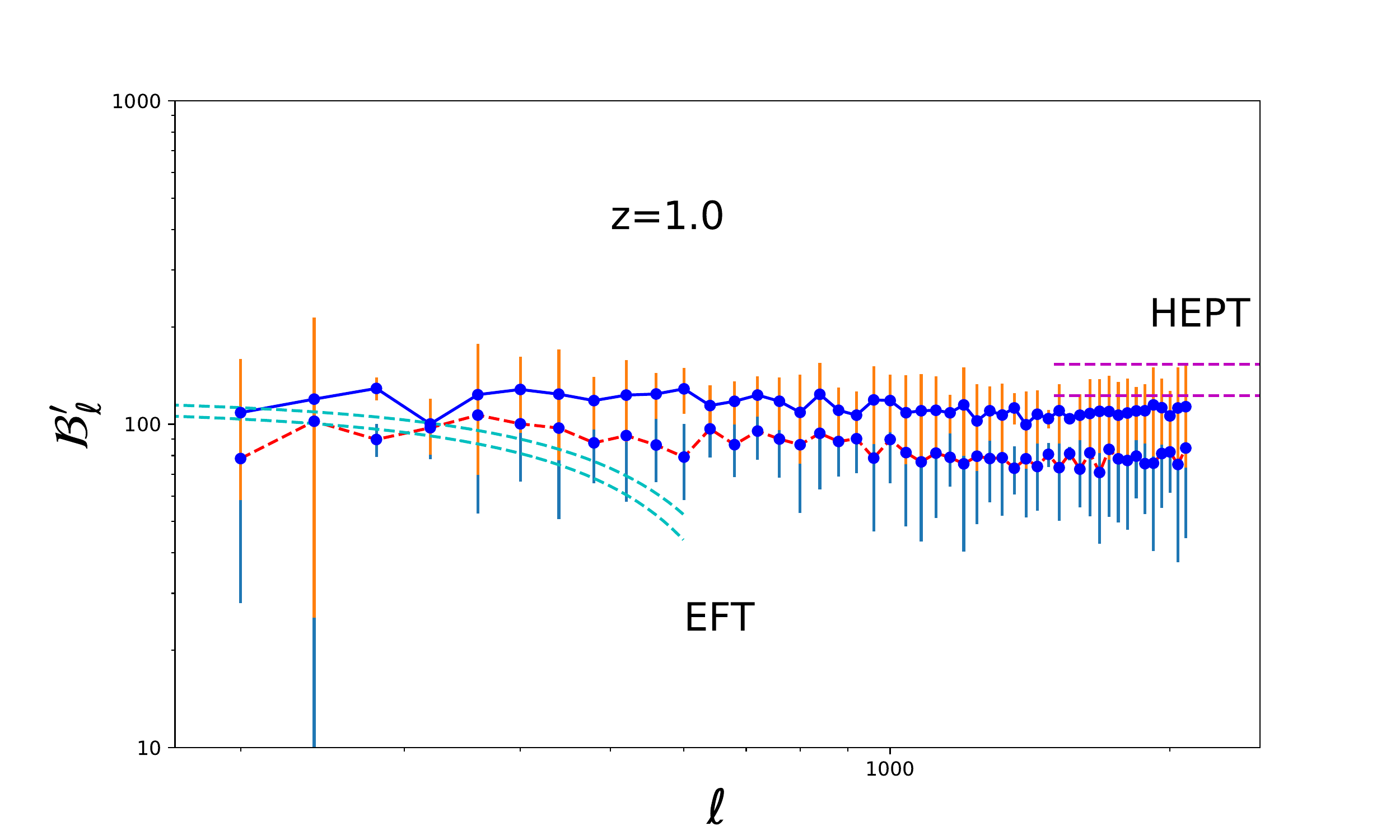}
    \caption{Same as Figure-\ref{fig:IB_z2d0} but for $z_s=1.0$. \dtom{The range of $\ell$ for which the EFT is
      valid decreases with decrease in source redshift $z_s$ as the light rays encounter higher level of
      non-linearity at all scales. Due to the lack of formal derivation of validity range of EFT, we have
      plotted the EFT predictions till its value becomes comparable to
    $50\%$ of the SPT predictions. Indeed this also depends on the values of the EFT parameters being used.}}
    \label{fig:IB_z1d0}
  \end{center}
\end{figure}
We present results of the numerical evaluation of the IB from simulated maps. This is done by first dividing the
all-sky maps into non-overlapping square patches; the next step is to 
cross-correlate the power spectrum estimated from these patches
with the average convergence $\bar\kappa$ from these patches.
In our study,
we have considered $18$ sub-divisions in the longitudinal $\theta$
direction and $9$ sub-divisions in the azimuthal $\phi$ direction
where the center of the individual patches are placed i.e. $\theta_{\rm S}=10^{\circ}$
and $\phi_{\rm S}=10^{\circ}$.
The Euclid flagship all-sky maps were constructed at a resolution
specified by the {\tt HEALPix}\footnote{{\href{http://healpix.sourceforge.net/gallery.php}{\tt Healpix}}}
resolution parameter $N_{\rm side}=2048$. We have considered various degradations
to $N_{\rm side}=1024$ and $N_{\rm side}=512$ to ascertain the pixelisation error.
The square patches were constructed at $1024\times 1024$ as well
as $512\times 512$ to investigate any residuals originating from mapping
from all-sky to patch-sky. We have considered convergence maps
at four different redshifts $z_s=0.5,1.0,1.5$ and $2.0$.
For estimation of the power spectrum, we have taken the
minimum and maximum value of harmonics to be $\ell_{\rm min}=200$
and $\ell_{\rm max}=2000$ with $30$ logarithmic bins. These choices are dictated by
the patch size and pixel size. 
%
\begin{figure}
  \begin{center}
     \includegraphics[width=100mm]{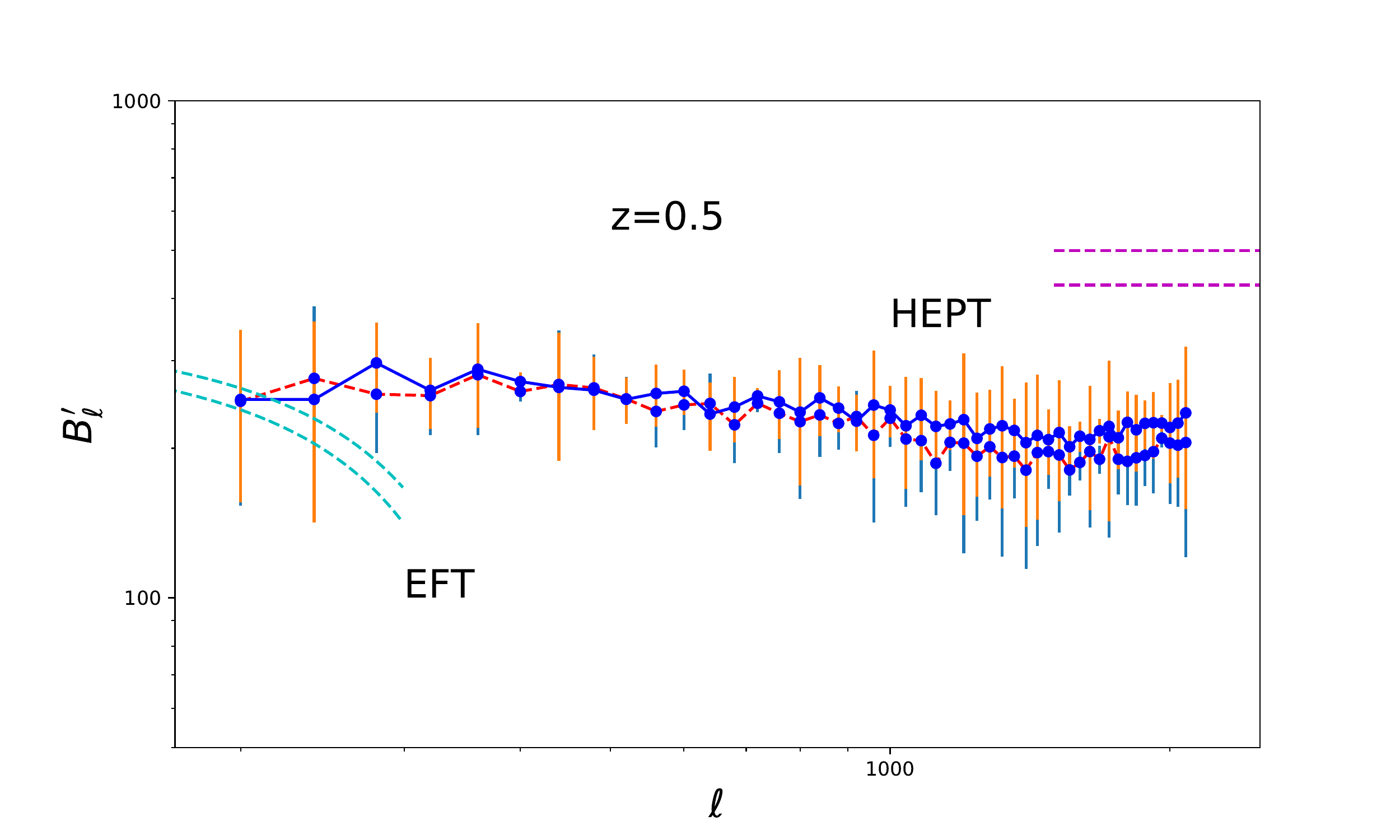}
    \caption{\dtom{Same as Figure-\ref{fig:IB_z2d0} but for $z_s=0.5$. At this particular source redshift the
        the light rays which are being lensed are probing highly nonlinear regime. This means the
        range of validity of EFT is rather limited. Indeed, the bias introduced by the
        finite size of the map are more pronounced too. In addition the theoretical predictions are lot more sensitive
        to the source redshift $z_s$. These effects combine to make comparison of theoretical predictions and
    numerical estimates from simulations rather difficult.}}
    \label{fig:IB_z0d5}
  \end{center}
\end{figure}
%
%
The results of our analysis for various redshifts are shown respectively in
Figure -\ref{fig:IB_z2d0},
Figure -\ref{fig:IB_z1d5}, Figure -\ref{fig:IB_z1d0} and
Figure -\ref{fig:IB_z0d5}. We have compared the theoretical expectations
from SPT against the results from simulations. For the highest redshift
maps the detection of non-Gaussianity is dominated
by the variance of the IB estimator and lower redshift they are dominated
by finite (volume) size effect.

We summarise our observations here: 
\begin{itemize}
\item
  {\bf Finite size of the catalog:}
  \xref{The higher-order statistics such as the IB are sensitive to
    the high-$\kappa$ tails of the PDF. This tail represents the rare events.
    A survey of finite-size, do not sample the
   rare events accurately. This results in a biased estimation of high-order
  statistics. This bias caused by the finite size of the catalogue primarily
  depends on the level of non-Gaussianity as with increasing
  non-Gaussianity the tail gets extended further.
  Such a bias also depends on the order of the statistcs being employed.
  The higher the order more sensitive it is to the tail as it is more sensitive
  to the rare events represented by the tail.}
  
At low redshift, the non-Gaussianity is more pronounced hence
effects due to finite size of the map bias the estimates.
This effect has been studied in great detail in the literature
for other non-Gaussianity estimators
namely the lower-order one-point and two-point cumulants
and their correlators \citep{Colombi1,Colombi2,MunshiBernardeau}. The IB probes the same underlying 
statistics as skewness namely the bispectrum. Thus, it is expected,
that it too will be affected by finite size of a catalog in a similar manner.
It can be reduced by increasing the size of the catalog. However, correction
to the small angle approximation needs to be included in analysing low-$\ell$  estimates
from such patches. Indeed, such
an effect also depends on the level of non-linearity. So, corrections
are more pronounced at lower redshift ($z_s=0.5$) where convergence maps
\xref{have higher variances} compared to the ones at higher redshift ($z_s=2.0$).
This results in a larger deviation from theoretical (SPT) predictions in Figure -\ref{fig:IB_z0d5}.
We have not generalized the available methods in the
literature \citep{bernardeau_review} to {\em correct} the
bias in our estimates at low $z$ which is left to a future work.
Most current results concern one-point statistics in real space.

The corrections from the curvature of sky start play an increasingly dominant
role for $\ell < 100$ \citep{flat1,flat2,flat3}.
To ensure that our results are not affected by any deviation
from flat-sky approximation we have only considered the results for $\ell>200$. 

\item
{\bf Approximate window function:}
We have presented our results
for two different patch sizes to investigate
the effect of the {\em window function}. Our numerical results
were evaluated using a square patch while the theoretical
results were derived using a circular window.
Such effects are more visible at higher redshifts as the
underlying non-Gaussianity is relatively low. \dtom{It is possible to
include a circular window when analysing the maps or
include the effect by direct integration when computing
theoretical predictions. However, this will not change the results
at a fundamental level.}

\item
{\bf Sample variance and error-bars:}
To have crude estimates of error bars,
we have used the patches generated for our study.
We haven't included \xref{galaxy shape} noise \xref{as well as the shot noise}  in our analysis. Inclusion of
noise would reduce the signal-to-noise (S/N) \xref{see \citep{Error_Munshi_Coles} for for more discussion}. 

\item
{\bf Comparison against theory:}
Computation of the IB depends on a particular model of bispectrum.
In the past various scenarios such as the
Standard (Eulerian) perturbation theory (SPT; at two and three
loops) (see Figure-{\ref{fig:loops}}), halo models,
separate Universe models and effective field theories (see Figure-{\ref{fig:eft}}) were used.
\xref{However, previous results were relevant mainly for 3D surveys (see \citep{PhD_chiang} and references therin).}
We have presented a number of approaches
and shown that as far as the theoretical predictions are considered
there is a huge variation in the underlying 3D bispectrum
depending on the underlying model, e.g.\ EFT results points to a lower
values of the IB for high-$k$ compared to SPT whereas, other
methods predict higher than SPT values. 
Ideally, to be consistent
the parameters defining a specific implementation of EFT
should be extracted from the N-body simulations from
which the convergence maps are obtained through ray-tracing simulations.
However, this is outside the scope of our current study.
Indeed, one can also
use an accurate {\em fitting functions} for the bispectrum to derive the squeezed limit
and the resulting IB \citep{Gil-Marin}. \dtom{Most such fitting functions include the
Hierarchal Ansatz (introduced before) as a high $k$ limiting situation.
We have tested the predictions for hierarchal ansatz coupled to HEPT
for the high $\ell$ regime and find reasonable agreement of the general trend. This is encouraging
as it proves that a full non-linear calculation may provide accurate
prediction for the entire range of $\ell$ values. However, we also notice significant
departure at the lower redshift}

\xref{We would like to also point out that the integrated bispectrum (equivalently the squeezed bispectrum)
has recently been studied by other authors using very different
simulations. The conclusion of these studies agree with our conclusions.
Many theoretical predictions based on fitting functions
actually fail to reproduce the results from numerical simulations accurately.
These fitting function \citep{Hyper, Gil-Marin} are interpolations
of perturbative results and the results in the highly nonlinear regime
based on suitable modification of HEPT which we test. It was found that the simulations
consistently produce results that are lower than the theoretical
predictions. This was found using different techniques \citep{Coulton18}
who analysed square-patches of the sky
using Fourier-transformation based approaches. Whereas \citep{Namikawa1}
studied not only equilateral and squuezed configurations
but also folded and isosceles configurations and showed that the
analytical estimates do not match numerical simulations and the squeezed configuration
is most severely affected among other configurations.}
 
\xref{Based on these studies a new fitting function was recently proposed
  in \citep{Halofit} which seems to produce better fit to simulation data.}

The presence of the bias due to finite volume
(area) correction makes a direct comparison difficult.
Thus, in this work we have restricted our comparison against
the SPT. Though, we have provided detailed derivations of
other methods of theoretical modeling. For SPT,
we have assumed a locally power law power spectrum
which simplifies theoretical analysis. We have used three spectral
index $n=-1.0,-1.5$ and $-2.0$ to the variation in theoretical
prediction.

\item
\xref{{\bf Beyond bispectrum to higher-order:}
The idea of the skew-spectrum has already been extended to
higher-order power spectrum which leads to two different
kurtosis spectra. In Appendix-\textsection\ref{sec:Tri} similar extensions is possible
in case of IB to higher-order which will
probe the collapsed and squeezed configurations of the trispectrum.
The collapsed spectra will correspond to the covariance
of the IB and the squeezed limit of the tripsectrum
will represent the position-dependent bispectrum.
These estimators can be useful in situations where
the bispectrum vanishes identically and the trispectrum
remains the dominant contribution to non-Gaussianity (e.g. kinetic Sunyaev-Zeldovich sky). In surveys with
high S/N it would be possible to extend such methods
beyond the kurtosis spectra to fifth-order or even higher.
Corresponding correlation functions in real space will also
be equally effective non-Gaussianity statistics for weak lensing surveys.
However, the effect of finite area covered by the survey
will increasingly play a dominant role and techniques must be
developed to correct such effect. We also provide the expressions
of doubly-squeezed trispectrum in the context of response function approach.}

\end{itemize}

\section{Conclusions and Future Prospects}
\label{sec:conclu}
We have used the IB to probe the gravity induced non-Gaussianity
from simulated weak lensing maps. The results derived are valid for generic projected surveys.
Using the small angle approximation we show how an estimator for the
IB can bypass many of the complexities associated with the estimation of the entire bispectrum. 
By further focusing on the squeezed limit the IB,
we can avoid many of the issues associated with estimation of bispectrum, 
and can compute the IB using ordinary power spectrum estimation techniques.
Indeed, this comes at a price as information is only available in the squeezed limit.
We discuss below some key-aspects of our study and point to future directions:

\begin{enumerate}

\item
  {\bf Tomography and Bayesian generalization:}
  Extending flat-sky results we also develop a full-sky
  estimator for the IB. These estimators take advantage of the existing 
  Pseuco-${\cal C}_{\ell}$ estimators developed for power spectrum analysis.
  A tomographic generalization can be implemented by correlating
  convergence maps from one bin against power spectrum from
  the same or different bins.
  We also point out that the estimators based on
  pseudo-${\cal C}_{\ell}$ can also be extended to a Bayesian version
  by using a Bayesian power spectrum estimator (see e.g. \citep{HamiltonSampling,Wandelt}).

\item
  {\bf Theoretical estimates:}
  We have developed analytical estimates of IB using various approximations
  to gravitational clustering including variants of halo-models, tree-level perturbation theory
  as well as using EFT.

\item
{\bf Test against \textit{Euclid} Flagship simulations:}
We have used the Euclid Flagship simulations to test our results using the convergence
or $\kappa$ maps. We have studied the impact of resolution, patch-size and level of non-Gaussianity
as a function of redshift $z$. \ssb{We compared the results against the theoretical predictions
  and founnd the existing theoretical models to over-predict the results obtained from simulation.}
We have presented a rather simplistic model for
error analysis. More realsitic analysis of error covariance will be presented elsewhere.

\item
{\bf Generalization of IB to shear $\gamma$ and other spin-2 objects:}
Our results can be
easily generalized to shear $\gamma$ maps and directly used to analyses the maps from
ongoing and planned weak lensing surveys. In this case one would replace the
pseudo-${\cal C}_\ell$ estimator for spin-$0$ scalar field with a spin-$2$
estimator for ``Electric'' $\rm E$ and ``Magnetic'' $\rm B$ power spectra. A correlation
of these spectra with the $\rm E$ and $\rm B$ maps will
provide a squeezed estimator for the pure $\rm EEE$, and $\rm BBB$ bispectrum or mixed
$\rm EEB$ and $\rm EBB$ type bispectrum. The first two estimators correspond to
the correlators $\rm \la{\bar E}{\cal C}^{EE}_\ell\ra$ and $\la \rm {\bar B}{\cal C}^{BB}_\ell \ra$.
The squeezed limits of mixed bispectra are constructed in a similar manner i.e.\
$\rm \la{\bar B}{\cal C}^{EE}_\ell\ra$ and $\rm\la{\bar E}{\cal C}^{BB}_\ell\ra$.
%
\item
{\bf Position-dependent correlation functions:} 
Correlations functions carry the same information as their Fourier counterparts.
In case of masks with complex topology it may be useful to compute the {\em position-dependent}
two-point correlation function. This will probe the squeezed limit of the three-point
correlation function (Munshi et al. (2019) in preparation). Such an approach can be specially
useful for surveys with small sky coverage.

\item
{\bf Relation to other estimators:}
The other estimators of bispectrum include the skew-spectrum which
can be useful as it too collapses the information content of the
bispectrum from three wave numbers in the harmonic domain
to just one. The skew-spectrum  is an important estimator for
the bispectrum though it mixes various shapes
and is not just sensitive to the squeezed limit.
The squeezed limit has interesting relationship to various
consistency relations discussed in the recent past \citep{Ke14a,KR,PP1}.

\item
{\bf Numerical implementation:}
Exploration of the entire configuration dependence of the bispectrum
requires elaborate computation. Both IB
as well as the skew-spectrum can also be seen as a method
of data compression beyond what can be achieved
by one-point estimators such as the skewness  probes specific aspects of
the bispectrum.

\item
  {\bf External data sets and {\em mixed} IB:}
  The estimator for IB can be generalized to investigate mixed
  bispectrum of two different data sets, e.g.\
  cross-correlating the power spectrum estimates from CMB secondaries, such as the all-sky
  $y$-parameter power spectrum estimates from e.g.\ Planck ${\cal C}^{yy}_\ell$,
  and cross-correlating
  against  $\bar\kappa$ estimated from the same patch of sky $\la\bar\kappa{\cal C}^{yy}_\ell\ra $,
  will provide information about $yy\kappa$ bispectrum
  in squeezed configurations. An estimator for $\kappa\kappa y$ bispectrum
  too can be constructed in an analogous manner
  $\la\bar \kappa{\cal C}^{yy}_{\ell}\ra$.
\end{enumerate}
Such extensions will be presented in future work.

%

\newpage
\bibliographystyle{utphys}
\bibliography{new_IB1}

\begin{thebibliography}{}
\bibitem{Planck1}
  Planck 2018 results. VI. Cosmological parameters,
  Planck Collaboration,
  [\href{https://arxiv.org/abs/1807.06209}{\tt arxiv/1807.06209}]
 \bibitem{MG1}
  Beyond the Cosmological Standard Model, A. Joyce, B. Jain, J. Khoury, M. Trodden, 2015, Phys. Rep., 568, 1
  [\href{hhttps://arxiv.org/abs/1407.0059}{\tt astro-ph/1407.0059}]
\bibitem{MG2}
  Modified Gravity and Cosmology, T. Clifton, P. G. Ferreira, A. Padilla, S. Skordis, 2012, Phys. Rep., 513, 1, 1
  [\href{hhttps://arxiv.org/abs/1101.1529}{\tt astro-ph/1106.2476}]
\bibitem{nu}
  Massive neutrinos and cosmology, J. Lesgourgues, S. Pastor, 2006, Phys. Rep., 429, 307,
  [\href{hhttps://arxiv.org/abs/1610.02956}{\tt astro-ph/1610.02956}]
\bibitem{DES}
  Cosmology from Cosmic Shear with DES Science Verification Data,
  The Dark Energy Survey Collaboration, T Abbott, F. B. Abdalla,
  S. Allam, et al., 2016, Phys. Rev. D, 94, 022001
  [\href{https://arxiv.org/abs/1507.05552}{\tt arxiv/1507.0552}]
\bibitem{KIDS}
  Gravitational Lensing Analysis of the Kilo Degree Survey,
  K. Kuijken, C. Heymans, H. Hildebrandt, et al., 2015, MNRAS, 454, 3500
  [\href{https://arxiv.org/abs/1507.00738}{\tt astro-ph/1507.00738}]
\bibitem{Euclid}
  \emph{Euclid} Definition Study Report,
  R. Laureijs, J. Amiaux, S. Arduini, et al. 2011, ESA/SRE(2011)12.
\bibitem{LSST_Tyson}
  LSST: a complementary probe of dark energy,
  J. A. Tyson, D. M. Wittman, J. F. Hennawi, D. N Spergel, 
  2003, Nuclear Physics B Proceedings Supplements, 124, 21
  [\href{https://arxiv.org/abs/astro-ph/0209632}{\tt astro-ph/0209632}]
\bibitem{WFIRST}
  National Research Council. 2010. New Worlds, New Horizons in
  Astronomy and Astrophysics. The National Academies Press.
  https://doi.org/10.17226/12951.
\bibitem{review}
  Cosmology with Weak Lensing Surveys,
  D. Munshi, P. Valageas, L. Van Waerbeke, A. Heavens, 2008,
  Phys. Rep, 462, 67
  [\href{https://arxiv.org/abs/astro-ph/0612667}{\tt arXiv/0612667}]
\bibitem{SDSSIII}  
  SDSS-III: Massive Spectroscopic Surveys of the Distant Universe, the Milky Way Galaxy, and
  Extra-Solar Planetary Systems, D. J. Eisenstein, D. H. Weinberg, E. Agol, et al., 2011, AJ, 142, 72
  [\href{https://arxiv.org/abs/1101.1529}{\tt astro-ph/1101.1529}]
\bibitem{WiggleZ}  
  The WiggleZ Dark Energy Survey: Survey Design and First Data Release, Drinkwater, M. J.,
  R. J. Jurek, C. Blake, et al., 2010, MNRAS, 401, 14
  [\href{hhttps://arxiv.org/abs/0911.4246}{\tt astro-ph/0911.4246}]
\bibitem{bias_review}
  Large-Scale Galaxy Bias,
  V. Desjacques, D. Jeong, F. Schmidt,
  2018, Phys. Rep. 733, 1
  [\href{https://arxiv.org/abs/1611.09787}{\tt arXiv/1611.09787}]
\bibitem{bernardeau_review}
  Large scale structure of the
  universe and cosmological perturbation theory,
  F. Bernardeau, S. Colombi, E. Gaztanaga, R. Scoccimarro, 2002, Phys.Rep. 367, 1
  [\href{http://lanl.arxiv.org/abs/astro-ph/0112551}{{\tt astro-ph/0112551}}]
 \bibitem[\protect\citeauthoryear{Castro}{2002}]{3D}
   Weak lensing analysis in three dimensions,
   P. G. Castro, A. F. Heavens, T. D. Kitching,
   2005, Phys Rev D, 72, 3516
   [\href{http://lanl.arxiv.org/abs/astro-ph/0503479}{\tt astro-ph/0503479}]
 \bibitem{higher1}
  Higher-order Statistics of Weak Lensing Shear and Flexion,
  D. Munshi, J. Smidt, A. Heavens, P. Coles, A. Cooray
  2011, MNRAS, 411, 2241
  [\href{http://lanl.arxiv.org/abs/1003.5003}{\tt astro-ph/1003.5003}]
  \bibitem{higher2}
  Higher-order Convergence Statistics for Three-dimensional Weak Gravitational Lensing,
  D. Munshi, A. Heavens, P. Coles,
  2011, MNRAS, 411, 2161
  [\href{hhttp://lanl.arxiv.org/abs/1002.2089}{\tt astro-ph/1002.2089}]
\bibitem{higher3}
  Higher Order Statistics for Three-dimensional Shear and Flexion,
  D. Munshi, T. Kitching, A. Heavens, P. Coles,
  2011, MNRAS, 416, 629
  [\href{http://lanl.arxiv.org/abs/1012.3658}{\tt astro-ph/1012.3658}]
\bibitem{MunshiBarber1}
  On the estimation of gravity-induced non-Gaussianities from weak lensing surveys,
  P. Valageas, D. Munshi, A. J. Barber,
  2005, MNRAS. 356, 386
  [\href{http://lanl.arxiv.org/abs/astro-ph/0402227}{\tt astro-ph/0402227}]
\bibitem{MunshiBarber2}
  Weak lensing shear and aperture-mass from linear to non-linear scales,
  D. Munshi, P. Valageas, A. J. Barber,
  2004, MNRAS 350, 77
  [\href{http://lanl.arxiv.org/abs/astro-ph/0309698}{\tt astro-ph/0309698}]
\bibitem{MunshiBarber3}
  From linear to non-linear scales: analytical and numerical predictions for the weak lensing convergence,
  A. J. Barber, D. Munshi, P. Valageas,
  2004, MNRAS, 347, 667
  [\href{http://lanl.arxiv.org/abs/astro-ph/0304451}{\tt astro-ph/0304451}]
\bibitem{Inflation}
  Testing Inflation with Large Scale Structure: Connecting Hopes with Reality,
  M. Alvarez et al.,
  [\href{https://arxiv.org/abs/1412.4671}{\tt arXiv/1412.4671}]
  \bibitem{MunshiJain1}
    Statistics of Weak Lensing at Small Angular Scales:
    Analytical Predictions for Lower Order Moments,
  D. Munshi, B. Jain,
  2001, MNRAS, 322, 107
  [\href{http://lanl.arxiv.org/abs/astro-ph/9912330}{\tt astro-ph/9912330}]
 \bibitem{MunshiBias}
   Probing The Gravity Induced Bias with Weak Lensing:
   Test of Analytical results Against Simulations,
   D. Munshi, 2000, MNRAS, 318, 145
   [\href{http://lanl.arxiv.org/abs/astro-ph/0001240}{\tt astro-ph/0001240}]
\bibitem{AlanBi}
  A New Approach to Probing Primordial Non-Gaussianity,
  D. Munshi, A. Heavens,
  2010, MNRAS, 401, 2406
  [\href{http://lanl.arxiv.org/abs/0904.4478}{\tt arxive/0904.4478}]
\bibitem{AlanTri}
  New Optimised Estimators for the Primordial Trispectrum,
  D. Munshi, A. Heavens, A. Cooray, J. Smidt, P. Coles, P. Serra,
  2011, MNRAS, 412, 1993
  [\href{http://lanl.arxiv.org/abs/0910.3693}{\tt arxive/0910.3693}]
\bibitem{MunshiJain2}
  The Statistics of Weak Lensing at Small Angular Scales:
  Probability Distribution Function,
  D. Munshi, B. Jain,
  2000, MNRAS, 318, 109
  [\href{http://lanl.arxiv.org/abs/astro-ph/9911502}{\tt astro-ph/9911502}]
\bibitem{MunshiBarber4}
  Analytical Predictions for Statistics of Cosmic Shear: Tests Against Simulations,
  P. Valageas, A. J. Barber, D. Munshi
  2004, MNRAS, 347, 654
  [\href{http://lanl.arxiv.org/abs/astro-ph/0303472}{\tt astro-ph/0303472}]
\bibitem{SDSSII}
  Position-dependent correlation function from the SDSS-III Baryon Oscillation Spectroscopic Survey Data Release 10 CMASS Sample,
  C.-T. Chiang, C. Wagner, A. G. Sánchez, F. Schmidt, E. Komatsu
  2015, JCAP, 09, 028
  [\href{http://lanl.arxiv.org/abs/1504.03322}{\tt astro-ph/1504.03322}]
 \bibitem{Lyman_bispec}
  The Lyman-$\alpha$ power spectrum - CMB lensing convergence cross-correlation,
  C.-T. Chiang, A. Slosar,
  2018, JCAP, 01, 012
  [\href{https://arxiv.org/pdf/1708.07512}{\tt arXiv/1708.07512}]
\bibitem{chiang2}
  Response approach to the squeezed-limit bispectrum: application to the correlation of quasar and Lyman-$\alpha$ forest power spectrum,
  C.-T. Chiang, A. M. Cieplak, F. Schmidt, A. Slosar,
  2017, JCAP, 06, 022
  [\href{https://arxiv.org/abs/1701.03375}{\tt arXiv/1701.03375}]
\bibitem{PhD_chiang}
   Position-dependent power spectrum: a new observable in the large-scale structure,
   C.-T. Chiang, 
   [\href{https://arxiv.org/abs/1508.03256}{\tt arXiv/1508.03256}]
\bibitem{Chiang_original}
  Position-dependent power spectrum of the large-scale structure:
  a novel method to measure the squeezed-limit bispectrum
  C.-T. Chiang, C. Wagner, F. Schmidt, E. Komatsu
  2014, JCAP, 05, 048 
  [\href{https://arxiv.org/abs/1403.3411}{\tt arXiv/1403.3411}]
\bibitem{Integrated}
  The Integrated Bispectrum and Beyond, 
  D. Munshi, P. Coles,
  2017, JCAP, 02, 010 
  [\href{https://arxiv.org/abs/1608.04345}{\tt arxive/1608.04345}]
\bibitem{MG_IB}
  The Integrated Bispectrum in Modified Gravity Theories,
  D. Munshi, 2017, JCAP, 01, 049
  [\href{http://lanl.arxiv.org/abs/1610.02956}{\tt arXiv/1610.02956}]
\bibitem{Bernardeau95}
  The angular correlation hierarchy in the quasilinear regime,
  F. Bernardeau,
  1995, A\&A, 301, 309
 [\href{https://arxiv.org/pdf/astro-ph/9502089.pdf}{\tt arXiv/9502089}]
\bibitem{Hyper}
  Hyperextended Cosmological Perturbation Theory:
  Predicting Non-linear Clustering Amplitudes
  R. Scoccimarro, J. A. Frieman
  1999, ApJ. 520, 35
  [\href{https://arxiv.org/pdf/astro-ph/9811184.pdf}{\tt astro-ph/9811184}]
\bibitem{MunshiRegan}
   Consistency Relations in Effective Field Theory,
   D. Munshi, D. Regan,
   2017, JCAP, 06, 042
   [\href{hhttp://lanl.arxiv.org/pdf/1705.07866}{\tt arXiv/1705.07866}]
\bibitem{Gil-Marin}
   An improved fitting formula for the dark matter bispectrum,
   H. Gil-Marín, C. Wagner, F. Fragkoudi, R. Jimenez, L. Verde
   2012, JCAP, 02, 047
   [\href{https://arxiv.org/abs/1111.4477}{\tt arXiv/1111.4477}]
\bibitem{chiang4}
  Separating the Universe into the Real and Fake,
  W. Hu, C.-T. Chiang, Y. Li, M. LoVerde,
  2016, Phs. Rev. D, 94, 023002 
  [\href{https://arxiv.org/abs/1605.01412}{\tt arXiv/1605.01412}]
\bibitem{chiang5}
  Scale-dependent bias and bispectrum in neutrino separate universe simulations,
  C.-T. Chiang, W. Hu, Y. Li, M. LoVerde,
  2018, Phys. Rev. D, 97, 123526 
  [\href{https://arxiv.org/abs/1710.01310}{\tt arXiv/1710.01310}]
\bibitem{Bartolo}
  Non–Gaussianity from Inflation: Theory and Observations,
  N. Bartolo, E. Komatsu, S. Matarrese, A. Riotto,
  2004, Phys.Rep., 402, 103
  [\href{https://arxiv.org/pdf/astro-ph/0406398}{\tt arXiv/0406398}]
\bibitem{chiang3}
  The halo squeezed-limit bispectrum with primordial non-Gaussianity:
  a power spectrum response approach,
  C.-T. Chiang,
  2017, Phys. Rev. D, 95, 123517 
  [\href{https://arxiv.org/abs/1701.03374}{\tt arXiv/1701.03374}]
\bibitem{MSS}
  Non-Linear Approximations to Gravitational Instability: A Comparison in the Quasi-Linear Regime,
  D. Munshi, V. Sahni, A. A. Starobinsky
  1994, ApJ., 436, 517
  [\href{https://lanl.arxiv.org/abs/astro-ph/9402065}{\tt arXiv/9402065}]
\bibitem{Horndeskii1}
  Second-order scalar-tensor field equations in a four-dimensional space,
  G. W. Horndeski,
  1974, Int. J. Theor. Phys. 10, 363
\bibitem{Horndeskii2}
  Covariant Galileon
  Deffayet, C.; Esposito-Farèse, G.; Vikman, A.
  PRD, 79, 084003
  [\href{https://arxiv.org/abs/1909.07366}{{\tt arxiv/1909.07366}}]
\bibitem{Sherwin}
    The Shift of the Baryon Acoustic Oscillation Scale: A Simple Physical Picture,
    B.D. Sherwin, M. Zaldarriaga, 2012, Phys. Rev. D 85, 103523
    [\href{http://lanl.arxiv.org/pdf/1202.3998}{\tt arXiv/1202.3998}]
     \bibitem{YinLi}
     Super-Sample Signal,
     Y. Li, W. Hu, M. Takada,
     Phys. Rev. D 2014, 90, 103530 [\href{https://arxiv.org/abs/1408.1081}{\tt arXiv/1408.1081}]
   \bibitem{Baldauf}
     Linear response to long wavelength fluctuations using curvature simulations,
     T. Baldauf, U. Seljak, L. Senatore, M. Zaldarriaga,
     2016, JCAP, 09, 007
     [\href{https://arxiv.org/pdf/1511.01465}{\tt arXiv/1511.01465}]
  \bibitem{Fabian}
    On Separate Universes,
    L. Dai, E. Pajer, F. Schmidt,
    2015, JCAP, 10, 059
    [\href{https://arxiv.org/pdf/1504.00351}{\tt arXiv/1504.00351}]
\bibitem{Bareira_response1}  
  Responses in Large-Scale Structure
  A. Barreira, F. Schmidt
  2017, JCAP, 06, 053
  [\href{https://arxiv.org/abs/1703.09212}{\tt arXiv/1302.6994}]
%
\bibitem{Planck_ng}
  Planck 2015 results. XVII. Constraints on primordial non-Gaussianity,
  Planck Collaboration,
  2016, A\&A, 594, 17
  [\href{http://lanl.arxiv.org/abs/1502.01592}{\tt arxive/1502.01592}]
\bibitem{Planck_mg}
  Planck 2015 results. XIV. Dark energy and modified gravity,
  Planck Collaboration,
  2016, A\&A, 594, 14
  [\href{http://lanl.arxiv.org/abs/1502.01590}{\tt arxive/1502.01590}]
\bibitem{DETF}
Report of the Dark Energy Task Force
A. Albrecht et al.
[\href{https://arxiv.org/abs/astro-ph/0609591}{\tt astro-ph/0609591}]
\bibitem{strings}
   The bispectrum of cosmic string temperature fluctuations including recombination effects,
   D. Regan, M. Hindmarsh,
   2015, JCAP, 10, 030
   [\href{https://arxiv.org/abs/1508.02231}{\tt arXiv/1508.02231}]
\bibitem{all_sky_sq}
  A Consistency Relation for the Observed Galaxy Bispectrum and the Local non-Gaussianity from Relativistic Corrections,
  A. Kehagias, A. Moradinezhad Dizgah, J. Norena, H. Perrier, A. Riotto,
   2015, JCAP, 08, 018
   [\href{http://lanl.arxiv.org/pdf/1503.04467}{\tt arXiv/1503.04467}]
\bibitem{Hu}
  Weak lensing of the CMB: A harmonic approach,
  W. Hu,
  2000, Phys.Rev. D, 62, 043007
  [\href{https://arxiv.org/abs/astro-ph/0001303}{\tt arXiv/0001303}]
\bibitem{Master}
  MASTER of the CMB Anisotropy Power Spectrum:
  A Fast Method for Statistical Analysis of Large and Complex CMB Data Sets,
  E. Hivon, K. M. Gorski, C. B. Netterfield, B. P. Crill, S. Prunet, F. Hansen,
  2002, ApJ, 567, 2
  [\href{https://arxiv.org/abs/astro-ph/0105302}{\tt arXiv/0105302}]
\bibitem{Marika}
  Flat-Sky Pseudo-Cls Analysis for Weak Gravitational Lensing,
  M. Asgari, A. Taylor, B. Joachimi, T. D. Kitching
  [\href{https://arxiv.org/abs/1612.04664}{\tt arXiv/1612.04664}]
\bibitem{hikage}
  Shear Power Spectrum Reconstruction using Pseudo-Spectrum Method,
  C. Hikage, M. Takada, T. Hamana, D. Spergel,
  MNRAS, 412, 65, 2011
 [\href{ http://lanl.arxiv.org/abs/1004.3542}{\tt arXiv/1004.3542}]
\bibitem{HamiltonSampling}
  Fast optimal CMB power spectrum estimation with Hamiltonian sampling,
  J. F. Taylor, M. A. J. Ashdown, M. P. Hobson,
  2008, MNRAS, 389, 1284
  [\href{http://lanl.arxiv.org/abs/0708.2989}{\tt arXiv/0708.2989}]
\bibitem{Wandelt}
  Methods for Bayesian power spectrum inference with galaxy surveys,
  J. Jasche, B. D. Wandelt,
  2013, ApJ, 779, 15
  [\href{https://arxiv.org/pdf/1306.1821}{\tt arXiv/1306.1821}]
\bibitem{halo}
  Halo Models of Large Scale Structure,
  A. Cooray, R. Sheth,
  2002, Phys.Rep., 372, 1
  [\href{https://arxiv.org/abs/astro-ph/0206508}{\tt arXiv/0206508}]
\bibitem{Takada1}
  Joint analysis of cluster number counts and weak lensing power spectrum to correct for the super-sample covariance
  M. Takada, D. N. Spergel
  2014, MNRAS. 441, 2456
[\href{https://arxiv.org/abs/1307.4399}{\tt arXiv/1307.4399}]
\bibitem{Takada2}
  Joint likelihood function of cluster counts and n-point correlation functions:
  Improving their power through including halo sample variance
  E. l Schaan, M. Takada, D. N. Spergel
  2014, Phys. Rev. D 90, 123523
  [\href{https://arxiv.org/abs/1406.3330}{\tt arXiv/1307.4399}]
\bibitem{Takada3}
  Power Spectrum Super-Sample Covariance,
  M. Takada, W. Hu
  2013, PRD 87, 123504 
  [\href{https://arxiv.org/abs/1302.6994}{\tt arXiv/1302.6994}]
\bibitem{EFT_bispec}
  The Bispectrum in the Effective Field Theory of Large Scale Structure,
  T. Baldauf, L. Mercolli, M. Mirbabayi, E. Pajer
  2015, JCAP, 05, 007
  [\href{ https://arxiv.org/abs/1406.4135}{\tt arXiv/1406.4135}]
\bibitem{EFT_power}
  The Effective Field Theory of Large Scale Structures at Two Loops
  J. J. M. Carrasco, S. Foreman, D. Green, L. Senatore
  2014, JCAP, 07, 057
  [\href{https://arxiv.org/abs/1310.0464}{\tt arXiv/1310.0464}]
\bibitem{BiHalo}
  BiHalofit: A new fitting formula of non-linear matter bispectrum
  [\href{https://arxiv.org/abs/1911.07886}{\tt arXiv/1911.07886}]
\bibitem{Baumann}
    Cosmological Non-Linearities as an Effective Fluid, 
    D. Baumann, A. Nicolis, L. Senatore, M. Zaldarriaga,
    2012, JCAP 1207, 051
    [\href{https://arxiv.org/abs/1004.2488}{\tt arXiv/1004.2488}]
\bibitem{Senatore}
    The Effective Field Theory of Cosmological Large Scale Structures,
    J. J. M. Carrasco, M. P. Hertzberg, L. Senatore,
    2012, JHEP 09, 082
    [\href{https://arxiv.org/abs/1206.2926}{\tt arXiv/1206.2926}]
\bibitem{EFTBi}
  The Bispectrum in the Effective Field Theory of Large Scale Structure,
  T. Baldauf, L. Mercolli, M. Mirbabayi, E. Pajer
  2015, JCAP, 05, 007
  [\href{hhttps://arxiv.org/pdf/1406.4135}{\tt arXiv/1406.4135}]
 \bibitem{Salman}
  The Coyote Universe Extended: Precision Emulation of the Matter Power Spectrum,
  K. Heitmann, E. Lawrence, J. Kwan, S. Habib, D. Higdon
  2014, ApJ, 780, 111
  [\href{hhttps://arxiv.org/abs/1304.7849}{\tt arXiv/1304.7849}]
\bibitem{roman}
  A Fitting Formula for the Non-Linear Evolution of the Bispectrum,
  R. Scoccimarro, H.M.P. Couchman, 2001, MNRAS, 325, 1312
  [\href{http://lanl.arxiv.org/abs/astro-ph/0009427}{\tt arXiv/0009427}]
\bibitem{Potter:17}
  PKDGRAV3: Beyond Trillion Particle Cosmological Simulations for the Next Era of Galaxy Surveys,
  D. Potter, J. Stadel, R. Teyssier,
  2017, Computational Astrophysics and Cosmology, Volume 4, Issue 1, article id.2,
  [\href{https://arxiv.org/abs/1609.08621}{\tt astro-ph/1609.08621}]
\bibitem{Potter:16}
  D. Potter, J. Stadel,
  PKDGRAV3: Parallel gravity code,
  2016,  Astrophysics Source Code Library, record ascl:1609.016,
  [\href{arXiv:1609.08621}{\tt arXiv/0711.1540}]
\bibitem{Fosalba:08}
  The onion universe: all sky lightcone simulations in spherical shells,
  P. Fosalba, E. Gaztañaga, F. J. Castander, J. Francisco, M. Manera,
  2008, MNRAS, 391, 435,
  [\href{hhttps://arxiv.org/abs/0711.1540}{\tt arXiv/0711.1540}]
\bibitem{Fosalba:15a}
  The MICE grand challenge lightcone simulation - I. Dark matter clustering
  The MICE Grand Challenge Lightcone Simulation I: Dark matter clustering
  P. Fosalba, M. Crocce, E. Gaztanaga, F. J. Castander
  2013, MNRAS, 448, 2987
  [\href{https://arxiv.org/abs/1312.1707}{\tt arXiv/1312.1707}]
\bibitem{Fosalba:15b}
  The MICE Grand Challenge light-cone simulation - III.,
  Galaxy lensing mocks from all-sky lensing maps,
  P. Fosalba, E. Gaztañaga, F. J. Castander, M. Crocce,
  2015, MNRAS, 447, 1319,
  [\href{https://arxiv.org/abs/1312.2947}{\tt arXiv/1312.2947}]
\bibitem{Hilbert:20}
  The Accuracy of Weak Lensing Simulations,
  S. Hilbert et al.,
  2020, MNRAS, 493, 305,
  [\href{https://arxiv.org/abs/1910.10625}{\tt arXiv/1910.10625}]
%
\bibitem{LamHui}
  Biased-estimations of the Variance and Skewness
  L. Hui, E. Gaztanaga
  1999, ApJ, 519, 622
  [\href{https://arxiv.org/abs/astro-ph/9810194}{\tt arXiv/9810194}]
  %
\bibitem{szapudi1}
Cosmic Statistics of Statistics
I. Szapudi, S. Colombi, F. Bernardeau
1999, MNRAS, 310, 428
[\href{https://arxiv.org/abs/astro-ph/9912289}{\tt arXiv/9912289}]
%
\bibitem{szapudi2}
Experimental Cosmic Statistics II: Distribution
I. Szapudi, S. Colombi, A. Jenkins, J. Colberg 
2000, MNRAS, 313, 725
[\href{https://arxiv.org/abs/astro-ph/9912238}{\tt arXiv/9912289}]
%
\bibitem{szapudi3}
Experimental Cosmic Statistics I: Variance
I. Szapudi, S. Colombi, A. Jenkins, Jörg Colberg
2000, MNRAS, 313, 711
[\href{https://arxiv.org/abs/astro-ph/9912236}{\tt arXiv/9912236}]
%
\bibitem{Error_Munshi_Coles}
Error Estimates for Measurements of Cosmic Shear
D. Munshi, P. Coles
2003, MNRAS, 338, 846
[\href{https://arxiv.org/abs/astro-ph/0003481}{\tt arXiv/9912236}]
%
\bibitem{Colombi1}
Effects of Sampling on Statistics of Large Scale Structure,
S. Colombi, I. Szapudi, A.S. Szalay,
1998, MNRAS, 296,  253
[\href{http://lanl.arxiv.org/pdf/9711087}{\tt arXiv/9711087}]
\bibitem{Colombi2}
Cosmic Error and the Statistics of Large Scale Structure,
I. Szapudi, S. Colombi,
1996, ApJ, 470, 131
[\href{http://lanl.arxiv.org/pdf/9711087}{\tt arXiv/9711087}]
\bibitem{MunshiBernardeau}
Scaling in Gravitational Clustering, 2D and 3D Dynamics,
D. Munshi, F. Bernardeau, A. L. Melott, R. Schaeffer,
1999, MNRAS, 303, 433
[\href{http://lanl.arxiv.org/pdf/9510030}{\tt arXiv/9510030}]
\bibitem{flat1}
  The Limits of Cosmic Shear,	
  T. D. Kitching, J. Alsing, A. F. Heavens, R. Jimenez, J. D.  McEwen, L. Verde,
  2017, MNRAS, 469, 2737
  [\href{https://lanl.arxiv.org/abs/1611.04954}{\tt arXiv/1611.04954}]
\bibitem{flat2}
  The effect of Limber and flat-sky approximations on galaxy weak lensing,
  P. Lemos, A. Challinor, G. Efstathiou,
  2017, JCAP, 05, 014
  [\href{https://lanl.arxiv.org/pdf/1704.01054}{\tt arXiv/1704.01054}]
\bibitem{flat3}
  Precision calculations of the cosmic shear power spectrum projection,
  M. Kilbinger,
  2017, MNRAS, 472, 2126
  [\href{https://arxiv.org/abs/1702.05301}{\tt arXiv/1702.05301}]
\bibitem[\protect\citeauthoryear{Coulton et al.}{2018}]{Coulton18}
  Constraining Neutrino Mass with the Tomographic Weak Lensing Bispectrum
  W. R. Coulton, J. Liu, M. S. Madhavacheril, V. Böhm, D. N. Spergel,
  [\href{https://arxiv.org/abs/1810.02374}{\tt arXiv/1810.02374}
 \bibitem{Namikawa1}
  The Weak Lensing Bispectrum Induced By Gravity
  D. Munshi, T. Namikawa, T. D. Kitching, J. D. McEwen, R. Takahashi, F. R. Bouchet, A. Taruya, B. Bose
  [\href{https://arxiv.org/abs/1910.04627}{\tt arXiv/1910.04627}]
\bibitem{Halofit}
BiHalofit: A new fitting formula of non-linear matter bispectrum
R. Takahashi, T. Nishimichi, T. Namikawa, A. Taruya, I. Kayo, K. Osato, Y. Kobayashi, M. Shirasaki
[\href{https://arxiv.org/abs/1911.07886}{\tt arXiv/0202090}]
%
\bibitem{Ke14a} Consequences of Symmetries and Consistency Relations in the
  Large-Scale Structure of the Universe for Non-local bias and Modified Gravity,
  A. Kehagias, J. Norena, H. Perrier, A. Riotto, 
  2014, Nuclear Physics B, 883, 83,
  [\href{https://arxiv.org/pdf/1311.0786}{\tt arXiv/1311.0786}]
\bibitem{KR}
  Symmetries and consistency relations in the large scale structure of the universe,
  A. Kehagias, A.  Riotto, 2013, Nuclear Physics B, 873, 514,
  [\href{https://arxiv.org/abs/1302.0130}{\tt arXiv/1004.3542}]
\bibitem{PP1}
  Ward identities and consistency relations for the large scale structure with multiple species
  M. Peloso, M. Pietroni, 2014, JCAP, 4, 011
  [\href{https://arxiv.org/abs/1310.7915}{\tt arXiv/1004.3542}]
\bibitem[\protect\citeauthoryear{Bernardeau \& Brax}{2011}]{BB11}
  Cosmological Large-scale Structures beyond Linear Theory in Modified Gravity
  F. Bernardeau, P. Brax
  2011, JCAP, 1106, 019
 [\href{ https://arxiv.org/abs/1102.1907}{\tt arXiv/1102.1907}]
\bibitem[\protect\citeauthoryear{Munshi, Sahni, Starobinsky}{1994}]{Staro}
  Non-Linear Approximations to Gravitational Instability: A Comparison in the Quasi-Linear Regime
  D. Munshi, V. Sahni, A. A. Starobinsky
  1994, ApJ, 436, 517
  \href{https://arxiv.org/abs/9402065}{\tt arXiv/9402065}]
\bibitem{NN}
  Cosmological parameter constraints for Horndeski scalar-tensor gravity
  J. Noller, A. Nicola
  Phys. Rev. D 99, 103502 (2019)
  [\href{https://arxiv.org/abs/1811.12928}{{\tt arxiv/1811.12928}}]
\bibitem[\protect\citeauthoryear{Bellini et al.}{2015}]{Verde}
  E. Bellini, R. Jimenezb, L. Verde
  2015, JCAP, 05, 057
  [\href{https://arxiv.org/abs/1504.04341}{\tt arXiv/1504.04341}]  
\bibitem[\protect\citeauthoryear{Gleyzes et al.}{2015a}]{BeyondHordensky1}
  Exploring gravitational theories beyond Horndeski
  J. Gleyzes, D. Langlois, F. Piazza, F. Vernizzi,
  2015, JCAP, 2,  018,
  [\href{https://arxiv.org/abs/1408.1952}{\tt arXiv/1408.1952}]
\bibitem[\protect\citeauthoryear{Gleyzes et al.}{2015b}]{BeyondHordensky2}
  Healthy theories beyond Horndeski
  J. Gleyzes, D. Langlois, F. Piazza, F. Vernizzi,
  2015, PRL, 114, 211101,
  [\href{https://arxiv.org/abs/1404.6495}{\tt arXiv/1404.6495}]
\bibitem[\protect\citeauthoryear{Hirano, Kobayashi, Tashiro, Yokoyama}{2017}]{Hirano}
  Matter bispectrum beyond Horndeski theories
  S. Hirano, T. Kobayashi, H. Tashiro, S. Yokoyama
  2018, Phys. Rev. D 97, 103517 (
  [\href{https://arxiv.org/pdf/1801.07885}{\tt arXiv/1801.07885}]
\bibitem[\protect\citeauthoryear{Dvali, Gabadadze, Porati}{2000}]{DGP}
  4D Gravity on a Brane in 5D Minkowski Space
  G. Dvali, G. Gabadadze, M. Porrati,
  2000, Phys. Rev. B, 485, 208,
 [\href{https://arxiv.org/abs/hep-th/0005016}{\tt arXiv/0005016}]  
\bibitem{Marco_Cristomi}
  Consistency relations for large-scale structure in modified gravity and the matter bispectrum
  M. Crisostomi, M. Lewandowski, F. Vernizzi
  [\href{https://arxiv.org/abs/1909.07366}{{\tt arxiv/1909.07366}}]
\bibitem[\protect\citeauthoryear{Bose \& Taruya}{2018}]{Bose1}
  The one-loop matter bispectrum as a probe of gravity and dark energy
  B. Bose, A. Taruya,
  2018, JCAP, 1810, 2018,, 019
  [\href{https://arxiv.org/abs/1808.01120}{\tt astro-ph/1808.01120}]
\bibitem{nu_review}
  Massive neutrinos and cosmology
  J. Lesgourgues, S. Pastor
  2006, Phys.Rept. 429, 307
%
\bibitem[\protect\citeauthoryear{Liu et al.}{2018}]{Liu17}
  MassiveNuS: Cosmological Massive Neutrino Simulations
  J. Liu, S. Bird, J. M. Z. Matilla, J. C. Hill, Z. Haiman, M. S. Madhavacheril,
  D. N. Spergel, A. Petri, 
  2018, JCAP, 03, 049
  [\href{https://arxiv.org/abs/1711.10524}{\tt arXiv/1711.10524}]
%
  [\href{https://arxiv.org/abs/astro-ph/0603494}{{\tt astro-ph/0603494}}]
 \bibitem[\protect\citeauthoryear{Ruggeri}{2018}]{Ruggeri}
    DEMNUni: Massive neutrinos and the bispectrum of large scale structures
  R. Ruggeri, E. Castorina, C. Carbone, E. Sefusatti
  2018, JCAP, 03, 003
  [\href{https://arxiv.org/abs/1712.02334}{\tt arXiv/1712.02334}]
  %
  \bibitem{Q_review}
  Quintessence: A Review
  S. Tsujikawa
  2013, Class.Quant.Grav., 30, 214003
  [\href{https://arxiv.org/abs/1304.1961}{{\tt astro-ph/1304.1961}}]
  %
\bibitem{Spherical_DE}
  Spherical collapse of dark energy with an arbitrary sound speed
  T. Basse, O. E. Bjalde1, Y. Y. Y. Wong
  [\href{https://arxiv.org/abs/1009.0010}{{\tt arxiv/1009.0010}}]
\bibitem[\protect\citeauthoryear{Sefusatti \& Vernizzi}{2018}]{Sefusatti}
  Cosmological structure formation with clustering quintessence
  E. Sefusatti, F. Vernizzi
  2011, JCAP, 1103, 047
  [\href{https://arxiv.org/abs/1101.1026}{\tt arXiv/1902.04877}]
  %
\bibitem[\protect\citeauthoryear{Wagner et al.}{2015}]{Wagner}
  The angle-averaged squeezed limit of nonlinear matter N-point functions
  C. Wagner, F. Schmidt, C.-T. Chianga, E. Komatsua,
  2015, JCAP, 08, 042
  [\href{https://arxiv.org/abs/1503.03487}{\tt arxiv/1503.03487}]
\bibitem{Valageas}
  Statistical properties of the convergence due to weak gravitational lensing by non-linear structures,
  P. Valageas,
  2000, A \& A, 356, 771
  [\href{https://arxiv.org/abs/astro-ph/9911336}{\tt astro-ph/9911336}]
\bibitem{Munshi2}
  Probing the Gravity Induced Bias with Weak Lensing: Test of Analytical results Against Simulations,
  D. Munshi,
  2000, MNRAS, 318, 145
  [\href{https://arxiv.org/abs/astro-ph/0001240}{\tt astro-ph/00001240}]
\bibitem{Taruya}
  Lognormal Property of Weak Lensing Fields,
  A. Taruya, M. Takada, T. Hamana, I. Kayo, T. Futamase,
  2002, ApJ, 571, 638
  [\href{https://arxiv.org/abs/astro-ph/0202090}{\tt astro-ph/0202090}]
%
%
\end{thebibliography}

\section*{Acknowledgment}
DM is supported by a grant from the Leverhulme Trust.
TDK is supported by Royal Society University Research Fellowship.
PF acknowledges support from MINECO through grant ESP2017-89838-C3-1-R,
the European Union H2020- COMPET-2017 grant Enabling Weak Lensing Cosmology,
and Generalitat de Catalunya through grant 2017-SGR-885. 
It's a pleasure for DM to acknowledge many helpful discussions
with Andrea Petri, Peter Taylor, Alexander Eggemeier
and Chi-Ting Chiang for useful comments.
It is our pleasure also to acknowledge constructive comments from P. Schneider
and V. F. Cardone. DM would also like to organisers of the
Euclid Theory Working Group Meeting (8th April - 9th, April 2019)
in Oxford where many of the related ideas were discussed.
\appendix
%

\section{Integrated Bispectrum in beyond $\Lambda$CDM cosmologies}
\label{sec:beyond}

In this paper we have derived the IB in the standard cosmological scenario  known also as the  $\Lambda$CDM model.
This model is based on General theory of Relativity (GR). 
The GR is an extremely successful theory of gravity.
However, it is at best an effective theory of gravity and
suffers from many fundamental problems including the cosmological constant problem.
Any extension or modifications of the GR would invariably include
new gravitational degrees of freedom. Such modifications of gravity will
also change the statistics of cosmological density distributions.
We will next consider some well known Modified Gravity (MG) theories
and derive the IB in such model.
The cosmological models with dark energy equation of state are considered
as an equivalent formulation of the MG theories.
We will also consider a specific model of dark energy (DE) equation of state known also
as the clustering quitessence models and compute the integrated bispectrum
in this scenarion. Finally, We will also consider cosmological models with massive neutrinos.

%
%
%
\subsection{Gamma $\gamma$ and Beta $\beta$ Models:}
%
%
An interesting model $\gamma$-model was considered by \cite{BB11}
as a precursor to more complicated modification of gravity.
This model is generated by modifying the Euler equation of the Euler-Continuity-Poisson equation.
Such modification of the Euler equation changes the force-law. However, continuity equation gurantees
the conservation of mass. In this model the gravitational field seen by massive particles denoted as $\phi^{\rm eff}$ is different
  from the gravitational potential that solved the Poisson equation $\phi$. These two
  potentials are different and related by $\phi^{\rm eff}(\bx,t) = (1+\epsilon(t))\phi(\bx,t)$ through
  parameter $\epsilon(t)$ in the sub-horizon scale. The model can be analysed using
  perturbation theory and in this parametrization the kernel $F_2$ in Eq.(\ref{eq:define_F2}) is
  modified to the following form:
  \ben
  && F_2(\bk_1,\bk_2) = {1\over 2}(1+\epsilon) +
  {1\over 2} {\mu}_{12} \left ( {k_1\over k_2} + {k_2 \over k_1} \right )  +
  {1\over 2}(1-\epsilon){\mu}_{12}^2; \quad \mu_{12} = {\bk_1\cdot\bk_2 \over k_1 k_2}.
  \label{eq:Brax}
  \een
  In general the paramter $\epsilon$ can be a function of scale factor $a$ or the wavelength $k$. however, it is interesting to note that
  for $\epsilon = {3/7}$ recover the expression given in Eq.(\ref{eq:define_F2}). The Lagrangian perturbation
  theory is often used to model quasilinear evolution of gravitational clustering  The Zel'dovich approximation
  is the linear order in Lagrangian perturbation theory. The bispectrum in the Zel'dovich approximation
  can be recovered from Eq.(\ref{eq:Brax}) $\epsilon=0$ \citep{Staro}. 
  \ben
  && \langle F_2 \rangle_{\rm 2D} = {{\epsilon + 3}\over 4}; \quad \langle F_2 \rangle_{\rm 3D} = {{\epsilon + 2}\over 3}.
  \een
  The corresponding expressions in the squeezed configuration takes the following form:
  \bes
  \ben
  && B_{\rm sq}= \left [ (1+2\epsilon) + 2(1-\epsilon)\musq_{} -\musq_{} {d\ln P_{\rm 3D}(k) \over d\ln k}\right ]P_{\rm 3D}(q_3)P(k); \\
  && B^{\rm 3D}_{\rm sq} = \left [ {4 \over 3}(2+\epsilon) -{1\over 3}(n+3) \right ] P_{\rm 3D}(q_3)P_{\rm 3D}(k); \;\; \\
  && B^{\rm 2D}_{\rm sq} = \left [ (3+\epsilon) -{1\over 2} (n+2) \right ]P_{\rm 3D}(q_{3\perp})P_{\rm 3D}(k_{\perp})
  \een
  \ees
  \noindent
  The 3D derivation mirrors closesly that of 2D.  
  The actual value of the parameter $\epsilon$ can be computed using the linearised Euler-Continuity-Poisson equation,
  and assuming a parametric form for the growth fcator $f=  d\ln D_+/d\ln a \approx \Omega^\gamma_{\rm M}$.

  In the $\beta$ model proposed by \citep{BB11} where the expression for the kernel $F_2(\bk_1,\bk_2)$ we have:
  \ben
  && F_2(\bk_1,\bk_2)  = \left (  {3\nu_2 \over 4} - {1\over 2}\right ) +
  {1\over 2} \mu_{12} \left [ {k_1\over k_2} + {k_2\over k_1} \right ]
  +  \left (  {3  \over 2} - {3\nu_2 \over 4}\right )\mu^2_{12} 
  \een
  where, the parameter $\nu_2$ can be related to the $\epsilon$ parameter in Eq.(\ref{eq:Brax})  $\epsilon = {3 \over 2} \nu_2 -2$.
  The parametric value for $\nu_2$  can be obtained in a manner similar to the $\gamma$ model. However, we would leave them
  unspeified. The corresponding squeezed limit is as follows:
  \bes
  \ben
  && B_{\rm sq}=  \left [{3}(\nu_2-1) + (6- 3\nu_2)\musq_{} - \musq_{} {d\ln P(k) \over d\ln k}\right ]P_{\rm 3D}(q_3)P_{\rm 3D}(k);\\
  && B^{\rm 3D}_{\rm sq}(k)=  \left [ 2\nu_2  -{1 \over 3}(n+3) \right ] P_{\rm 3D}(q_{3})P_{\rm 3D}(k_{}); \\
  && B^{\rm 2D}_{\rm sq}(k_\perp)= \left [ \left ( {3 \over 2}\nu_2 +1 \right ) -{1 \over 2}(n+2) \right ] P_{\rm 3D}(q_{3\perp})P_{\rm 3D}(k_{\perp}).
  \een
  \ees
  In these models the $\nu_2$ can in general be a function of $z$ as well as wave-number $k$ \cite{BB11}.
%
%
 \subsection{Horndeskii Theories of Gravity}
%
 The Horndeski theoriy provide a minimal extension of GR as a single new scalar dof is incorporated \citep{Horndeskii1,Horndeskii2}.
 Many well known scalar-tensor theories which respect Lorentz symmetry and do not suffer
 from higher-derivative ghosts can be considered special cases of this particular model.
 This particular model is probably remains one of the most studied theories of gravity.
 Cosmological consequences of Hordenskii theories have been investigated in great detail (see e.g. \citep{NN} and the references therein).
 Among different paramtetrization for Hordenskii theories we will consider the one given in  \citep{Verde}
 for bispectrum:
 \bes
 \ben
 && F_2(\bk_1,\bk_2) = c(z) + \left ({k_1\over k_2} + {k_2\over k_1} \right )\mu_{12}  -\left ( 1 - {1\over 2} c(z) \right )(1 - 3\mu^2_{12}).\\
 && F_2(\bk_1,\bk_2) = \left ( {3c(z) \over 2} -1 \right ) + \mu_{12} \left ( {k_1\over k_2} + {k_2\over k_1} \right )
 + \left ( 3 - {3c(z) \over 2}\right )\mu^2_{12}.
 \een
 \ees
 The bispectrum in the squeezed limit has the following expressions in 3D and 2D. 
 \ben
 && B^{\rm 3D}_{\rm sq}(k_\perp)= \left [2c(z) -{1\over 3}(n+3) \right ]P_{\rm 3D}(k_\perp)P_{\rm 3D}(q_\perp) \\
 && B^{\rm 2D}_{\rm sq}(k)= \left [\left ({3\over 2}c(z)+1 \right ) -{1\over 2}(n+2) \right ]P_{\rm 3D}(k)P_{\rm 3D}(q_{3}).
 \een
 These results reduces to their EdS values if we set $c(z)=34/21$.
 \subsection{Beyond Horndeskii Theories or DHOST Theories}
 %
 Indeed Hordenskii theories are often used as a test bed for possible departure from GR and
 their possible cosmological consequences. However, the Hordenskii theories have also been extended
 by considering theories theories that are also known as the degenerate higher-order scalar tensor theories
 or DHOST theories. The simplest example in the context of non-degenerate
scenarios are also known as the Galeyzes-Langlois-Piazza-Venizzi or GPLV theories \citep{BeyondHordensky1,BeyondHordensky2}.
The second-order kernel in these scenario
include a scale dependent additional term which changes the bispectrum \citep{Hirano}
which can be constrained using the staistics discussed here.
\bes
\ben
&& F_2(\bk_1,\bk_2) = \kappa_s(z) \alpha_s(\bk_1,\bk_2) - {2 \over 7}\lambda(z)\gamma(\bk_1,\bk_2) \label{eq:Hordenski_F_2}\\
&& \alpha_s(\bk_1,\bk_2) = 1 + {1 \over 2} \mu_{12} \left ({k_1\over k_2} + {k_2\over k_1} \right );
\quad \gamma(\bk_1,\bk_2) = 1- \mu^2_{12};\;\; \mu_{12} =    {(\bk_1\cdot\bk_2)^2 \over k_1^2 k_2^2}
\label{eq:Hordenski_alphabeta}
\een
\ees
On substitution of  $\alpha_s$ and $\gamma$ in Eq.(\ref{eq:Hordenski_F_2}) we arrive at:
\ben
F_2(\bk_1,\bk_2) =  \left ( \kappa_s(z) - {2 \over 7} \lambda(z) \right ) + {\kappa_s(z) \over 2} \mu_{12}
\left ( {k_1 \over k_2} + {k_2 \over k_1 } \right ) + {2 \over 7} \lambda \mu^2_{12}
\label{eq:Dhost_F2}
\een
Taking angular averages we can see $\alpha_s(\bk_1,\bk_2) = 1$ and $\gamma(\bk_1,\bk_2)= 2/3$ which leads us
respectively in 3D and 2D to:
\ben
&& \langle F_2 \rangle_{\rm 3D} = \kappa_s(z) - {4 \over 21} \lambda(z) \quad {\rm in\;3D}; \quad
\langle F_2 \rangle_{\rm 2D} = \kappa_s(z) - {1 \over 7} \lambda(z); \quad {\rm in\; 2D}
\een
Notice that $\kappa_s(z)$ is a free parameter that describes the theory.
It is not linked to the convergence $\kappa$ defined previously.
We use this notation to be consistent with the existing literature.
To differentiate from the convergence $\kappa$ we have used an extra subscript $_s$.
The corresponding squeezed bispectrum respectively in 3D and 2D takes the following form:
\bes
\ben
&& B^{\rm 3\rm D}_{sq} = 2 \left [ {3 \over 2} \kappa_s(z) -{16 \over 21} \lambda(z)  - {\kappa_s(z)} {n \over 6}  \right ]P_{\rm 3D}(k)P_{\rm 3D}(q_3)
\label{eq:ep1} \\
&& B^{\rm 3\rm D}_{sq} = 2 \left [ {3 \over 2} \kappa_s(z) -{2 \over 7} \lambda(z) - {\kappa_s(z)} {n \over 4}  \right ]P_{\rm 3D}(k_{\perp})P_{\rm 3D}(q_{3\perp}).
\label{eq:ep2}
\een
\ees
The data from future surveys can be used to constraint the variation of IB and hence the functional form of $\kappa_s(z)$.

In Ref.\citep{Marco_Cristomi} the following equivalent parametrization for the kernel $F_2$ was introduced:
\ben
&& F_2(\bk_1,\bk_2) = A_{\alpha} + {2\over 3} A_{\gamma} + A_{\alpha} {\mu\over 2} \left ( {k_1\over k_2} + {k_2 \over k_1} \right )
-A_{\gamma} \left ( \mu^2 -{1\over 3}  \right )
\een
In general the parameters $A_{\alpha}$, $A_{\gamma}$ are time dependent.
\bes
\ben
&& B^{\rm 2D}_{sq}(k) = 2\left [ {3 \over 2} A_{\alpha} + A_{\gamma} - {1 \over 4} A_{\alpha} n \right ] P_{\rm 3D}(k_\perp) P_{\rm 3D}(q_{3\perp}) \label{eq:nu1}\\
&& B^{\rm 3D}_{sq}(k_\perp) = 2\left [ {3 \over 2} A_{\alpha} + {4 \over 3}A_{\gamma} - {1 \over 6} A_{\alpha} n \right ] P_{\rm 3D}(k_{}) P_{\rm 3D}(q_{3})  \label{eq:nu2}
\een
\ees
For this model we have $\langle F_2 \rangle = A_{\alpha} + {2\over 3} A_{\gamma} $
%
\subsection{The Normal-branch of Dvali, Gabadadze, Porrati (nDGP) model:}
The normal branch of  Dvali, Gabadadze, Porrati model \citep{DGP} known also as the nDGP is a
    prototypical example that involve Vainshtein screening.
    The bispectrum in this model
corresponds to the case $\kappa_s=1$ in Eq.(\ref{eq:Dhost_F2}).
\ben
&& \kappa_s(z)=1 ; \quad\quad \lambda(z)= \left ( 1- {7\over 2} {D_2(z) \over D^2_1(z)} \right )
\een
Here $D_2(z)$ and $D_1(z)$ are the second- and first-order growth factors
that can be computed by numericaly solving the equations governing growth of perturbations \citep{Bose1}.
%
\subsection{Massive Neutrinos}


A small but non-negligible fraction of the cosmological matter density is
provided by massive neutrinos \citep{nu_review}. The massive neutrinos are known to have
significant thermal distribution and a different cosmological
evolutionary history in comparision to the cold dark matter.
The thermal dispersion in verlocity results in a damping of perturbation
below a length scale also known as the free-streaming length scale.
This will be probed by future surveys with a very high degree of accuracy.
In the long run cosmological surveys are expected to provide
an upper limit to the sum of the neutrino masses.. This will be very
useful when jointly considered with the lower limits from
neutrino-osccilation experiments.

The neutrinos decouple and free-stream with a large thermal velocities.
The redshift $z_{nr}$ at wshich neutrinos become relativistic
depend on their mass eigenstate $m_i$: $1+ z_{nr} = 1980 \left [ {m_{\nu,i}/1{\rm eV} }\right ]$
The fractional contribution to the total matter density is denoted as $f_{\nu}$
which can be expressed as
\ben
f_{\nu} \equiv {\Omega_{\nu} \over \Omega_{M}} = {1 \over \Omega_{M,0} h^2} {\sum_i M_{\nu,i} \over 93.14 \rm eV}.
\een
In future it will also be interesting to consider the effect of neutrino mass
on bispectrum when simulated all-sky lensing maps for such cosmologies will be available \citep{Liu17,Coulton18}.
The total matter distribution thus can be wriiten in terms of the cold dark matter perturbation $\delta_{cdm}$
and the fluctuations in the neutrino density distribution $\delta_{\nu}$.
\ben
&& \delta_m = f_c\delta_c + f_{\nu}\delta_{\nu}; \quad \ f_c+f_\nu=1.
\een
The resulting matter power spectrum $P_{mm}(k)$ and bispectrum $B_{mmm}(\bk_1,\bk_2,\bk_3)$ can be expressed as \citep{Ruggeri}:
\bes
\ben
&& P_{mm}(k) = f_c^2 P_{cc}(k) + 2f_{\nu}f_c P_{\nu c}(k) + f^2_{\nu} P_{\nu\nu}(k)\\
&& B_{mmm} = f_c^3 B_{ccc} + f_c^2 f_{\nu} B_{cc\nu} + f_c f^2_{\nu} B_{\nu\nu c} + f^3_{\nu} B_{\nu\nu\nu}.
\een
\ees
Here $P_{cc}$ and $P_{\nu\nu}$ represent the power spectrum cold dark matter and the neutrino componentm where as the $P_{\nu c}$
is the cross spectra between them. We will drop the suffix 3D to avoid cluttering. 
We will only consider the linear order perturbation in $\delta_\nu$ and ignore all higher order contributions which implies $B_{\nu\nu\nu} =0.$
For  $B_{ccc}$ the expression in the squeezed limit is exactly same as derived before.
\bes
\ben
&& B^{\rm 3D, sq}_{ccc} =  \left [ {68 \over 21} -{1 \over 3}{ d k^3 \,P_{cc}(k) \over d\ln k } \right ]P_{cc}(k)P_{cc}(q_3); \\
&& B^{\rm 2D, sq}_{ccc} = \left [ {24 \over 7} -{1 \over 2}{ d k^2 \,P_{cc}(k) \over d\ln k  } \right ]P_{cc}(k_\perp)P_{cc}(q_{3\perp}). 
\een
\ees

We will next consider the mixed terms $B_{\nu\nu c}$ 
These contributions in terms of $\delta_c$ and $\delta_{\nu}$ can be expressed as:
\bes
\ben
&& B_{cc\nu}(\bk_1,\bk_2,\bk_3) =  \langle\delta_c(\bk_1)\delta_{c}(\bk_2)\delta_{\nu}(\bk_3)\rangle +  {\rm cyc.perm.}; \\
&& B_{\nu\nu c}(\bk_1,\bk_2,\bk_3) =  \langle\delta_\nu(\bk_1)\delta_{\nu}(\bk_2)\delta_{c}(\bk_3)\rangle. + {\rm cyc.perm.}. 
\een
\ees
In the above equations the cyc. perm. represent  cyclic permutations of the wave vectors $\bk_1,\bk_2$ and $\bk_3$.

To evaluate $B_{\nu\nu c}$ we expand the terms  perturbatively. Employing tree level perturbation theory, the contributions from  $B_{\nu\nu c, 112}$ are
from these terms:
\ben
B_{\nu\nu c} =  B_{\nu\nu c, 112}(\bk_1,\bk_2,\bk_3) +  B_{\nu\nu c, 112}(\bk_2,\bk_3,\bk_1) + B_{\nu\nu c, 112}(\bk_1,\bk_2,\bk_3).
\een
In our notation, $B_{\nu\nu c, 112}(\bk_1,\bk_2,\bk_3) \equiv \langle \delta_\nu^{(1)}(\bk_1)\delta_\nu^{(1)}(\bk_2)\delta_c^{(2)}(\bk_3)\rangle$
and similarly for the other terms. In tems of the second-order kernels $F_2(\bk_1,\bk_2)$ we have:
\ben
&& B_{\nu\nu c,112}(\bk_1,\bk_2,\bk_3)= 2F_2(\bk_1,\bk_2) P_{\nu c}(k_1) P_{\nu c}(k_2)  \, ; 
\een
The other terms can be recovered by cyclic permutation of the wave number. In the squeezed limit we have:
\bes
\ben
&& B^{\rm 3D, sq}_{\nu\nu c} =  \left [ {68 \over 21} -{1 \over 3}{ d \ln k^3 \,P_{\nu c}(k) \over d\ln k } \right ]P_{\nu c}(k)P_{\nu c}(q_3); \\
&& B^{\rm 2D, sq}_{\nu\nu c} = \left [ {24 \over 7} -{1 \over 2}{ d \ln k^2 \,P_{\nu c}(k) \over d\ln k  } \right ]P_{\nu c}(k_\perp)P_{\nu c}(q_{3\perp}). 
\een
\ees
Finally we turn to $B_{cc \nu}$. The perturbative contributions are as follows:
\ben
 B^{\rm 3D, sq}_{\nu\nu c}(\bk_1,\bk_2,\bk_3) = 2[ F_2(\bk_1,\bk_2) P_{cc}(k_1)P_{c\nu}(k_2) + {\rm cyc.perm.} ]
 \een
Going through an elaborate algebraic manipulation we arrive at the squeezed limi:
\ben
&& B^{\rm 3D, sq}_{\nu\nu c}(\bk_1,\bk_2,\bk_3) = \left [ {34 \over 21} -{1\over 6} {d \ln k^3 P_{cc}(k) \over d \ln k } \right ]P_{cc}(k)P_{cc}(q_3) \nonumber \\
&& \hspace{2cm} + \left [ {34 \over 21} -{1\over 6} {d k^3 P_{c\nu}(k) \over d \ln k } \right ]P_{c\nu}(k)P_{cc}(q_3)
\een
The corresponding 2D expression can be derived by replacing ${34/21}$ with $6/7$ and the wave vectors $\bk$ and $\bq$
with their components perpendicular to libe of sight i.e. $\bk_\perp$ and $\bq_\perp$. 
%
%
%
%
%
%
%
%
%
%
%
%
%
%
%
%
%
%
%
%
\subsection{Clustering Quintessence}
%
%
The quintessence \citep{Q_review}
is the most popular dynamics of dark enegy in which the
potential energy of a single scalar field drives the accelerated
expansion of the Universe.  The quintessence model is different compared
to the cosmological constant scenario as the tempral dependence of the observables
can have a different temporal evolution. The scalar field in
most quintessence models is considered homegeneous and
is typically minimally coupled. The sound speed of the scalar field
in these models is same as the speed of light which prevents
any clustering below the horizon scale. However, extensions
of such models with vanishing or lower than speed of light
have also been considered. These models are known as the clustering quintessence models\citep{Sefusatti, Spherical_DE}.
The future large scale structe surveys
can be used to differentiate the two scenarios. We use the results derived above to derive the changes in
the bispectrum in squeezed limit in these models.
We quote the expression of the kernel $F_2$ from Ref.\citep{Sefusatti}:
\bes
\ben
&& {D_+ \over a} = {5 \over 2} \Omega_{\rm M} \left [ {\Omega_M}^{4/7} + {3 \over 2} \Omega_M
  + \left ({1 \over 70}- {1+w \over 4} \right ) \Omega_Q \left ( 1 + {\Omega_{\rm M} \over 2} \right )\right ]^{-1} \\
&& F_2(\bk_1,\bk_2) = {\nu_2 \over 2} +
(1 -\epsilon){\mu_{}} \left ( {k_1\over k_2} + {k_2 \over k_1}\right ) 
- {1 \over 2} \left ( 1 - \epsilon -{\nu_2 \over 2}\right ) \left [1 - 3  {\mu^2_{}} \right ];
\een
\ees
Here, $\Omega_Q$ and $\Omega_M$ are the density paramter related to Quintessence and dark matter.
The corresponding linear growth rates are denoted by $D_{Q+}$ and $D_+$.
The paramter $\epsilon = {\Omega_Q \over \Omega_M} {D_{Q,+} \over D_{+}}$ and $\nu_2$
can also be expressed in terms of $\Omega_Q$ and  $\Omega_M$ and are function of redshift $z$.
We will treat them as free paramters and derive a formal expression for the IB in these
models.
\bes
\ben
&& B^{\rm 3D}_{sq}(k) = 2\left [ \nu_2 - (1-\epsilon) +  - {1\over 3}(1-\epsilon)n \right ]P_{\rm 3D}(k)P_{\rm 3D}(q_3)\\
&& B^{\rm 2D}_{sq}(k_\perp) = 2\left [ {5 \over 6} \nu_2 - {1\over 2} (1-\epsilon)  - {1\over 2}(1-\epsilon)n \right ]
P_{\rm 3D}(k_\perp)P_{\rm 3D}(q_{3\perp}).
\een
\ees
The numerical factor within the parenthesis can be derived using 2D spherical dynamics \citep{Bernardeau95}.
Similar approximation can also be used to simplify Eq.(\ref{eq:ib_za_approx}) using
the 2D spherical dynamics in the context of ZA.

\subsection{The General Expressions}
The cases derived above are special cases of the following expressions expression for the bispectrum:
\ben
F_2(\bk_1,\bk_2) = \left [ a + b \mu_{12} \left ({k_1 \over k_2} + {k_1 \over k_2} \right )
+ c \mu^2_{12} \right ] P_{\rm 3D}(k_1)P_{\rm 3D}(k_2); 
\een
In the case of GR we have $a = {5/7}$, $b= 1/2$ and $c={2/7}$.
By going through tediuous but straight forward algebra we arrive at the following expression:
\ben
B_{sq}(k) = 2 P_{\rm 3D}(k)P_{\rm 3D}(q_3) \left [ 2a - b + 2c \musq
  - b  {d\ln P(k) \over d\ln k} \musq
  \right ]
\een
We use the fact that $\langle \muq \rangle =0$. For 3D we have $\langle \musq  \rangle  = {1 / 3}$ and for 2D $\langle \musq \rangle  = {1/2}$
which results in the following expression:
\bes
\ben
&& B^{\rm 3D}_{sq}(k) = 2\left [2a -b +{2 \over 3}c - {b\over 3}n \right ]P_{\rm 3D}(k)P_{\rm 3D}(q_3); \\
&& B^{\rm 2D}_{sq}(k_\perp) = 2\left [2a -b + c - {b\over 2}n \right ]P_{\rm 3D}(k_\perp)P_{\rm 3D}(q_{3\perp});
\een
\ees
The varius cases discussed before are results based on specific choices of parameters $a,b$ and $c$.

\section{A (Very) Brief Summary of Standard Perturbation Theory Results}
\label{sec:brief}
We will briefly quote some results from Standard (Eulerian) Perturbation Theory (SPT)
that are relevant in our context.
The perturbative expansions of the density field $\delta$ and $\Theta$ (divergnce of peculiar velocity) can be 
expressed in terms of kernels  $F_n$ and $G_n$: 
\ben
&& \delta(\bk) = \delta^{(1)}(\bk)+\delta^{(2)}(\bk)+ \cdots; \nonumber \\
&& \delta^{(n)}(\bk) = \int d^3 \bq_1 \cdots \int  d^3 \bq_n F_n(\bq_1,\cdots, \bq_n) \delta(\bq_1)\cdots \delta(\bq_n).\\
&& \Theta(\bk) = \Theta^{(1)}(\bk)+\Theta^{(2)}(\bk)+ \cdots; \nonumber \\
&& \Theta^{(n)}(\bk) = \int d^3 \bq_1 \cdots \int  d^3 \bq_n G_n(\bq_1,\cdots, \bq_n) \delta(\bq_1)\cdots \delta(\bq_n).
\een
The expressions for the $n$th order kernels
$F_n$ and $G_n$ for $\delta$ and $\Theta$ respectively are Ref.\citep{review}:
\ben
&& F_n({\bf q}_1, \cdots, {\bf q}_n) =
\sum_{m=1}^{n-1}{G_m({\bf q}_1,\cdots, {\bf q}_m) \over (2n+3)(n-1)}
[(2n+1)\alpha({\bf k}_1,{\bf k}_2) F_{n-m}({\bf q}_{m+1},\cdots,{\bf q}_{n})\nonumber \\
&& \hspace{4cm} + 2\beta({\bf k}_1,{\bf k}_2)G_{n-m}({\bf q}_{m+1}, \cdots, {\bf q}_n)].
\label{eq:kerF} \\
&& G_n({\bf q}_1, \cdots, {\bf q}_n) =
\sum_{m=1}^{n-1}{G_m({\bf q}_1,\cdots, {\bf q}_m) \over (2n+3)(n-1)}
[3\alpha({\bf k}_1,{\bf k}_2) F_{n-m}({\bf q}_{m+1},\cdots,{\bf q}_{n}) \nonumber \\
&& \hspace{4cm} + 
2n\beta({\bf k}_1,{\bf k}_2)G_{n-m}({\bf q}_{m+1}, \cdots, {\bf q}_n)].
\label{eq:kerG}
\een
Here $F_1=1$ and $G_1=1$ and the functions $\alpha$ and $\beta$ are defined as:
\ben
\alpha({\bf k}_1,{\bf k}_2) \equiv {{\bf k}_{12} \cdot {\bf k}_1 \over k_1^2};
\;\;\;
\beta({\bf k}_1,{\bf k_2}) \equiv k^2_{12} {{\bf k}_1 \cdot {\bf k}_2 \over 2k_1^2 k_2^2}.
\label{eq:alphabeta}
\een
We have defined the following quantities above:
\ben
{\bf k}_1 = {\bf q}_1+ \cdots + {\bf q}_m; \;\;\; 
{\bf k}_2 = {\bf q}_{m+1} + \cdots + {\bf q}_n; \;\;\;
{\bf k}= {\bf k}_1 + {\bf k}_2.
\een
The vertices $F_n$ for the lowest order Lagrangian Perturbation Theory (LPT) or
ZA take the following form:
\ben
&& F_n({\bf q}_1,\cdots,{\bf q}_n) = {1 \over n!} {{\bf k}\cdot {\bf q}_1 \over q_1^2}\cdots {{\bf k}\cdot {\bf q}_n \over q_n^2};\quad\quad {\bf k} \equiv {{\bf q}_1+\cdots + {\bf q}_n}.
\label{eq:za_approx}
\een
The angular averages of the kernels are the tree-levels amplitudes or the vertices as defined below:
\ben
\label{eq:nu}
&& \nu_n \equiv n! \int {d\oh_1 \over 4\pi}\cdots \int {d\oh_n \over 4\pi}
F_n(\bk_1,\cdots \bk_n); \\
&& \mu_n \equiv n! \int {d\oh_1 \over 4\pi}\cdots \int {d\oh_n \over 4\pi}
G_n(\bk_1,\cdots \bk_n).
\label{eq:mu}
\een
Using Eq.(\ref{eq:kerF}) and Eq.(\ref{eq:kerG}) the second and third order kernels are defined as follows:
\bes
\ben
&& F_2({\bf k}_1,{\bf k}_2) \equiv {5 \over 7} + {1 \over 2}\left ( {k_2 \over k_1} + {k_1 \over k_2} \right )\mu_{12}
+ {2 \over 7} \mu^2_{12}
\label{eq:define_F2} \\
&& F_3({\bf k}_1,{\bf k}_2, {\bf k}_3) = {7 \over 18} \alpha(\bk_1,\bk_2)
\left [ F_2({\bf k}_2,{\bf k}_3)+G_2({\bf k}_1,{\bf k}_2) \right ]
\label{eq:define_F3}\nonumber\\
&& \hspace{2.4cm} +{2 \over 18} \beta(\bk_1,\bk_2)
%
%
\left [ G_2({\bf k}_2,{\bf k}_3)
+ G_2({\bf k}_1,{\bf k}_2) \right ].\\
&& G_2({\bf k}_1,{\bf k}_2) \equiv {3 \over 7} + {1 \over 2}\left ( {k_2 \over k_1} + {k_1 \over k_2} \right )\mu_{12}
+ {4 \over 7} \mu^2_{12}
\een
\ees
\noindent
Using the fact that in 3D the angular averages of $\alpha$ and $\beta$ 
are respectively $\langle\alpha \rangle =1$ and $\langle \beta \rangle ={1 \over 3}$ we obtain:
\bes
\ben
&& \nu_2\equiv 2\langle F_2 \rangle = 2\left [{5 \over 7}+{2 \over 7}{1 \over 3}\right ]= {34 \over 21}; \quad\quad
\mu_2 \equiv 2\langle G_2 \rangle = 2\left[ {3 \over 7}+{4 \over 7}{1 \over 3}\right ]= {26 \over 21};\quad \\
&& \nu_3\equiv 6\langle F_3 \rangle = 6\left[ {7\over 18}\left ( {17\over 21}+{13 \over 21}\right ) + {4 \over 18}\cdot{1 \over 3}\cdot{13 \over 21} \right] = {682 \over 189}.
\label{eq:vertices}
\een
\ees

Following recursion relation can be derived using Eq.(\ref{eq:kerF}) and Eq.(\ref{eq:kerG}) that is useful in
evaluation of $\nu_n$ and $\mu_n$ results quoted above:
\bes
\ben
&& \nu_n = \sum_{m=1}^{n-1}{n\choose m} {\mu_m \over (2n+3)(n-1)} 
\left [ (2n+1)\nu_{n-m} + {2 \over 3}\mu_{n-m} \right ];\\
&& \mu_n = \sum_{m=1}^{n-1}{n\choose m} {\mu_m \over (2n+3)(n-1)}
\left [ 3\nu_{n-m} + {2 \over 3}n\mu_{n-m} \right ].
\een
\ees
For 2D we use $\langle \alpha \rangle =1$ and $\langle \beta \rangle ={1 \over 2}$;
in this case we have  $\nu_2 \equiv 2{\langle F_2 \rangle}= {12\over 7}$, $\mu_2 \equiv 2{\langle G_2 \rangle}= {10\over 7}$.
Using these expression the recursion relations can easily be derived.

The perturbative bispectrum $B^{\rm PT}(\bk_1,\bk_2,\bk_3)$ and 
trispectrum  $T^{\rm PT}(\bk_1,\bk_2,\bk_3,\bk_4)$ take the following forms:
\ben
&& B^{\rm PT}_{\rm 3D}({\bf k}_1,{\bf k}_2,{\bf k}_3) = 2F_2({\bf k}_1,{\bf k}_2)P_{\rm 3D}(k_1)P_{\rm 3D}(k_2) + 2\; {\rm perm.}; \label{eq:2Dbispec}\\
&& T^{\rm PT}_{\rm 3D}({\bf k}_1,{\bf k}_2,{\bf k}_3,{\bf k}_4) =  
4 \left [ F_2({\bf k}_{13},-{\bf k}_1) F_2({\bf k}_{13},{\bf k}_2)  P_{\rm 3D}(k_{13})P_{\rm 3D}(k_2)P_{\rm 3D}(k_2) + 11\; {\rm perm.} \right ] \nonumber \\
&& \quad\quad\quad\quad + 6 \left [ F_3({\bf k}_1,{\bf k}_2,{\bf k}_3) P_{\rm 3D}(k_1)P_{\rm 3D}(k_2)P_{\rm 3D}(k_3) + 3 {\rm perm.} \right ]
\een
The contribution to the trispectrum have a two different topological structure. For a diagrammatic representation
of these terms see \citep{bernardeau_review}

The terms involving the kernel $F_3$ are represented by a {\em star} diagram, wheras the ones without the kernel $F_3$ are represented as snake diagrams.
Their contribution to the trispectrum are thus also known as the {\em star} or {\em snake} contributions respectively.

The Eulerian perturbation theory breaks down on scales where the average of the two-point correlation function reaches unity $\bar\xi_2 \approx 1$.
In the highly nonlinear regime the Hierarchical Ansatz (HA) is used which has a long history \citep{bernardeau_review}. In this 
approach the angular dependence of the kernels $F_n$ and $G_n$ are assumed to be independent of their angular dependence.
The actual values of the vertices $\nu_n$ and $\mu_n$ depends on specific models. The HA in 2D was discussed in \citep{MunshiBernardeau}.
%
%
%
%
%
%
%
%
%
\section{Weak Lensing Trispectrum}
\label{sec:Tri}
In this section we will consider two different limiting configurations for the
trispectrum (a) Collapsed and (b) Squeezed configuration.
We will derive the weak lensing trispectrum in both of these limiting configuratons.
We will be following the discussion in \citep{Integrated}. However, our focus here is projected or 2D survey.
Though the results are derived keeping weak lensing surveys in mind they will be
of interest to any projected survey.
%
%
\subsection{Collapsed Limit}
  \label{sec:tri_collapsed}
  %
  %
  In the collapsed configuration, one of the diagnoal of the quadrilateral representing
  the trispectrum approaches zero. In this limit, the trispectrum includes contributions only from 
{\em snake} ($F_2^2$) diagrams.
\ben
&& B_3(\bk_1,\bk_2,\bk_3,\bk_4) =
\la\delta(\bk_1)\delta^{(2)}(\bk_2)\delta^{(2)}(\bk_3)\delta(\bk_4)\ra_c
+ \la\delta(\bk_2)\delta^{(1)}(\bk_2)\delta^{(2)}(\bk_3)\delta(\bk_4)\ra_c \nonumber\\
&& \hspace{2cm}+\la\delta(\bk_1)\delta^{(2)}(\bk_2)\delta^{(2)}(\bk_4)\delta(\bk_3)\ra_c
+ \la\delta(\bk_2)\delta^{(1)}(\bk_2)\delta^{(2)}(\bk_4)\delta(\bk_3)\ra_c.
\label{eq:bi}
\een
Following  Eq.(\ref{eq:kerF}), the second-order kernel $\delta^{(2)}(\bk)$:
\ben
\delta^{(2)}(\bk) = \delta_{\rm 3D}(\bk-\bk_{ab})\int\int  {\rm F}_2(\bk_1,\bk_2)\delta^{(1)}(\bk_1)\delta^{(1)}(\bk_2) d^3\bk_1 \, d^3\bk_2 ;\quad \bk_{12} = \bk_1+\bk_2.
\een
Here $\delta_{3\rm D}$ is the 3D Dirac delta-function. Taking an ensemble average leads us to the following expression:
\ben
\la\delta(\bk_1)\delta^{(2)}(\bk_2)\delta^{(2)}(\bk_3)\delta(\bk_4)\ra_c = 
F_2(-\bk_2,\bk_{12})F_2(-\bk_4,-\bk_{12})P_{\rm 3D}(\bk_1)P_{\rm 3D}(\bk_{12})P_{\rm 3D}(\bk_4).
\een
We combine the contributions from all four {\em snake} or $F_2^2$ terms in Eq.(\ref{eq:bi}):
\ben
&& T_{3D}(\bk_1,\bk_2,\bk_3,\bk_4) =
\rP(\bk_{12}) \left [ F_2(-\bk_1,\bk_{12}) \rP(\bk_1) + F_2(-\bk_2,\bk_{12}) {\rP}(\bk_2) \right ]\nonumber \\
&& \hspace{3cm}\times \left [ F_2(-\bk_3,\bk_{34}) \rP(\bk_3) + F_2(-\bk_4,\bk_{34}) \rP(\bk_4)\right ].
\een
The weak lensing Trispectrum $T^\kappa$ is defined as follows:
\ben
\la\kappa({\bl}_{1})\kappa({\bl}_{2})\kappa({\bl}_{3})\kappa({\bl}_{4})\ra 
:= (2\pi)^{2}\delta_{\rm 2D}({\bl}_{1234})T^{\kappa}({\bl}_{1},{\bl}_{2},{\bl}_{3}, {\bl}_4).
\label{eq:fat_sky_trispec}
\een
In our notation ${\bl}_{1\cdots n} = \bl_1+\cdots+\bl_n$.
In terms of 3D trispectrum $T_{\rm 3D}$ the weak lensing trispectrum $T^{\kappa}$ is defined as follows:
\ben
&& T^{\kappa}({\bl}_{1},{\bl}_{2},{\bl}_{3},{\bl}_{3}) = \int_0^{r_s} \xd r\, {\omega^4(r) \over d_A^6(r)}
T_{\rm 3D}\left ({{\bl}_{1} \over d_A(r)},{{\bl}_{2} \over d_A(r),},
{{\bl}_{3} \over d_A(r)}, {{\bl}_{4} \over d_A(r)}; r \right).
\een
This is similar to the definition of power spectrum and bispectrum defined before.

We are now ready to derive the expression for the collapsed trispectrum.
The results will be of practical use in estimation of covariance of 
local power spectrum estimates from survey patches.  
We start with the definition of the local power spectrum in a sub-volume as given in Eq.(\ref{eq:definePS}).
Next, we compute the covariance between the power spectrum at different 
mode $l$ and $l^{\prime}$:
\ben
&& \la \hat \rP^{\kappa}({\bf l},{\bm\theta}_0)\hat \rP^{\kappa}({\bf l}^{\prime},{\bm\theta}_0) \ra_c
= {1 \over \alpha^2}\int {d^3 {\bf l}_1\over (2\pi)^2}
\int {d^3 {\bf l}_2\over (2\pi)^2}\int {d^3 {\bf l}^{\prime}_1\over (2\pi)^2}
\int {d^3 {\bf l}^{\prime}_2\over (2\pi)^2}\nonumber \\
&& \hspace{1cm}\times \la\kappa(\bkl-\bql_1)\kappa(-\bkl-\bql_2)
\kappa(\bkl^{\prime}-\bql^{\prime}_1)\kappa(-\bkl^{\prime}-\bql^{\prime}_2)\ra \nonumber \\
&& \hspace{1cm}\times W_{\rL}(\bql_1)W_{\rL}(\bql_2)W_{\rL}(\bql_1^{\prime})W_{\rL}(\bql_2^{\prime})
\exp[-i{\bf r}_{\rL}\cdot(\bql_{12}+\bql^{\prime}_{12})].
\een
We use the following definition of collapsed trispectrum:
\ben
&& \la\kappa(\bl-\bl_1)\kappa(-\bl+\bl_1+\bl_2)
\kappa(\bl-\bl_1)\kappa(-\bl-\bl^{\prime}_1 + {\bf l}^{\prime}_2 )\ra_c \nonumber \\
&& =(2\pi)^2\delta_{\rm 2D}(\bl_{12}+\bl^{\prime}_{12})
{T}^{\kappa}[{\bf l}-{\bf l}_1, -{\bf l}+{\bf l}_1+{\bf l}_2, 
{\bf l}^{\prime}-{\bf l}_1^{\prime},-{\bf l}^{\prime}+{\bf l}^{\prime}_1+{\bf l}^{\prime}_2]. 
\label{eq:tri1}
\een
In the collapsed limit the trispectrum takes the following form:
\ben
\lim_{\bl_i\rightarrow 0} {T}^{\kappa}[{\bf l}-{\bf l}_1, -{\bf l}+{\bf l}_1+{\bf l}_2, 
{\bf l}^{\prime}-{\bf l}_1^{\prime},-{\bf l}^{\prime}+{\bf l}_1^{\prime}+{\bf l}^{\prime}_2] 
\stackrel{\text{collapsed}}{\approx} {T}^{\kappa}[\bkl, -\bkl, \bkl^{\prime}, -\bkl^{\prime}].
\een
Next, to simplify further, we express the 3D delta function $\delta_{3\rD}$ in Eq.(\ref{eq:tri1}) as a convolution
of two 3D delta function:
\ben
\delta_{3\rD}(\bql_{12}+\bql^{\prime}_{12}) = \int d^3 \bql_3\, 
\delta_{3\rD}(\bql_{12}+\bql_3)\,
\delta_{3\rD}(\bql^{\prime}_{12}-\bql_3).
\een
Using these $\delta_{3\rD}$ functions to collapse the $\bl_2$ and $\bl^{\prime}_2$ integrals:
\ben
&& \la \hat \rP^{\kappa}(\bkl,\bm\theta_0)\hat \rP^{\kappa}(\bkl^{\prime},\bm\theta_0)\ra_c={1 \over \alpha^2} 
\int{d^3 \bql_1 \over (2\pi)^3} \int {d^3 \bql^{\prime}_1\over(2\pi)^3}
\int {d^3 \bql_3\over(2\pi)^3}\nonumber \\
&& \times {T}^{\kappa}[\bkl-\bql_1,-\bkl-\bql_1-\bql_3, 
\bkl^{\prime}-\bql^{\prime}_1,-\bkl^{\prime}-\bql^{\prime}_1+\bql_3]\nonumber  \\
&& \times W_L(\bql_1)W_L(-\bql_1-\bql_3)W_L(\bql_1^{\prime})W_L(\bql^{\prime}_1-\bql_3).
\een
After a tedious but straightforward algebraic manipulation, we get:
\ben
&& {T}^{\kappa}[{\bf l}-{\bf l}_1, 
-{\bf l}+{\bf l}_1+{\bf l}_3, 
{\bf l}^{\prime}-{\bf l}_1^{\prime},-{\bf l}^{\prime}+{\bf l}_1^{\prime}-{\bf l}_3] \nonumber \\
&& \vspace{2cm }=  {R_3}\, {P}^{\kappa}(l){P}^{\kappa}(l^{\prime}){P}^{\kappa}(l_3)
\left [{13 \over 7}+ 
{8 \over 7}\left({{\bf l}\cdot{\bf l}_3 \over l\, l_3}\right)^2
-\left({{\bf l}\cdot {\bf l}_3 \over l\, l_3}\right)^2 {d\ln {P}_{\rm 3D}(l)\over d\ln l}\right ]
      [{\bf l}\rightarrow {\bf l}^{\prime}].
      \label{eq:collapsed_expr}
\een
Here we have defined the projection coefficient $R_3$ which is analogus to $R_2$ defined in Eq.(\ref{eq:define_normIB}):
\ben
R_3  = \int_0^{r_s} \xd\,r {w^4(r)\over d_A^{6+3n}(r)} \bigg / \left ( \int_0^{r_s} \xd\,r{ w^2(r) \over d_A^{2+n}(r)} \right )^3,
\label{eq:def_R3}
\een
We emphasize that, the calculations are done using a flat-sky approximation and power-law power spectrum
is assumed before. The expression in the second bracket in Eq.(\ref{eq:collapsed_expr})
is obtained by replacing ${\bf l}$ with ${\bf l}^{\prime}$.
Next, we perform the angular averaging in the Fourier space and finally upon normalisation we get the following expression:
\ben
&&{\cal T}_{}^{\kappa, \rm coll}(l,l^{\prime}) \equiv
{1 \over P^{\kappa}(l)} {1 \over P^{\kappa}(l^\prime)}{1 \over \sigma_L^2}\int {d\varphi_{l}\over 2\pi}\int {d\varphi_{{l}^{\prime}}\over 2\pi}
\, \nonumber \\
&& \times  {T}^{\kappa}[{\bf l}-{\bf l}_1, 
-{\bf l}+{\bf l}_1+{\bf l}_3, 
{\bf l}^{\prime}-{\bf l}_1^{\prime},-{\bf l}^{\prime}+{\bf l}_1^{\prime}-{\bf l}_3]
\een
The factorisation of the expression in terms of products of two factors that depend individually either on ${\bl}$ or ${\bl}^{\prime}$
allows us to perform the respective angular integration independently. 
Finally, assuming a local power-law for the power spectrum $P_{3D}(k)\propto k^n$, we get:
\ben
&& {\cal T}^{\kappa,\rm coll}(l,l^{\prime})
\equiv  R_3 \, 
\left [2\nu_2 - {1 \over 2} (n+2)\right ]
\left [2\nu_2 - {1 \over 2} (n^{\prime}+2)\right ]; \nonumber \\
&& {d\ln l^2 {P}(l) \over d\ln l} = (n+2); \quad\quad  
\sigma_\rL^2 = {1 \over \alpha^2} \int l\,dl\, { P}^{\kappa}(l)\, W^2_\rL(l).
\label{eq:B21}
\een
\cb{The factorization is a result of flat-sky approximation as well as assumption of a locally power-law profile for the power spectrum.
  The indices $n$ and $n^{\prime}$ are the logartihmic slope of the power spectrum respectively at the wave-numbers $l$ and $l^{\prime}$.
The amplitude $\nu_2={12/7}$ is defined in Eq.(\ref{eq:vertices}).}
Higher order contributions will be ${\cal O}(\sigma_L^4)$. For a reasonable 
big sub-volume such contribution will be negligible.

To recover the results derived for the HA valid in the non-linear regime
we have to set $n=-2$.
%
%
%
\subsection{{Trispectrum In A Doubly Squeezed Configuration}}
\label{sec:tri_squeezed}

In this section we will compute the {\em projected} weak lensing trispectrum in squeezed limit. This is an extension of
the results involving squeezed bispectrum presented earlier. We will consider the contributions from snake ($F_2^2$) and  star ($F_3$)
diagrams.


\noindent
%
\underline{\bf The Contributions From Snake Diagrams:}
%
The following six {\em snake} terms of the total twelve terms contribute in the leading order in the squeezed configuration: 
\ben
\label{eq:snake}
&&\lim_{\bq\rightarrow 0} {T}_{\rm 3D}(\bq,\bk_2,\bk_3,\bk_4) \stackrel{\text{snake}}{=} \lim_{\bq \rightarrow 0}\rP_{\rm 3D}(\bq)\Big \{ P_{\rm 3D}(\bk_2)P_{\rm 3D}(\bk_4)
F_2(-\bk_2,-\bk_4)\left [ F_2(-\bq, \bk_2) + \bk_2\rightarrow \bk_4 \right ]\nonumber\\
&& \hspace{3cm} + \rP_{\rm 3D}(\bk_3)\rP_{\rm 3D}(\bk_4) F_2(-\bk_3,-\bk_4)\left [ F_2(-\bq, \bk_3) + \bk_3 \rightarrow \bk_4 \right ] \Big \} \nonumber \\
&& \hspace{3cm} + \rP_{\rm 3D}(\bk_2)\rP_{\rm 3D}(\bk_3) F_2(-\bk_3,-\bk_2)\left [ F_2(-\bq, \bk_2) + \bk_2\rightarrow \bk_3 \right ] \Big \}\delta_{3\rm D}(\bk_{234}).
\een
Thus the configuration from the snake diagrams in the squeezed trispectrum
takes the form of a bispectrum with a {\em different} vertex amplitude ${\rm F}^{\rm sq}_2$.
\ben
&& {B}^{\prime}_{\rm 3D}(\bk_2,\bk_3,\bk_4) = {\rm F}^{\rm sq}_2(\bk_1,\bk_2)\rP_{\rm 3D}(k_2)\rP_{\rm 3D}(k_3) + \rm cyc. perm.;\\
&& {\rm F}^{\rm sq}({\bk_2,\bk_4})= {\rm F}_2(\bk_2,\bk_4) \left [ F_2(-\bq, \bk_2) + \bk_2 \rightarrow \bk_4 \right ]
\een
For the hierarchical model the vertices are independent of angles between various wave-vectors 
 $F_2(\bk_1,\bk_2)=\nu_2$. In this limit the squeezed trispectrum takes a simpler form
and can be expressed in terms of the hierarchical amplitudes:
\ben
\lim_{\bq\rightarrow 0}{T}_{\rm 3D}(\bq,\bk_2,\bk_3,\bk_4) = 2\nu_2\, \rP_{\rm 3D}(q)\,{B}^{\prime}_{\rm 3D}(\bk_2,\bk_3,\bk_4)
\een
To deduce the doubly squeezed limit $\{\bq, \bq^{\prime}\}\rightarrow 0$ of Eq.(\ref{eq:snake}) we relabel $\bk_2$ with $\bq^{\prime}$.
In this limit $\bk_3=-\bk_4$ which we relabel as $\bk$ and $-\bk$:
\ben
&&\lim_{\bq,\bq^{\prime}\rightarrow 0} {T}_{3\rm D}(\bq,\bq^{\prime},\bk,-\bk) \stackrel{\text{snake}}{=} 
\rP_{\rm 3D}(q)\rP_{\rm 3D}(q^{\prime})\rP_{\rm 3D}(k) \nonumber \\
&& \hspace{3cm} \times\{F_2(\bq,\bk)F_2(-\bk,\bq^{\prime}) + \bq \leftrightarrow \bq^\prime \}.
\label{eq:contribution_snake}
\een
\noindent
\underline{\bf The Contributions From Star Diagrams:}
%
%
Next, we consider the star or $F_3$ contributions.
The following four terms represent the star contributions to trispectrum:
\ben
T_{\rm 3D}(\bk_1,\bk_2,\bk_3,\bk_4) \stackrel{\text{star}}{=} \la \delta^{(3)}(\bk_1)\delta(\bk_2)\delta(\bk_3)\delta(\bk_4)\ra_c + \rm cyc.perm.
\label{eq:tri1_star}
\een
The expression for $\delta^{(3)}(\bk)$ is expressed in terms of the kernel ${\rm F}_3$ defined in Eq.(\ref{eq:define_F3}):
\ben
&& \delta^{(3)}(\bk) = \delta_{3\rm D}(\bk-\bk_{abc})\int d^3\bk_1\,\delta({\bf k}_1) \int d^3\bk_2\,\delta({\bf k}_2) 
\int d^3\bk_3\,\delta({\bf k}_3)\; {\rm F}_3(\bk_1,\bk_2,\bk_3); \nonumber \\
&& \hspace{3cm} \bk_{123}= \bk_1+\bk_2+\bk_3.
\een
We will consider the following following configuration to compute the squeezed limit:
\ben
&& \lim_{\bq_i\rightarrow 0} {T}_{\rm 3D}(\bk_1-\bq_1,\bk_2-\bq_2,\bk_3-\bq_3, -\bq_4)\delta_{\rm 3D}(\bk_{123})\delta_{\rm 3D}(\bq_{1234});\nonumber \\
&& \hspace{3cm} \approx \lim_{\bq_4\rightarrow 0}{T}_{\rm 3D}(\bk_1,\bk_2,\bk_3, -\bq_4)\delta_{\rm 3D}(\bk_{123}).
\een
The Dirac's $\delta_{\rm 3D}$ function in the Fourier domain $\delta_{\rm 3D}(\bk_{1234})$
reduces to $\delta_{\rm 3D}(\bk_{123})$ in the limit $\bq_4 \rightarrow 0$. 
This effectively reduces the trispectrum to a bispectrum to a bispctrum.
The terms that contribute are:
\ben
    {T}_{\rm 3D}(\bk_1,\bk_2,\bk_3,-\bq) \stackrel{\text{star}}{=}
{P}_{\rm 3D}(q)\left [{F}_3(\bk_1,\bk_2,-\bq){P}_{\rm 3D}(k_1){P}_{\rm 3D}(k_2) + {\rm cyc.perm.}\right ].
\een
We have relabeled $\bq_4$ to $\bq$ to simplify the notation.
Of the four terms listed in Eq.(\ref{eq:tri1_star}) only three survive as 
the contribution from the term ${\rm F}_3(\bk_1,\bk_2,\bk_3)$ vanishes due to the presence of
the factor $\delta_{\rm 3D}(\bk_{123})$.
Next, we relabel $\bk_3=\bq^{\prime}$ and to take the doubly squeezed limitwe impose $\bq^{\prime} \rightarrow 0$. 
In the limit $\{\bq, \bq^{\prime}\}\rightarrow 0$ in Eq.(\ref{eq:tri1_star}) $\bk_4=-\bk_3$ which we denote as $\bk=-\bk$ :
\ben
&&\lim_{\bq,\bq^{\prime}\rightarrow 0} {T}_{\rm 3D}(\bq,\bq^{\prime},\bk,-\bk) \stackrel{\text{star}}{=}
\rP_{\rm 3D}(q)\rP_{\rm 3D}(q^{\prime})\rP_{\rm 3D}(k)
\left [F_3(\bq,\bq^{\prime},\bk)+F_3(\bq,\bq^{\prime},-\bk)  \right ].
\label{eq:contribution_star}
\een
%
%
\underline{\bf Total Contribution:}
%
%

Combining expressions from Eq.(\ref{eq:contribution_snake}) 
and Eq.(\ref{eq:contribution_star}) we get in the limit $\{\bq, \bq^{\prime}\}\rightarrow 0$
\ben
&& \lim_{\bq,\bq^{\prime}\rightarrow 0} {T}_{\rm 3D}(\bq,\bq^{\prime},\bk,-\bk) =
P_{\rm 3D}(q)P_{\rm 3D}(q^{\prime})P_{\rm 3D}(k) \Big [F_3(\bq,\bq^{\prime},\bk)+F_3(\bq,\bq^{\prime},-\bk) \nonumber \\
&& \hspace{2cm} + \{F_2(\bq,\bk)F_2(-\bk,\bq^{\prime}) + \bq \leftrightarrow \bq^{\prime} \}\Big ].
\label{eq:total}
\een


Next, the trispectrum in the squeezed configuration is obtained
by expanding ${T}_{\rm 3D}(\bq,\bq^{\prime},\bk,-\bk)$ and keeping the second order terms in $q/k$ and $q^{\prime}/k$.
Both the $F_2^2$ or the snake-terms as well as the $F_3$ or the star-terms suffer from IR divergences but they cancel each other,
\ben
&& T_{\rm 3D} = \int {d\varphi_q \over 2\pi} \int {d\varphi_{q^\prime} \over 2\pi}
    {T}_{\rm 3D}(\bq_\perp,\bq_\perp^\prime,\bk_\perp,-\bk_\perp) \nonumber \\
&& \stackrel{\text{tree}}{=} \left [ {872 \over 49}  - {71 \over 28}{d\ln P_{\rm 3D}(k_\perp) \over d \ln k_\perp}
  + {1\over 4}{k_\perp^2 \over P_{\rm 3D}(k_\perp)}{d^2 {P}_{\rm 3D}(k_\perp) \over dk_\perp^2}
  \right ] P_{\rm 3D}(q_\perp)P_{\rm 3D}(q_\perp^{\prime}).
\een
To arrive at this result all angular averaging is performed in 2D as before. The vectors $k$, $q$ and $q^\prime$
are considered 2D. In effect this replaces all vectors with its line of sight projections i.e. we
write $\bq =\bq_{\parallel}+ \bq_{\perp}$ and assume $\bq_{\parallel}\ll \bq_{\parallel}$. Hence, $\bq \approx \bq_{\perp}$
and similarly for $\bq$ and $\bk$. This is a result of our small angle approximation.
We will next use the following expression:
\ben
    {k_\perp^2 \over P_{\rm 3D}(k_\perp)}{d^2 P_{\rm 3D}(k_\perp)\over dk_\perp^2} =
    \left [{d^2\ln P_{\rm 3D}(k_\perp) \over d (\ln k_\perp)^2} -{d\ln P_{\rm 3D}(k)\over d\ln k} +
      \left ( d\ln P_{\rm 3D}(k)\over d\ln k\right)^2\right ].
\een
To convert the above Lagrangian trispectrum ${T}_{\rm 3D}^{}$ to Eulerian frame denoted as ${T}_{\rm 3D}^{\rm E}$ we use the following transformation in Eq.(4.1) of Ref.\citep{Wagner} :
\ben
{T}_{\rm 3D}^{\rm E} = {T}^{}_{\rm 3D} - 2\,f_2\,{B}_{\rm 3D}; \quad f_2={17 \over 21}.
\een
The subscript 3D in $T_{\rm 3D}$ can be confusing. Though the 3D expressions are used the angular averaging
is done in 3D hence the numerical values are not the same.
On simplification we arrive at the following expression:
\ben
 {T}_{\rm 3D}^{\rm E} = &&  \Bigg [ {872 \over 49} - {78 \over 28}\left [ {d\ln k_\perp^2 P_{\rm 3D}(k_\perp)\over d\ln k_\perp} -2 \right ]+
   {1 \over 4}\left [{d\ln k_\perp^2 P_{\rm 3D}(k_\perp)\over d\ln k_\perp} -
     2 \right ]^2 + {1 \over 4}{d^2\ln k_\perp^2 P_{\rm 3D}(k_\perp)\over d(\ln k_\perp)^2}  \nonumber \\
   && -2 {17 \over 21}\left ( {24\over 7} - {1\over 2}{d\ln k_\perp^2 P_{\rm 3D}(k_\perp) \over d \ln k_\perp} \right )  \Bigg ]
 P_{\rm 3D}(q_\perp)P_{\rm 3D}(q_\perp^{\prime}).
\label{eq:double}
\een
Notice that for $n=-2$ for $T_{3D}$ and  $T^E_{3D}$ the above expression reduces to:
\bes
\ben
\label{eq:r_3}
&& {T}_{\rm 3D}^{\rm E} \stackrel{\text{n=-2}}{=}  T^{\rm E}_{\rm 3D} -2\cdot {17 \over 21}\cdot {24 \over 7} = {922\over 49}
\een
\ees
As an aside, we notice that using HA which is valid in the highly nonlinear regim we recover:
\ben
\lim_{\bq,\bq^{\prime}\rightarrow 0}  T_{\rm 3D}[\bq_\perp,\bq_\perp^{\prime},\bk_\perp,-\bk_\perp] \stackrel{\text{\rm HA}}{=}
(4\nu_2^2+2\nu_3) \rP_{\rm 3D}(q_\perp)\rP(q^{\prime})P_{\rm 3D}(k_\perp).
\label{eq:my_sq}
\een

Next we use the doubly squeezed trispectrum derived in Eq.(\ref{eq:double}) to compute the weak lensing trispectrum in the same limit:
\ben
&& T^{\kappa}({\bl}_{q},{\bl}_q^{\prime},{\bl}_{},-{\bl}_{}) = \int_0^{r_s} \xd r\, {\omega^4(r) \over d_A^6(r)}
T_{\rm 3D}\left ({{\bl}_{q} \over d_A(r)},{{\bl}^{\prime}_{q} \over d_A(r)},
{{\bl}_{} \over d_A(r)}, -{{\bl}_{} \over d_A(r)}; r \right).
\label{eq:kappa_double_sq}
\een
In addition to the flat-sky approximation,
if we assume a power law power spectra $\rP_{\rm 3D}(k)\propto k^n$ then we have ${d\ln k^2 P_{\rm 3D}(k)\over d\ln k} = (n+2)$ 
and the term involving the second derivative vanishes. 
%
%
\bes
\ben
&& {\cal T}^\kappa(l) = {1\over P^\kappa(l_q)} {1\over P^{\kappa}(l_q^\prime)} {1\over P(l)}{T}^{\kappa}_{} = R_3
\left [ {1194 \over 49} - {53 \over 14} (n+2) + {1 \over 4} (n+2)^2 \right ] \\ 
&& {\cal T}^\kappa(l) = {1\over P^\kappa(l_q)} {1\over P^{\kappa}(l_q^\prime)} {1\over P(l)}{T}^{\kappa,E}_{} =  R_3 \left [ {922 \over 49} - {125 \over 42} (n+2) + {1 \over 4} (n+2)^2 \right] 
\een
\ees
The two expression correspond to whether we use the Lagrangian or Eulerian 3D trispectrum to compute the weak lensing trispectrum in Eq.(\ref{eq:kappa_double_sq}).
The prefactor is given in Eq.(\ref{eq:def_R3}). 

In the literature these quantites are also known as the response functions. The ${\cal T}^\kappa(l)$ are thus
the third-order response function for projected weak lensing convergence maps.
The second-order response function is ${\cal B}^{\prime}$  defined in Eq.(\ref{eq:define_normIB}).
It is possible to extend
these hierarchy beyond third-order to an arbitrary order. The results will be presented elsewhere.
%
\section{Approximate Line-of-Sight Integration}
\label{sec:los}
%
In this appendix we discuss how the line-of-sight integrals can be approximated that leads to
separation of projection effects and the dynamical effect.
It was shown in \citep{MunshiJain1,Valageas} that the scaled convergence field $\kappa/|\kappa_{\rm min}|$ (see Eq.(\ref{eq:how2computeIB}) for a definition)
statistically behaves exactly like the underlying density contrast $\delta$:
Thus $\kappa_{\rm min}$ is the minimum value $\kappa$ can reach when $\delta=-1$ (representing the void regions)
along the entire line-of-sight in Eq.(\ref{eq:def_kappa}). Notice $\kappa_{\rm min}$ doesn't depend on
smoothing and it is a global quantity.
This was checked in great detail against 
simulations as was found to be extremely successful in describing cumulants, cumulant correlators
\citep{Munshi2} as well as the Minkowski Functionals \citep{Taruya}.

In the context of one-point probability density function (PDF),
the normalized cumulant $S_3$ of the underlying density contrast $\delta$ defined as
$S_3:=\la \delta^3\ra_c/\la \delta^2\ra_c^2$
is related to that of $\kappa$ defined as $S^\kappa_3:=\la \kappa^3\ra_c/\la\kappa^2\ra_c^2$.
The exact relation can be derived using the line-of-sight integration. 
However, using the $\delta \rightarrow \kappa/|\kappa_{\rm min}|$ mapping an approximate
but very accurate and simple result can be obtained $S_3^\kappa = S_3/|\kappa_{\rm min}|$.
Similarly, for the two-point probability density function, the third-order statistics
${\rm C}_{21} := \la \delta_1^2\delta_2 \ra_c/\la\delta_1\delta_2\ra_c$ (in our notation $\delta_1 := \delta(\bx_1)$)
for $\kappa$ and $\delta$, denoted as ${\rm C}_{21}$ and ${\rm C}^{\kappa}_{21}$,  can be related
in an analogous manner i.e. ${\rm C}^{\kappa}_{21} = {\rm C}_{21}/|\kappa_{\rm min}|$.

This is consistent with our derivation for the IB which is also a third-order statistics. 
We have used the mapping $\delta \rightarrow {\kappa/|\kappa_{\rm min}|}$ and resulting
$R_2 \rightarrow 1/|\kappa_{\rm min}|$ in simplifying the theoretical predictions in
\textsection{\ref{sec:model}}. To see how
this approximation can simplify the analytical calculation in \textsection\ref{sec:model}
we note that if we approximate the numerator and the denominator
of Eq.(\ref{eq:ib_def}) defining $R_2$ as $\int_0^{r_s} dr f(r) \approx r_sf(r_c)/2 $ ($r_c$ is the value
of $r$ for which the integrand $f(r)$ reaches its maximum value) we get
$R_2 \approx {1 /[2r_c w(r_c)]}$ which is same as $R_2 \approx {1 / |\kappa_{\rm min}|}$.
So Eq.(\ref{eq:define_normIB}) can be approximated as:
\ben
{\cal B}^{\prime}(l_{}) \approx {1 \over |\kappa_{\rm min}|} \left [ {24 \over 7} - {1 \over 2} {d\ln l_{}^2 P_{\rm 3D}(l_{}) \over d\ln l_{}} \right ].
\een

\end{document}